\documentclass[10pt]{article}
\pdfoutput=1

\newlength{\abstractwidth}
\usepackage{changepage}
\usepackage{cite}
\usepackage{showlabels}
\usepackage{enumitem}
\usepackage{ifpdf}
 \usepackage{lmodern}
\usepackage[T1]{fontenc}
\usepackage[ansinew]{inputenc}
\usepackage{amsthm}
\usepackage{amsmath}
\usepackage{relsize}
\usepackage{amssymb}
\usepackage{comment}
\usepackage{tikz}
\usetikzlibrary{tikzmark}
\usepackage{graphicx}
\usepackage{color}
\definecolor{darkblue}{cmyk}{0.9,0.9,0,0}
\definecolor{darkgreen}{rgb}{0,0.55,0}
\definecolor{vert}{rgb}{0.1367,0.543,0.1367}
\usepackage[colorlinks=true,linkcolor=darkblue,citecolor=darkblue,urlcolor=darkblue]{hyperref}
\usepackage{epsfig}
\usepackage{epstopdf}
\usepackage{mathtools}
\usepackage{graphicx}

\usepackage{empheq} 
\usepackage{url}

\makeatletter
\long\def\@makecaption#1#2{
  \vskip\abovecaptionskip
  \sbox\@tempboxa{{\captionfonts #1: #2}}
  \ifdim \wd\@tempboxa >\hsize
    {\captionfonts #1: #2\par}
  \else
    \hbox to\hsize{\hfil\box\@tempboxa\hfil}
  \fi
  \vskip\belowcaptionskip}
\makeatother

\def\llangle{\langle\!\langle}
\def\rrangle{\rangle\!\rangle}
\def\tls{{\tilde\lambda}_{\mathsf{S}}}
\def\ii{{instanton-anti-instanton~}}
\def\ints{\int_{\Re s=\half} ds\,}
\def\np{{non-perturbative~}}
\def\sl{SL(2,\Z)}
\def\tl{\tilde\lambda}

\def\ls{\l_{{\mathsf{S}}}}
\def\fp{f_{\rm p}}

\def\fnp{f_{\rm np}}
\def\s{\sigma}

\def\ni{\noindent}

\def\L{\Lambda}
\def\vol{{\rm vol}}

\def\im{\rm Im~}
\def\z{\zeta}

\renewcommand{\thanks}[1]{\footnote{#1}}
\newcommand{\starttext}{
\setcounter{footnote}{0}
\renewcommand{\thefootnote}{\arabic{footnote}}}

\newcommand{\bea}{\begin{eqnarray}}
\newcommand{\eea}{\end{eqnarray}}
\newcommand{\be}{\begin{eqnarray}}
\newcommand{\ee}{\end{eqnarray}}
\newcommand{\bma}{\begin{matrix}}
\newcommand{\ema}{\cr\end{matrix}}
\newcommand{\<}{\langle}
\renewcommand{\>}{\rangle}

\def\cB{{\cal B}}

\def\cE{{\cal E}}
\def\cF{{\cal F}}
\def\cG{{\cal G}}

\def\cM{{\cal M}}
\def\cN{{\cal N}}

\def\cQ{{\cal Q}}

\def\cT{{\cal T}}

\def\cV{{\cal V}}

\def\mM{\mathfrak{M}}

\def\mf{\mathfrak{f}}

\def\ZZ{{\mathbb Z}}
\def\RR{{\mathbb R}}

\def\CC{{\mathbb C}}
\def\HH{{\mathbb H}}

\def\Re{{\rm Re \,}}
\def\Im{{\rm Im \,}}

\def\half{{1\over 2}}
\def\p{\partial}

\def\a{\alpha}

\def\b{\beta}
\def\g{\gamma}
\def\G{\Gamma}
\def\l{\lambda}

\def\s{\sigma}

\def\({\left(}
\def\){\right)}
\def\[{\left[}
\def\]{\right]}

\def\<{\langle}
\def\>{\rangle}

\def\min{{\rm min}}

\def\qb{\overline q}

\def\min{{\rm min}}

\def\Z{\mathbb{Z}}
\def\R{\mathbb{R}}

\def\qb{\bar q}
\def\tb{\bar\tau}

\def\x{\times}
\def\cB{\mathcal{B}}
\DeclareMathOperator*{\Res}{Res}

\def\bul{$\bullet$~}

\def\1{{\rm 1-loop}}

\def\c{\cite}

\def\cM{\mathcal{M}}

\def\zb{\bar z}
\def\cG{\mathcal{G}}
\def\cN{\mathcal{N}}

\def\vs{\vskip .1 in}

\def\G{\Gamma}
\def\p{\partial}
\def\o{\over}
\def\g{\gamma}
\def\D{\Delta}
\def\rar{\rightarrow}

\def\eqr{\eqref}

\def\O{{\cal O}}
\def\ra{\rangle}
\def\la{\langle}
\def\ssec{\subsection}
\def\sssec{\subsubsection}
\def\sec{\section}
\def\i{\infty}
\def\foot{\footnote}
\newcommand{\es}[2] {\begin{equation} \label{#1} \begin{split} #2 \end{split} \end{equation}}
\newcommand{\e}[2] {\begin{equation} \label{#1} #2 \end{equation}}

\def\t{\tau}

\def\tb{\bar\tau}

\newcommand\calo{\mathcal{O}}
\DeclareMathOperator{\iim}{Im}
\DeclareMathOperator{\rre}{Re}

\DeclareMathOperator{\vvol}{vol}
\DeclareMathOperator{\Discc}{Disc}
\DeclareMathOperator{\LLi}{Li}
\DeclareMathOperator{\csch}{csch}

\usepackage{varioref}
\usepackage{makeidx}
\makeindex

\usepackage[english]{babel}
\usepackage{caption}
\usepackage{subfig}

\usepackage{tabularx}
 \renewcommand{\baselinestretch}{1.03}
 
\begin{document}
\starttext
\thispagestyle{empty}

\begin{flushright}
\end{flushright}

\vskip 1in

\begin{center}

{\Large \bf Harnessing S-Duality in $\mathcal{N}=4$ SYM}
\vskip .07 in
{\Large \bf \&}
\vskip .07 in
{\Large \bf Supergravity as $SL(2,\mathbb{Z})$-Averaged Strings}

\vskip 0.3in

{ Scott Collier$^{\circ}$, Eric Perlmutter$^{\bullet}$} 
   
\vskip 0.15in

{\small $\circ$ Princeton Center for Theoretical Science, Princeton University, Princeton, NJ 08544, USA}

{\small $\bullet$ Universit\'e Paris-Saclay, CNRS, CEA, Institut de Physique Th\'eorique, 91191, Gif-sur-Yvette, France}

\vskip 0.15in

{\tt \small scott.collier@princeton.edu, perl@ipht.fr}

\vskip 0.4in

\begin{abstract}
\vskip 0.1in

We develop a new approach to extracting the physical consequences of S-duality of four-dimensional $\mathcal{N}=4$ super Yang-Mills (SYM) and its string theory dual, based on $SL(2,\mathbb{Z})$ spectral theory.

We observe that CFT observables $\mathcal{O}$, invariant under $SL(2,\mathbb{Z})$ transformations of a complexified gauge coupling $\tau$, admit a unique spectral decomposition into a basis of square-integrable functions. This formulation has direct implications for the analytic structure of $\mathcal{N}=4$ SYM data, both perturbatively and non-perturbatively in all parameters. These are especially constraining for the structure of instantons: $k$-instanton sectors are uniquely determined by the zero- and one-instanton sectors, and Borel summable series around $k$-instantons have convergence radii with simple $k$-dependence. In large $N$ limits, we derive the existence and scaling of non-perturbative effects, in both $N$ and the 't Hooft coupling, which we exhibit for certain $\mathcal{N}=4$ SYM observables. An elegant benchmark for these techniques is the integrated stress tensor multiplet four-point function, conjecturally determined by \cite{Dorigoni:2021guq} for all $\tau$ for $SU(N)$ gauge group; we derive and elucidate its form, and explain how the $SU(2)$ case is the simplest possible observable consistent with $SL(2,\mathbb{Z})$-invariant perturbation theory.

These results have ramifications for holography. We explain how $\langle\mathcal{O}\rangle$, the ensemble average of $\O$ over the $\mathcal{N}=4$ supersymmetric conformal manifold with respect to the Zamolodchikov measure, is cleanly isolated by the spectral decomposition. We prove that the large $N$ limit of $\langle\mathcal{O}\rangle$ equals the large $N$, large 't Hooft coupling limit of $\mathcal{O}$. Holographically speaking, $\langle\mathcal{O}\rangle = \O_{\rm sugra}$, its value in type IIB supergravity on AdS$_5 \times S^5$. This result, which extends to all orders in $1/N$, embeds ensemble averaging into the traditional AdS/CFT paradigm. The statistics of the $SL(2,\mathbb{Z})$ ensemble exhibit both perturbative and non-perturbative $1/N$ effects. We discuss further implications and generalizations to other AdS compactifications of string/M-theory.

\end{abstract}                                            
   
\end{center}

\newpage

\pagenumbering{roman}

\baselineskip .12 in
\setcounter{tocdepth}{2}

\tableofcontents

\setlength{\textheight}{8.7in}

\newpage

\pagenumbering{arabic}
\setcounter{page}{1}

\numberwithin{equation}{section}

\baselineskip=14pt
\setcounter{equation}{0}
\setcounter{footnote}{0}

\sec{Introduction}\label{sec:intro}

\quad \, This paper pursues two intertwined endeavors. 

The first is to understand how to extract the full implications of S-duality for the observables of superconformal field theories, focusing specifically on four-dimensional $\cN=4$ super Yang-Mills theory.  

The second is to reframe the AdS/CFT Correspondence, namely the original duality between $\cN=4$ super Yang-Mills and type IIB string theory on AdS$_5 \x S^5$, by applying new lessons from S-duality. 

The $\cN=4$ super Yang-Mills (SYM) theory is, at risk of stating the obvious, a beautiful theory from myriad points of view. Maximal supersymmetry imposes rigid structure and regularity of CFT data, yet provides a route to computing certain observables that coincide with those in less supersymmetric theories. For a finite number of colors $N$, perturbative gauge theory calculations are complemented by modern bootstrap and supersymmetric localization methods, providing rigorous and sometimes exact results for local and non-local quantities. In the large $N$ 't Hooft limit, new symmetries emerge, leading to integrability solutions at the planar level and beyond. Not to be forgotten, the holographic correspondence with type IIB string theory on AdS$_5 \x S^5$ furnishes our most explicit definition of a theory of quantum gravity, shedding light on both sides of the duality.

$\cN=4$ SYM also enjoys S-duality \cite{Montonen:1977sn,Witten:1978mh,Osborn:1979tq,Argyres:2006qr}. For simply-laced gauge group $G$ parameterized by the complexified gauge coupling $\t$, this is, up to global identifications, a self-duality: namely, an invariance of the theory under $(P)SL(2,\Z)$ transformations of $\t$.\foot{The global aspects only affect S-duality of non-local observables. For simply-laced $G$, extended objects are invariant under congruence subgroups of $SL(2,\Z)$ \c{Aharony:2013hda}. For non-simply-laced $G$, even local observables are not $SL(2,\Z$)-invariant, but are instead invariant under either a congruence subgroup of $SL(2,\Z)$ or a Hecke triangle group, depending on the choice of $G$ \c{Argyres:2006qr}. We restrict our attention to $SL(2,\Z)$-invariant observables in simply-laced $\cN=4$ SYM, though our techniques may be directly generalized \c{Iwaniec2002SpectralMO}.} Holographically, this symmetry appears as the $\sl$ symmetry of type IIB string theory in AdS$_5 \x S^5$ --- a background that preserves the $\sl$ symmetry of flat space string theory --- where $\t$ is dual to the axio-dilaton. 

There have been numerous checks of S-duality against independent calculations in $\cN=4$ SYM or in type IIB string theory. All have succeeded. The existence of this symmetry, while no less remarkable, seems without question. What are the full implications of S-duality for $\cN=4$ SYM observables? In the presence of symmetries in any physical theory, one should incorporate their effects at the outset of calculations. This will be our approach: to efficiently process S-duality invariance of $\cN=4$ SYM observables, reducing these data to their dynamical content. 

Viewing $SL(2,\Z)$-invariant $\cN=4$ SYM observables $\O(\t)$ --- such as conformal dimensions, OPE coefficients, or correlation functions of superconformal primary operators --- as non-holomorphic functions invariant under $\sl$ transformations of $\t$, we have very limited ``real-world'' information about the form of these functions. Being exactly marginal, $\t$ parameterizes a one-complex dimensional conformal manifold, $\cM$, that preserves the full $\cN=4$ supersymmetry. In terms of real parameters,
\e{}{\t = {\theta\o 2\pi} + {4\pi i\o g^2}}
where $g$ is the Yang-Mills coupling and $\theta$ is the topological theta angle. While perturbative results near $\t= i\i$ (possibly with instanton backgrounds \c{Bianchi:1999ge,Alday:2016tll,Alday:2016jeo,Alday:2016bkq}) give some information, the modular structure of $\O(\t)$ is inherently non-perturbative. S-duality has been input into the construction of interpolating functions across moduli space \cite{Sen:2013oza,Beem:2013hha,Alday:2013bha,Chowdhury:2016hny}, but (as noted there) those interpolating functions lacked a principle for eliminating ambiguity in the chosen function space. There is a shining recent exception to this general paucity of data which we discuss, then derive, at length below. 

Our idea is to introduce certain methods from the mathematics literature that we argue are perfectly tailored to the $\cN=4$ SYM context. There exists a robust {\it spectral theory of $\sl$}, applicable to quantities that are square-integrable on the $\sl$ fundamental domain, which we call $\cF = \mathbb{H}/SL(2,\Z)$. As we will explain, $\cN=4$ SYM observables $\O(\t)$ are in this category. Employing a unique spectral decomposition into an $\sl$-invariant eigenbasis fully incorporates the S-duality symmetry. Determining $\O(\t)$ then boils down to computing its spectral overlaps; specializing this mathematical toolkit to this physical context, the analytic structure of the spectral overlaps is in turn very strongly constrained by the consistency of weak coupling perturbation theory. 

With this framework in place, relatively simple calculations lead to a wealth of information at finite $N$, described in further detail below. This is especially true as regards the structure of instantons. The ensuing calculations look rather different than existing approaches to $\cN=4$ SYM observables. We gain confidence by applying the general formalism to the integrated correlator introduced in \cite{Binder:2019jwn} and studied in detail in \cite{Dorigoni:2021bvj,Dorigoni:2021guq} --- a unique observable that is conjecturally known {\it exactly} as a function of $N$ and $\t$ --- leading to a derivation of their result and a crisp accounting of many of its properties.

Applying this approach at large $N$ opens yet other doors. One may develop the 't Hooft limit of large $N$ and fixed $\l \coloneqq g^2 N$. At $\l\gg1$, the theory is famously dual to type IIB supergravity on AdS$_5 \x S^5$, endowed with a prescribed series of stringy $\a'$ corrections. Perhaps surprisingly, the spectral method has something fundamental to say about holography: in particular, it suggests a new picture of AdS/CFT that unifies the traditional holographic paradigm for UV complete theories with recent ideas on ensemble averaging in lower-dimensional AdS/CFT. 

In the current setting, the ensemble in question is what we call the {\it $\sl$ ensemble}, the space of $\cN=4$ SYM theories living on the conformal manifold $\cM$. The central result, elaborated upon in the description of Section \ref{sec:averageAndSugra} below, can be simply stated: {\it the large $N$ limit of ensemble-averaged $\cN=4$ SYM is the strong coupling limit of planar $\cN=4$ SYM.} This holds at the level of individual observables $\O(\t)$. In bulk terms, type IIB supergravity on AdS$_5 \x S^5$ is both a low-energy limit of type IIB string theory, and the average of type IIB string theory over moduli space. Unlike ensemble-averaged dualities in lower dimensions, the duality between $\cN=4$ SYM and AdS$_5 \x S^5$ string theory applies for every microscopic instance of $\cN=4$ SYM: the ensemble average is an {\it emergent} description of the strongly coupled, planar limit. This equivalence permits a satisfying embedding of various developments involving wormholes, factorization, and ensemble statistics into the quintessential holographic correspondence.

\begin{figure}
\centering
\begin{tikzpicture}[scale=1]
{
  \node at (0,0) {\includegraphics[scale=0.3]{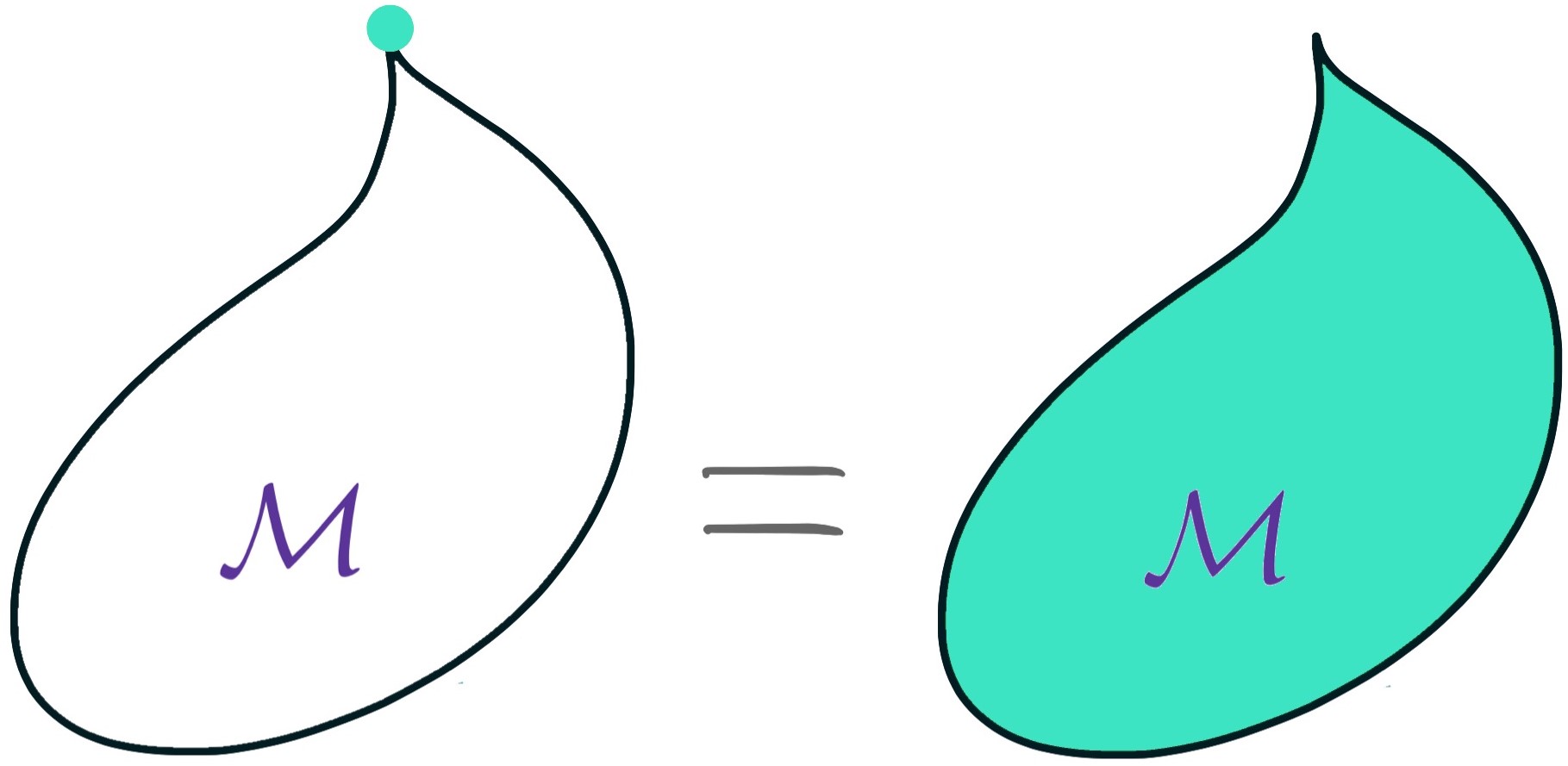}};
  \node at (-3.25,-3.0) {$(\lambda\to\infty)$};
  \node at (2.5,-3.0) {$(N\to\infty)$};
}
\end{tikzpicture}
\caption{A cartoon depicting the two equivalent field theory duals of type IIB supergravity on AdS$_5\times S^5$, phrased in terms of the conformal manifold $\cM$: as the limit of large 't Hooft coupling (depicted by the point approaching the cusp of $\cM$) of planar $\cN = 4$ SYM (left), and as the large $N$ limit of the ensemble average (denoted by the shading) of $\cN=4$ SYM (right). On the left it is understood that the $N\to\infty$ limit is taken first.}\label{fig:MFig}
\end{figure}

Let us now give a slightly more detailed description of our results. 

\begin{figure}[t]
\centering
{
\subfloat{\includegraphics[scale=.45]{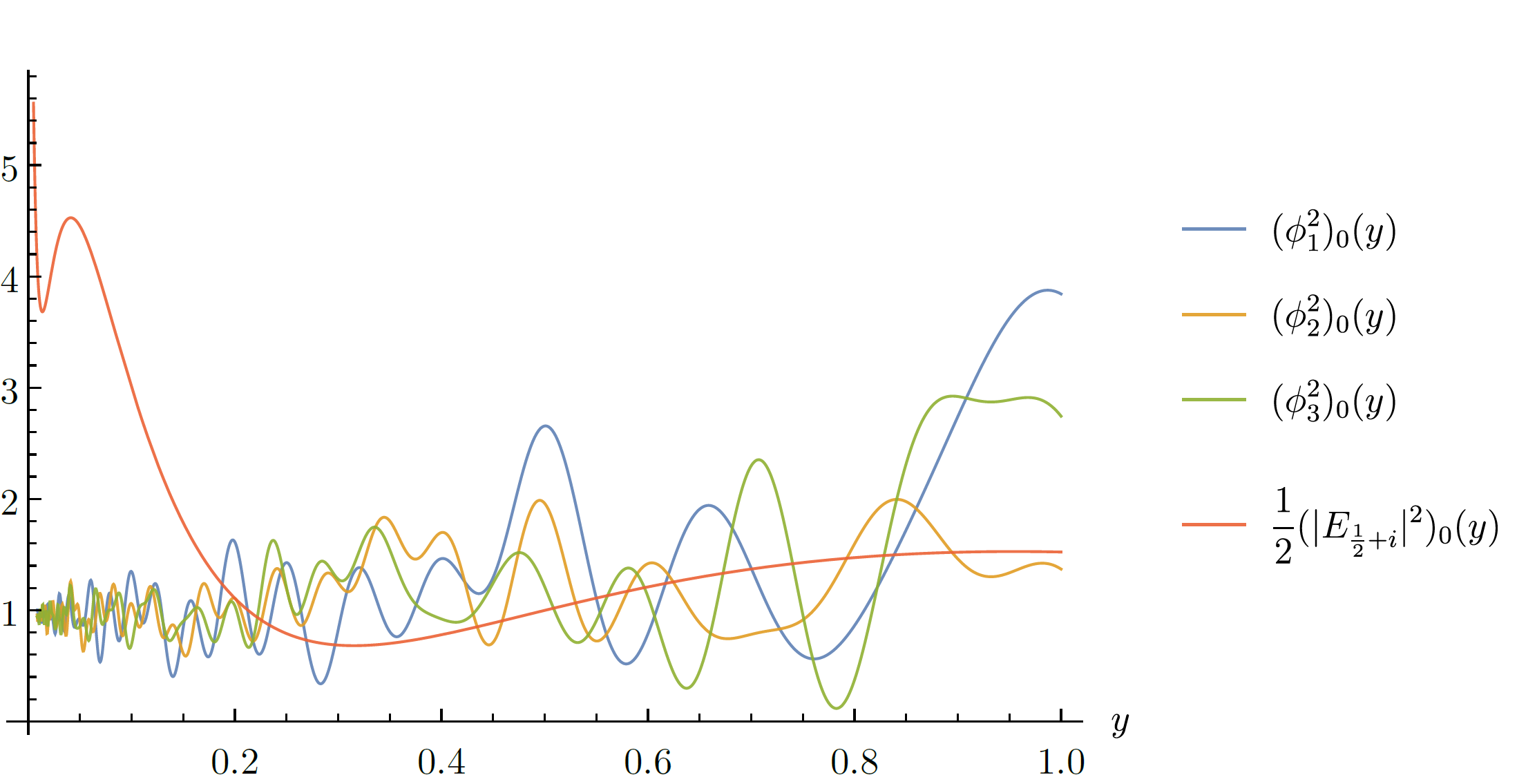}}
}
\caption{A plot of the zero mode of the square of the Eisenstein series $E_{\half+i}(\tau)$ (rescaled by a factor of $1/2$) and of the first three even Maass cusp forms, defined as $(\phi_i^2)_0(y) \coloneqq \int_{-\half}^\half dx\, \phi_i(\tau)^2$, for $0 < y \leq 1$. The data for the cusp forms was obtained using the first 400 Fourier coefficients, known numerically to over 100 digits, available at \c{lmfdb}.
}
\label{fig2}
\end{figure}

In {\bf Section \ref{sec:SL2ZSpectral}}, we begin with a brief, physicist-oriented introduction to the theory of harmonic analysis of $\sl$. The $L^2(\cF)$ eigenspace contains a continuous subspace spanned by non-holomorphic Eisenstein series $E_s(\t)$ with $s=\half+i \mathbb{R}$, and a discrete subspace spanned by an infinite set of Maass cusp forms $\phi_n(\t)$ labeled by $n\in\Z_{\geq 0}$ (where $\phi_0$ is constant). Both are eigenfunctions of the Laplacian on the upper-half plane. The former are well-understood, while the latter are wild objects, of prevailing interest to mathematicians, exhibiting several signals of chaos. We depict this wildness with a numerical plot; see Figure \ref{fig2}. 

In {\bf Section \ref{sec:N4Spectral}}, we develop the spectral decomposition of observables in $\cN=4$ SYM. We explain how, and why, the results of the previous section apply to this setting. A non-perturbatively well-defined observable $\O(\t) = \O(\g\t)$, with $\g\in SL(2,\Z)$, admits the following spectral decomposition:
\begin{equation}\label{sdint}
  \mathcal{O}(\tau) = \overline \calo +  {1\over 4\pi i }\int_{\rre s = \half}ds\, (\calo,E_s)E_s(\tau)  + \sum_{n=1}^\infty (\calo,\phi_n)\phi_n(\tau).
\end{equation}
where $(\cdot\,,\cdot)$ denotes the Petersson inner product. The first term, $\overline \calo$, is a constant, the {\it modular average} of $\O$, defined as the integral of $\O(\t)$ over $\cF$ with the $SL(2,\R)$-invariant measure,
\e{}{\overline\O \coloneqq \vvol(\cF)^{-1}\int_{\cF} {dx dy\o y^2} \O(\t)\,,\quad \t \coloneqq x+iy}
The basic statement of the spectral decomposition is that determining $\O(\t)$ thus reduces to computing its average $\overline \calo$ and the spectral overlaps $(\calo,E_s)$ and $(\calo,\phi_n)$. The spectral overlap $(\calo,E_s)$ is a meromorphic function of the spectral parameter $s$, whose analyticity in the complex $s$ plane is subject to stringent constraints from the consistency of the weak-coupling expansion, well-definedness of the CFT observable in the 't Hooft limit, and the Eisenstein series itself. As an example, the modified overlap $\{\calo,E_s\}\coloneqq (\calo,E_s)/\Lambda(s)$, where $\Lambda(s)$ is the completed Riemann zeta function defined in (\ref{eq:LambdaDefinition}), must satisfy the functional equation $\{\calo,E_s\} = \{\calo,E_{1-s}\}$.

Without further computation one observes two important consequences. The first is a redundancy in instanton physics: $k$-instanton effects are fully determined by $k=0,1$ instanton effects, because the latter can be used to ``invert'' this expansion.\foot{We develop a Fourier expansion $\O(\t) = \O_0(y) + \sum_{k=1}^\i2\cos(2\pi k x)\, \O_k(y)$, where $k$ is the total instanton number.} The second is a suggestive fact about the spectral decomposition. For any $N$ and any $\O(\t)$, we can define the ensemble average $\<\O\>$, as the integral of $\O(\t)$ over $\cM$ with respect to the Zamolodchikov measure on the conformal manifold. Thanks to maximal supersymmetry, {\it the modular average, $\overline \calo$, equals the ensemble average, $\<\O\>$}. Therefore, $\<\O\>$ appears in a very clean way in the spectral decomposition of $\O(\t)$. This foreshadows some results to follow.

In {\bf Section \ref{sec:integratedCorrelatorI}}, we introduce an object that will be a touchstone for our methods throughout: the integrated correlator $\cG_N(\t)$. This is a four-point function of the $\O_{\bf 20'}$ operator in the $SU(N)$ theory, integrated over space with a measure that preserves supersymmetry. Accordingly, $\cG_N(\t)$ may also be represented as a supersymmetric localization integral. In a beautiful paper \cite{Dorigoni:2021guq}, the authors conjectured and thoroughly tested an expression for $\cG_N(\t)$, valid for all $N$ and $\t$, given by a two-dimensional lattice sum over one-dimensional integrals (see \eqr{dgwmain}). Its $N$-dependence is moreover fixed recursively by a ``Laplace difference equation'' (see (\ref{laplacediffeq})). It is a rare situation, even in $\cN=4$ SYM, to have an exact expression for any observable at finite $N$, whose functional complexity (e.g. the number of integrals) is independent of $N$. The evidence collected in \cite{Dorigoni:2021guq} for their conjecture, reviewed later, is abundant. Here we will give the first concrete application of our spectral methods to $\cN=4$ SYM, by writing the spectral decomposition of $\cG_N(\t)$. For $SU(2)$, for example, 
\es{cg2int}{\{\cG_2,E_s\} = {\pi\o \sin\pi s}s(1-s) (2s-1)^2\,,\qquad (\cG_2,\phi_n)=0}
The result for all $N$ is extremely simple, with no cusp form overlap. This demonstrates both how neatly the $\t$-dependence is couched in the eigenbasis, and the non-genericity of $\cG_N(\t)$ in the space of possible observables. 

In {\bf Section \ref{sec:instantonsN4AnalyticStructure}}, we develop the general theory of spectral decomposition of $\cN=4$ SYM observables. We explain how perturbative expansions $g^2\rar 0$, i.e. $y\rar\i$, may be developed by contour deformation of \eqr{sdint}. Because the cusp forms $\phi_n(\t)$ have the special property that their $k=0$ Fourier modes vanish, they do not appear in perturbation theory. This simplifies matters. Indeed, a consistent perturbative expansion --- namely, no logarithms or fractional powers of $g^2$ --- implies that the key player $(\O,E_s)$ must take a simple functional form (\ref{form}), with rigid analytic structure (see the discussion surrounding (\ref{averageFromOverlap})). In particular, there is a clear separation between perturbative and non-perturbative parts of the overlap, i.e. those which contribute power law terms $\sim y^{-n}$ and those which contribute instanton-anti-instanton terms $\sim (q\qb)^n$, where $q\coloneqq e^{2\pi i \t}$. 

We then proceed to derive a bevy of useful results for general $\O(\t)$, first in the simplifying case that $(\O,\phi_n)=0$, later re-integrating $\phi_n(\t)$. Many observables in perturbative gauge theory have Borel summable expansions. We introduce the {\it $\sl$ Borel transform,} which is a Borel transform specifically tailored to the resummation of $\sl$-invariant functions. Using this we show that not only are instantons redundant as described earlier, but the radius of convergence of the $\sl$ Borel transform of the perturbative series around $k$ instantons admits a universal, merely quadratic dependence on $k$ --- see (\ref{Rk}). We also construct $\O(\t)$ as a (regularized) $\sl$ Poincar\'e sum of its zero mode. This representation is equivalent to the spectral decomposition. A non-vanishing $(\calo,\phi_n)$ introduces arithmetic quantum chaos \cite{sarnak} to the $(k\geq1)$-instanton sectors, thus isolating the chaotic parts of the instanton data. We provide a sharp diagnostic (\ref{cuspdiag}) for the presence of cusp forms in terms of the radius of convergence of the Borel transform of the expansion around $k\geq 1$ instantons, relying on a result of Kim and Sarnak \c{kimsarn} toward proving the Ramanujan-Petersson conjecture; in the case that the cusp forms give factorially-divergent contributions to perturbation theory, (\ref{Rk}) generalizes to \eqr{cuspdiag2}, in turn providing a two-way diagnostic.

All of this is then applied to $\cG_N(\t)$, which we can immediately derive and explain as a flagship demonstration of these results. The lattice-integral representation of \cite{Dorigoni:2021guq} is nothing but the Poincar\'e sum representation described above, and the integral kernel (called $B_N$) is the $\sl$ Borel transform of the $y\rar\i$ perturbative expansion of the zero mode, $\cG_{N,0}(y)$. Furthermore, the result \eqr{cg2int} for $\cG_2(\t)$ is seen to be the {\it simplest} possible pair of spectral overlaps consistent with $\sl$-invariant weak coupling perturbation theory. We mean ``simplest'' in the mathematical, and hopefully uncontroversial, sense that $(2s-1)^2$ is the simplest non-constant, entire function that is even in $s\rar 1-s$ (a condition required by the functional equation for the overlap $\{\cG_2,E_s\}$). The overlaps for $SU(N)$ are then determined by the recursion relation of \cite{Dorigoni:2021guq}, or by direct analysis at fixed $N$.  Overall, this illuminates $\cG_N(\t)$ as a truly special object in the space of $\cN=4$ SYM observables, and (in our view) illustrates the clarifying value of the spectral decomposition. 

In {\bf Section \ref{sec:largeN}}, we write the general form of the spectral overlaps at large $N$. 

In {\bf Section \ref{sectH}}, we treat the 't Hooft limit of large $N$ and fixed $\l\coloneqq g^2N$. In this limit, the cusp forms $\phi_n(\t)$ are suppressed non-perturbatively in $N$. Combining the results of Section \ref{sec:instantonsN4AnalyticStructure} with the double scaling leads to the most general form of the $1/\l$ expansion at strong coupling (see (\ref{O0strong2})). We then observe a powerful consequence of S-duality:\footnote{Subject to a technical assumption tested against examples.} the convergence of the weak-coupling expansion $\lambda \ll 1$ of a CFT observable $\O(\t)$ directly leads to the existence of non-perturbative corrections both at strong coupling $\lambda \gg 1$, and at large $N\gg 1$ and finite $\l$. A remarkable imprint of $\sl$ invariance is that in both cases the strength of the non-perturbative corrections is set by the radius of convergence of the weak coupling expansion (see (\ref{nplargeth}) and (\ref{thfull})). While the $\l\gg1$ effects appear at the non-perturbative scale $\sim e^{-\sqrt{\lambda}}$, representing fundamental string worldsheet instantons in AdS$_5 \times S^5$, S-duality implies that the $N\gg1$, fixed $\l$ effects appear at the non-perturbative scale $\sim e^{{-\sqrt{\ls}}}$, where $\ls = (4\pi N)^2/\lambda$ may be thought of as an ``S-dual 't Hooft coupling''; these effects are present, and non-perturbative in $N$, in the ordinary 't Hooft limit. These reflect D-string instanton effects in AdS$_5\times S^5$. We apply both of these predictions to our prototypical example $\cG_N(\tau)$. The non-perturbative effects at strong coupling in the 't Hooft limit are consistent with results previously derived by \cite{Dorigoni:2021guq}. The computation of the D-string instanton effects to $\cG_N(\t)$ from resurgence of the weak-coupling expansion is more novel.

In general CFT, the question of whether the $1/N$ expansion of CFT observables is asymptotic, and whether non-perturbative corrections are needed, is open. The analysis in Section \ref{sectH} shows that under certain general conditions in $\cN = 4$ SYM, the answer to both is affirmative, and holographically implies the non-Borel summability of string perturbation theory on AdS$_5\times S^5$. 

In {\bf Section \ref{sec:VSC}}, we treat the ``very strongly coupled'' (VSC) limit of large $N$ and fixed $g$ \c{Azeyanagi:2013fla}. This interesting regime has recently been studied in \cite{Binder:2019jwn,Chester:2019jas,Chester:2020vyz,Alday:2021vfb}. Unlike in the 't Hooft limit, at finite coupling $\sl$ invariance remains manifest, so the $1/N$ expansion of $\O(\t)$ involves $\sl$-invariant functions at every order. The constraints on the spectral decomposition in the VSC limit are similar to (and can be thought of as inherited from) the 't Hooft limit. As an example, we explore the VSC limit of an integrated correlator different from $\cG_N(\t)$, which we call $\cF_N(\t)$. This one, studied in \cite{Chester:2020dja,Chester:2020vyz} at large $N$, is also supersymmetric. But as quickly becomes evident upon examining its explicit form, $\cF_N(\t)$ is substantially more intricate than $\cG_N(\t)$. Nevertheless, we are able to leverage previous results to determine its spectral overlap --- including its overlap with the cusp forms, $(\cF_N,\phi_n)$ --- to the first few orders in $1/N$.  

In {\bf Section \ref{secfiniteN}}, we pause to flesh out a particularly nice implication of the large $N$ results for {\it finite} $N$ physics. Specifically, we prove that if that strong 't Hooft coupling expansion of a CFT observable $\calo(\t)$ contains integer powers of $1/\lambda$, then it receives non-perturbative, instanton-anti-instanton corrections at finite $N$. This gives an easy diagnostic of finite $N$ non-perturbative physics. As an application, we show that unprotected conformal dimensions in $\cN=4$ SYM receive non-perturbative, instanton-anti-instanton corrections.

In {\bf Section \ref{sec:averageAndSugra}}, we return to the 't Hooft limit and examine the $\l\gg1$ limit in the context of holography. The ensemble average $\<\O\>$ now makes its star turn. This quantity must be computed at finite $N$,\foot{The 't Hooft double-scaling limit focuses on the cusp of the conformal manifold, obscuring the underlying $\sl$ invariance.} but admits a subsequent $1/N$ expansion. From the spectral decomposition, the $\l\gg1$ expansion of $\O(\t)$ takes a general form given in (\ref{O0strong4}). Focusing on the leading term of order $N^2$ for simplicity, that result is 
\e{O0strongint}{\O(\l\gg1) \approx N^2\(\llangle\O^{(0)}\rrangle+ \sum_{m=0}^\i \mathsf{a}_m^{(0)} \l^{-{3+m\o 2}} \)}
where $\mathsf{a}_m^{(0)}$ are coefficients. The quantity $\llangle\O^{(0)}\rrangle$ is the leading large $N$ limit of $\<\O\>$, i.e. $ \lim_{N\rar\i} N^{-2}\<\O\> = \llangle\O^{(0)}\rrangle$. This equation is familiar, but uncanny: it is the strong coupling expansion of $\O$, but with the ensemble average as the leading term! One deduces that at leading order in large $N$,
\es{infavgint}{\O(\l\rar\i) = \<\O\>}
Of course, by the usual AdS/CFT dictionary, $\O(\l\rar\i) = \O_{\rm sugra}$, its value in AdS$_5 \x S^5$ supergravity. This leads to the holographic reformulation
\e{sugraavg}{\O_{\rm sugra} = \<\O\>}
This relation extends to all orders in $1/N$ in a sense prescribed in (\ref{O0strong4}) and \eqr{infavg3}. We thus have the following picture. On the one hand, holographic duality between $\cN=4$ SYM and AdS$_5 \x S^5$ string theory works as it always has. On the other, semiclassical AdS$_5 \x S^5$ supergravity has two equivalent descriptions via type IIB string theory: first, as a low-energy limit; and second, as an $\sl$ average.

In {\bf Section \ref{sec:SL2ZStatistics}} and {\bf Section \ref{sec:remarks}}, we unfurl some consequences of this. 

If AdS$_5 \x S^5$ supergravity (i.e. strongly coupled planar $\cN=4$ SYM) is also a large $N$ limit of an ensemble average, one is motivated to study the statistics of the {\it $\sl$ ensemble} --- that is, the distribution of $\cN=4$ SYM observables over moduli space $\cM$. We initiate this study in {\bf Section \ref{sec:SL2ZStatistics}}. For general observables, the variance in the $\sl$ ensemble admits an expression in terms of the squares of spectral overlaps:
\es{varint}{\cV(\calo) &\coloneqq \<\O^2\> - \<\O\>^2\\
&= \vvol(\cF)^{-1}\left({1\over 4\pi i}\int_{\rre s = \half} ds\, |(\calo,E_s)|^2 + \sum_{n=1}^\infty (\calo,\phi_n)^2\right)}
All observables that vary over the conformal manifold $\cM$, supersymmetric or not, necessarily have nonzero variance. At large $N$, as shown in \eqr{varlargeN}, the variance is parametrically suppressed compared to the squared average,
\e{varsupp}{{\cV(\calo)\over\langle\calo\rangle^2}  \sim {1\o N}~.}
This suppression implies, and quantifies, a sense in which gravity is self-averaging. In particular, {\it the $SL(2,\Z)$ ensemble is self-averaging at large $N$, and the large $N$ average is AdS$_5 \x S^5$ supergravity.} That is, in any member of the large $N$ ensemble, observables $\O(\t)$ are well-approximated by their supergravity values, $\O_{\rm sugra}$, up to terms of order $1/N$. Using general properties of the large $N$ spectral overlaps, we also find certain non-perturbative contributions, scaling as positive powers of $e^{-4\sqrt{\pi N}}$.

Our result exhibits a role for spacetime wormholes even within the conventional holographic paradigm. (See Figure \ref{figworm}.) A trademark feature of recently formulated low-dimensional holographic dualities \cite{Saad:2019lba,Stanford:2019vob,Afkhami-Jeddi:2020ezh,Maloney:2020nni} involving ensemble-averaged boundary duals is the important role played by spacetime wormholes, which leads to the non-factorization of observables with multiple distinct boundaries, as the averaging induces correlations between the boundaries.  In a UV complete realization of holography, any wormhole contributions to the semiclassical bulk path integral must be supplemented by other contributions which restore factorization of multi-boundary observables. In our context, the effect of the ensemble average is to project out these UV-sensitive details; nevertheless, we identify, via the spectral decomposition, the natural appearance of wormhole-type contributions to (unaveraged) multi-boundary observables. The spectral decomposition at large $N$ also provides a clear incarnation of the ``half-wormholes'' of \cite{Saad:2021rcu} in $\cN=4$ SYM observables.

We then ask how these results extend to the general holographic context. While we set out a spectrum of possibilites, the most well-motivated is that large $N$ averages over more general U-duality symmetries of string theory (i.e. generalized S-duality symmetries of CFTs), or over exactly marginal directions dual to the string coupling, may also localize onto supergravity. It is of clear interest to explore these possibilities further.

In {\bf Section \ref{sec:conclusion}}, we conclude by highlighting some pertinent future directions. 

{\bf Two appendices} contain some technical details complementing the main text. We highlight {\bf Appendix \ref{app:inhomogeneous}}, which gives a self-contained and general treatment of a class of functions appearing in the integrated correlators and in other string theory contexts from the spectral point of view. 

\sec{$SL(2,\Z)$ Spectral Theory}\label{sec:SL2ZSpectral}

We will begin with a very brief review\foot{See \cite{Terras_2013} for a readable introduction.} of the spectral theory of the Laplacian on the fundamental domain domain of $SL(2,\mathbb{Z})$,
\begin{equation}
  \mathcal{F} = \mathbb{H}/SL(2,\mathbb{Z}) = \left\{\tau = x+i y \in\mathbb{H} \,\bigg| - \half\leq x \leq \half, |\tau| \geq 1\right\}.
\end{equation}
Functions $f(\tau)$ defined on the upper half-plane $\mathbb{H}$ that are invariant under the $SL(2,\mathbb{Z})$ action
\begin{equation}
  f(\gamma\tau) = f(\tau),\quad \gamma \tau = {a\tau+b\over c\tau+d},\quad \gamma = \begin{pmatrix} a & b \\ c & d \end{pmatrix}\in SL(2,\mathbb{Z})
\end{equation}
can be thought of as being defined solely on the fundamental domain $\mathcal{F}$, as any point outside $\cF$ can be mapped to a point inside it by a suitable $SL(2,\mathbb{Z})$ transformation. This space is equipped with the hyperbolic metric
\begin{equation}
  ds^2 = {dx^2+dy^2\over y^2}
\end{equation}
and the corresponding hyperbolic Laplacian, which acts on scalar functions as\foot{We note that this is defined with a minus sign compared to the Laplacian that often appears in the literature. We have defined it this way so that its spectrum of square-integrable functions is non-negative.}
\begin{equation}
  \Delta_\tau = -y^2(\partial_x^2+\partial_y^2).
\end{equation}
The Laplacian is self-adjoint with respect to the Petersson inner product on the space $L^2(\mathcal{F})$ of square-integrable modular-invariant functions
\begin{equation}
  (f,g) \coloneqq \int_{\mathcal{F}}{dxdy\over y^2} f(\tau)\overline{g(\tau)}.
\end{equation}
Square-integrability means that the norm defined with respect to the Petersson inner product is finite,
\begin{equation}
  (f,f) < \infty.
\end{equation}

Any square-integrable $\sl$-invariant function can be expanded in a complete eigenbasis of the Laplacian, which includes three distinct components:
\e{}{L^2(\cF) = L^2_{\rm const}(\cF) \oplus L^2_{\rm cont}(\cF) \oplus L^2_{\rm disc}(\cF)}
The expansion is given in \eqr{eq:RoelckeSelberg}, but let us first present the eigenbasis. The most trivial is the constant
\begin{equation}
  \Delta_\tau\nu_0 = 0, \quad \nu_0 = \vvol(\mathcal{F})^{-\half} = \sqrt{3\over \pi}.
\end{equation}
Next, there is a continuous branch spanned by the real-analytic Eisenstein series
\begin{equation}
  \Delta_\tau E_s(\tau) = \mu(s) E_s(\tau), \quad \mu(s) \coloneqq s(1-s), \quad \rre (s)   = \half.
\end{equation}
The Eisenstein series are defined by a Poincar\'e series, a sum over $PSL(2,\mathbb{Z})$ orbits
\begin{equation}\label{eq:eisensteinPoincare}
  E_s(\tau) = \sum_{\gamma\in\Gamma_\infty\backslash PSL(2,\mathbb{Z})}\im(\gamma\tau)^s,
\end{equation}
where $\Gamma_\infty$ is the subgroup of $PSL(2,\mathbb{Z})$ of upper triangle matrices that fix $y$. This sum converges for $\rre(s)>1$. However, the Eisenstein series admits a meromorphic continuation to the entire complex $s$ plane via its Fourier decomposition
\begin{equation}\label{eq:eisensteinFourier}
  E_s(\tau) = y^{s} + \varphi(s)y^{1-s} + \sum_{k=1}^\infty 4 \cos(2\pi k x){\sigma_{2s-1}(k)\over k^{s-\half}\Lambda(s)}\sqrt{y}K_{s-\half}(2\pi k y),
\end{equation}
where $\sigma_n(x) = \sum_{p|x}p^n$ is the divisor function,
\begin{equation}\label{eq:LambdaDefinition}
  \Lambda(s) \coloneqq \pi^{-s}\Gamma(s)\zeta(2s) = \Lambda(\half-s)
\end{equation}
is the completed Riemann zeta function, and
\begin{equation}\label{varphidef}
  \varphi(s) \coloneqq {\Lambda(1-s)\over\Lambda(s)}.
\end{equation}
In the following application of spectral theory to observables in $\cN = 4$ SYM, the Fourier mode number $k$ will represent the total instanton number. It will be important in what follows that the meromorphic continuation of the Eisenstein series satisfies the functional equation
\begin{equation}
  E^*_s(\tau) = E^*_{1-s}(\tau)
\end{equation}
where we've defined
\begin{equation}
  E_s^*(\tau) \coloneqq \Lambda(s)E_s(\tau).
\end{equation}
This is manifest from the Fourier decomposition (\ref{eq:eisensteinFourier}). Note from (\ref{eq:eisensteinFourier}) that the Eisenstein series behave perturbatively at the cusp $y=\infty$,
\begin{equation}
  E_s^*(\tau) \sim \Lambda(s)y^s + \Lambda(1-s)y^{1-s}\qquad (y\to\infty)\,.
\end{equation} 
A final comment about the Eisenstein series is that it has a simple pole at $s=1$ with a constant residue,
\begin{equation}\label{Es1res}
  \Res_{s=1}E_s^*(\tau) = \half
\end{equation}
In Appendix \ref{app:inhomogeneous} we mention a few details about the higher-order $s\rar 1$ behavior.

Finally, there is a discrete branch of the eigenspectrum of the Laplacian spanned by the Maass cusp forms. The cusp forms $\nu_n(\tau)$, infinite in number, are labelled by a parameter $t_n$ that specifies their eigenvalue, unbounded from above:
\begin{equation}
  \Delta_\tau\nu_n(\tau) = \mu_n\nu_n(\tau),\quad \mu_n \coloneqq \left({1\over 4}+ t_n^2\right), \quad {0} < t_1 <t_2 < \ldots
\end{equation}
The $t_n$ are a set of sporadic positive real numbers. The cusp forms admit a Fourier decomposition that is similar to that of the Eisenstein series\foot{We have here decomposed only the ``even'' cusp forms, invariant under $x \rar -x$, ignoring the ``odd'' cusp forms which are also infinite in number. Henceforth we treat only real observables $\O(\t)$, so all cusp forms are taken to be even.}
\begin{equation}\label{eq:cuspFourier}
  \nu_n(\tau) = \sum_{k=1}^\infty a_k^{(n)}\cos(2\pi k x)\sqrt{y}K_{it_n}(2\pi k y),
\end{equation}
where the Fourier coefficients $a_k^{(n)}$ are yet another set of sporadic real numbers that obey many interesting proven and conjectured statistical properties (partly summarized in e.g. \cite{Sarnak_1987, sarnak, hejrack, sarnakk}). From the Fourier decompositions (\ref{eq:eisensteinFourier}) and (\ref{eq:cuspFourier}) it is manifest that the cusp forms $\nu_n(\tau)$, like the Eisenstein series $E_s^*(\tau)$, are real-valued. However, {\it unlike} the Eisenstein series, the cusp forms have no zero mode,
\begin{equation}\label{cuspzero}
  \int_{-\half}^\half dx \, \nu_n(\tau) = 0\,,
\end{equation} 
and decay exponentially at the cusp $y=\infty$, 
\e{}{\nu_n(\tau) \sim e^{-2\pi y} \qquad (y\rar\i)\,.}
There are two conventional normalizations of the cusp forms in the literature. In the ``$L^2$ norm'', one rescales the cusp forms so that they have unit norm with respect to the Petersson inner product
\begin{equation}\label{eq:L2vsHeckeNorms}
  \phi_n(\tau) \coloneqq {\nu_n(\tau)\over\sqrt{(\nu_n,\nu_n)}}, \quad (\phi_n,\phi_n) = 1 \qquad \text{($L^2$ norm)}
\end{equation}
In this paper we will mostly make use of the $L^2$ norm in order to minimize notation, but some profound statistical properties of the cusp form Fourier coefficients are most naturally stated in the ``Hecke norm'', which sets the first Fourier coefficient to unity,
\begin{equation}\label{a1nhecke}
  a_1^{(n)} = 1 \qquad \text{(Hecke norm)}
\end{equation}
We will utilize one such property, on boundedness of Fourier coefficients with prime $k$ (the Ramanujan-Petersson conjecture), in Subsection \ref{seccusp}. 

To give a feel for these functions, we have shown a plot involving Eisenstein series and cusp forms in Figure \ref{fig2}. This puts the chaos of the cusp forms in rather sharp relief: in contrast to the power-law-dominated falloff of the Eisenstein series, the zero modes of the squares of the cusp forms have an irregular set of local extrema that varies with $n$.\foot{Since $\cF$ is two-dimensional, two distinct cusp forms intersect at some nonempty set of points as $\t$ varies over $\cF$. However, for three cusp forms, one would expect {\it no} triple intersections, particularly for chaotic functions. This is confirmed by the plot.} The reader is directed to \c{hejrack} for a lovely exposition of Maass cusp forms with more pictures. 

We would like to emphasize that even the existence of the (infinitely many) Maass cusp forms is a non-trivial fact, given that $\mathcal{F}$ is noncompact. Their existence was first established by Selberg, as an application of the Selberg trace formula (see e.g. \cite{1983} for a review). Indeed, the resulting version of Weyl's law establishes that the number of square-integrable eigenfunctions grows asymptotically linearly with the eigenvalue \cite{MR3287209}
\begin{equation}
\begin{aligned}
  N(\mu) &\coloneqq \left(\text{\# of Maass cusp forms with eigenvalue ${1\over 4}+t_n^2 \leq \mu$}\right) \\
  &\sim {\vvol(\cF)\over 4\pi}\mu,\quad \mu\to\infty
\end{aligned}
\end{equation} 
It is the finiteness of $\vvol(\cF)$ which allows the cusp forms to exist.

Any square-integrable $\sl$-invariant function $\calo(\tau)$ then admits a Roelcke-Selberg decomposition into this complete eigenbasis:
\begin{equation}\label{eq:RoelckeSelberg}
  \mathcal{O}(\tau) = \overline \calo +  {1\over 4\pi i }\int_{\rre s = \half}ds\, (\calo,E_s)E_s(\tau)  + \sum_{n=1}^\infty (\calo,\phi_n)\phi_n(\tau).
\end{equation}
The first term $\overline \O$ is the constant term in the spectral decomposition. The spectral overlap $(\calo,E_s)$ can be written in terms of a Mellin transform of the zero mode of $\calo$, known as the Rankin-Selberg (RS) transform \cite{rankin_1939,selberg1940bemerkungen}, which inherits many interesting properties, including a meromorphic continuation in the complex $s$ plane, from the Eisenstein series. For $\rre s = \half$, this is given by
\begin{equation}
  (\calo,E_s) = R_{1-s}[\calo]
\end{equation}
where
\begin{equation}
\begin{aligned}\label{eq:rankinSelberg}
  R_s[\calo] &\coloneqq \int_{\cF}{dxdy\over y^2} \calo(\tau)E_s(\tau) \\
  &= \int_0^{\infty} dy \, y^{s-2}\calo_0(y)\\
  &= \mM\left[{\calo_0(y)\over y};s\right]
\end{aligned}
\end{equation}
Here $\calo_0(y) = \int_{-\half}^\half dx\, \calo(\tau)$ is the zero mode of $\calo$ and $\mM[f;s]\coloneqq \int_0^\infty dy \, y^{s-1}f(y)$ is the Mellin transform. To arrive at this form, we have made use of the fact that the Eisenstein series is a Poincar\'e series as in (\ref{eq:eisensteinPoincare}) and ``unfolded'' the integral \cite{rankin_1939,selberg1940bemerkungen}. The RS transform also satisfies the same functional equation as the Eisenstein series
\begin{equation}
  \Lambda(s)R_{s}[\calo] = \Lambda(1-s)R_{1-s}[\calo].
\end{equation}
From this we see that the modified spectral overlap
\begin{equation}
  \{\calo,E_s\} \coloneqq {(\calo,E_s)\over\Lambda(s)}= \{\calo,E_{1-s}\}
\end{equation}
is invariant under the reflection $s \to 1-s$. 

The Mellin transform only converges for $\rre s>1$ (and indeed the form of the Eisenstein series as a Poincar\'e sum is only valid in this half-plane); however, the RS transform inherits a meromorphic continuation to the entire $s$ plane from the Eisenstein series. In particular, its only singularity in the half-plane to the right of the critical line $\rre s = \half$ is a simple pole at $s=1$ with a constant residue given by
\begin{equation}\label{modavgres}
  \Res_{s=1} R_s[\calo] = \vvol(\cF)^{-1}\int_{\cF} {dx dy\over y ^2}\calo(\tau) = \overline\calo.
\end{equation}
We see that the residue of the RS transform at $s=1$ encodes the {\it modular average} of the observable $\calo(\t)$ over the fundamental domain of $SL(2,\mathbb{Z})$, which is also the constant term in its spectral decomposition.

\sec{Spectral Decomposition in $\cN=4$ SYM}\label{sec:N4Spectral}
Having introduced some $\sl$ spectral technology, we transition to the physics by recalling the necessary salient features of four-dimensional $\cN=4$ SYM. We consider $\cN=4$ SYM with arbitrary simply-laced gauge group $G$, only later specifying $G=SU(N)$ when treating certain examples. This theory has a one-complex-dimensional $\cN=4$ supersymmetry-preserving conformal manifold, $\cM$.\foot{The $\cN=1$-preserving conformal manifold is larger, see \c{Aharony:2002hx} for a thorough exposition.} $\cM$ may be parameterized by $\t$, the complexified gauge coupling,
\e{}{\t = {\theta\o 2\pi} + {4\pi i\o g^2}}
S-duality of $\cN=4$ SYM with simply-laced $G$ means invariance of the theory under $SL(2,\Z)$ transformations of $\t$. This means, in turn, that local CFT observables $\O(\t)$ are $SL(2,\Z)$ invariant,
\e{}{\O(\g\t) = \O(\t)\,,\quad \g\in SL(2,\Z)\,.}
We would like to emphasize that the term ``CFT observables'' includes only those that are non-perturbatively well-defined. The spectrum of the dilatation operator, OPE coefficients among dilatation eigenstates, and four-point functions of superconformal primary operators are familiar examples of $\sl$-invariant quantities.\foot{On the other hand, ``the anomalous dimension of the Konishi operator'' is {\it not} non-perturbatively well-defined, as it makes reference to elementary fields in the $\cN=4$ Lagrangian. However, the anomalous dimension of the lowest-dimension $SU(4)_R$-singlet scalar, being agnostic about the ``composition'' of the operator,  is well-defined and must be $\sl$ invariant. Indeed, we will show in Section \ref{modstring} that any quantity which diverges in the large $N$ 't Hooft limit at $\l\gg1$ {\it cannot} be the limit of an $\sl$-invariant observable; this includes all single-trace anomalous dimensions.}

Reality of the coupling $g$ implies that $\t\in\HH$. S-duality invariance then implies that we may restrict $\t$ to the fundamental domain, whereupon $\O(\t)$ becomes a modular-invariant function of $\t\in\cF$. A natural measure on $\cM$ is determined by the Zamolodchikov metric, 
\e{}{d\mu_\cM \coloneqq  g_{\mu\nu}^Z( x)dx^\mu dx^\nu}
where $g_{\mu\nu}^Z(x)$ is defined in general as the matrix of two-point functions of exactly marginal operators. In $\cN=4$ SYM, the Zamolodchikov metric is, owing to maximal supersymmetry, exactly equal to the Poincar\'e metric on $\HH$, so the measure is the $SL(2,\R)$-invariant measure on $\HH$:
\e{zmetric}{d\mu_\cM = {dx dy\o y^2}}
This is special, and useful. In the notation of Section \ref{sec:SL2ZSpectral}, we have $y\coloneqq 4\pi/g^2$, so $g\rar 0$ at fixed $\theta$ approaches the cusp $y\rar\i$ at fixed $x$.  

Our observation is the following. Consider an $\sl$-invariant observable $\O(\t)$. On general grounds, CFT observables $\O(\t)$ are finite for all $\t$ in the interior of $\cF$, as singularities can only occur at boundaries of the moduli space. Moreover, the theory at the cusp $y\rar\i$ is the free theory, where $\O(\t)$ converges to its (finite) free value. Therefore, 
\e{}{\O(\t)\in L^2(\cF)}
This implies that $\O(\t)$ admits the Roelcke-Selberg spectral decomposition into the $\sl$-invariant eigenbasis described in Section \ref{sec:SL2ZSpectral}:
\e{Ospec}{\boxed{\O(\t) = \overline{\O} + {1\o 4\pi i}\ints \{\O,E_s\} \,E^*_s(\t) + \sum_{n=1}^\i (\O,\phi_n)\, \phi_n(\t)}}

\ni This is a simple observation, but it is a far-reaching one. We will spell out its consequences in detail. However, some of them are visible from the fundamental structure of the spectral decomposition alone. 

Let us develop a Fourier expansion of $\O(\t)$, invariant under $ x \rar -x$:
\e{Okfourier}{\O_k(\t) = \O_0(y) + \sum_{k=1}^\i2\cos(2\pi k x)\, \O_k(y)}
$k$ labels the total instanton number. Instanton-anti-instanton pairs contribute integer powers of $q\qb = e^{-4\pi y}$, where $q \coloneqq  e^{2\pi i \t}$, with total instanton number zero. The mode $\O_k(y)$ is obtained from \eqr{Ospec} by inserting the $k^{\rm th}$ Fourier modes of $E_s(\t)$ and $\phi_n(\t)$.

Without performing any detailed computation, one sees that, under a certain assumption stated below, {\it the $k=0$ and $k=1$ sectors uniquely determine all $k>1$ sectors.} The proof is as follows. The overlap functions $(\O,E_s)$ and $(\O,\phi_n)$ in the spectral decomposition may be determined by the $k=0$ and $k=1$ parts of $\O(\t)$, respectively. The former claim is manifest from the definition of $(\O,E_s)$ as the RS transform \eqr{eq:rankinSelberg}, which involves $\O_0(y)$ only; note that $\phi_n$ has no $k=0$ component, simplifying matters. Having fixed $(\O,E_s)$ from $\O_0(y)$, one then fixes $(\O,\phi_n)$ by ``inverting'' the Bessel function \eqr{eq:cuspFourier} using orthogonality relations \cite{Whittaker}. In this second step, we are assuming that the cuspidal eigenspectrum is non-degenerate, which is an unproven but widely-held property of $SL(2,\mathbb{Z})$ (e.g. \cite{cusp82,hejh,sque}).\foot{This same argument was used in \c{Benjamin:2021ygh}, where these mathematical tools were applied to 2d CFT partition functions, to prove a notion of ``spectral determinacy'' for 2d CFTs, in which the instanton number $k$ was instead interpreted as the spin $j$. We note that the inversion of the cusp forms must be performed approximately, rather than exactly, because the chaotic spectrum of eigenvalues $t_n$ is (interestingly!) not analytically known. That instantons with $k>1$ are fully determined by $k\leq 1$ is irrespective of that, as it relies only on the {\it existence} of the spectral decomposition and the structure of the Fourier expansions of the basis elements.}  

In this sense, instantons are ``redundant'' in $\cN=4$ SYM. This surprising simplicity naively suggests that the non-perturbative structure of instantons --- for instance, whether the weak coupling expansions are convergent, asymptotic, Borel summable, etc, or whether there exist non-perturbative instanton-anti-instanton corrections in powers of $q\qb$ --- should be the same for all $k$. We will show that, properly understood, this is correct. Moreover, in favorable cases, the overlap with cusp forms vanishes --- i.e. $(\O,\phi_n)=0$ for all $n$ --- in which case $\O(\t)$ is fully determined by the zero-instanton sector alone. We will provide such examples in $\cN=4$ SYM.

Let us now point out a second general consequence of \eqr{Ospec}. Recall that the first term is the modular average of $\O(\t)$ over $\cF$. In a general CFT with conformal manifold $\cM$ coordinatized by moduli $x^\mu$, one may define the {\it ensemble average} of $\O(x)$ as
\e{}{\<\O\> \coloneqq  \int d\mu_\cM \,\O(x)}
where the integral is supported on the fundamental domain of $\cM$. In $\cN=4$ SYM, the Zamolodchikov metric is simply the Poincar\'e metric, cf. \eqr{zmetric}, and the fundamental domain is $\cF$. Therefore, {\it the ensemble average is the modular average}:
\e{}{\boxed{\<\O\> = \overline \O}}
$\overline \O$ is, we recall from \eqr{modavgres}, determined as a residue of the RS transform of $\O(\t)$ at $s=1$, and is insensitive to the cusp form contributions to $\O(\t)$. So we see that the spectral decomposition makes manifest the ensemble average of $\O(\t)$, and gives a straightforward way to compute it.

\sec{Integrated Correlator I: Warmup}\label{sec:integratedCorrelatorI}

Before exploring the spectral decomposition of generic observables $\O(\t)$, we introduce $\cG_N(\t)$, the integrated correlator studied by Dorigoni, Green and Wen \cite{Dorigoni:2021bvj,Dorigoni:2021guq}. In addition to being a rare and beautiful object, $\cG_N(\t)$ will act throughout as an affirmative benchmark for our methods.  

\ssec{Definition}

$\cG_N(\t)$ is the four-point function of the $\O_{\mathbf{20'}}$ superconformal primary operator in the $SU(N)$ SYM theory, integrated over Euclidean space with the following supersymmetric measure:
\e{GNDef}{\cG_N(\t) = - {8\over \pi} \int_0^{\infty} dr \int_0^{\pi} d\theta {r^3 \sin^2\theta \over u^2} \cT_{N}(u,v;\t) \,,}
where 
\e{uvrtheta}{u = {x_{12}^2 x_{34}^2 \over x_{13}^2 x_{24}^2} = 1+r^2-2r\cos\theta\,,\qquad v = {x_{14}^2 x_{23}^2 \over x_{13}^2 x_{24}^2} = r^2}
To be precise, $\cT_{N}(u,v;\t)$ is the ``dynamical'' part of the $\O_{\mathbf{20'}}$ four-point function (after stripping off a power-law prefactor) that is not determined by $\cN=4$ superconformal Ward identities, in the conventions of \c{Dorigoni:2021guq}. We indicate the $\t$-dependence explicitly to emphasize it. That $\cG_N(\t)$ is a protected quantity follows from its alternative formulation as derivatives of a mass-deformed free energy on $S^4$ \c{Binder:2019jwn}
\e{}{\cG_N(\t) = -{1\o4}\D_\t \p^2_m \log Z_{S^4}(\t;m)\big|_{m=0}}
where $Z_{S^4}(\t;m)$ may be computed from localization as a matrix integral over the Cartan of $SU(N)$
\e{}{Z_{S^4}(\t;m) \coloneqq  \int d^{N-1} a \(\prod_{i<j} {a_{ij}^2 H^2(a_{ij})\o H(a_{ij}-m)H(a_{ij}+m)}\) |Z_{\rm inst}(\t)|^2 e^{-{2\pi y}\sum_i a_i^2}}
where $a_{ij} \coloneqq  a_i-a_j$ and $H(x)\coloneqq G(1+x)G(1-x)$, with $G(x)$ the Barnes $G$-function \c{Pestun:2007rz,Russo:2013kea}.

The authors of \cite{Dorigoni:2021guq} conjectured the following form for $\cG_N(\t)$ for all $N$ and $\tau$:\foot{Our conventions differ slightly from \cite{Dorigoni:2021guq}. We normalize our Eisenstein series such that $E_s(\t)\big|_{\rm there} = {2\o \G(s)}E^*_s(\t)\big|_{\rm here}$. Also, in the formulas to follow, $c_s^{(N)}|_{\rm here} = (-1)^{s+1}{2\o \G(s)} c_s^{(N)}|_{\rm there}$. Finally, $y_{\rm here} = \Im(\t) = y_{\rm there}/\pi$.}
\e{dgwmain}{\cG_{N}(\t) =\frac{1}{2} \sum_{(m,n)\in\Z^2}  \int_0^\infty d\xi \,B_N(\xi)
\exp\(- \pi \xi \,\frac{|m+n\tau|^2}{y}\) \,.}
The function $B_{N}(\xi)$ is a rational function of $\xi$, 
\e{}{B_{N}(\xi) = \frac{ \cQ_{N}(\xi)}{(1+\xi)^{2N+1}} \,,}
where $\cQ_N(\xi)$ is a polynomial of order $2N+1$ determined by the following Laplace difference equation for $\cG_N(\t)$: 
\e{laplacediffeq}{-\left(\Delta_\tau +2\right)\cG_N(\t) = N^2\Big[\cG_{N+1}(\t) -2 \cG_N(\t)+\cG_{N-1}(\t)\Big]-N \Big[  \cG_{N+1}(\t)-\cG_{N-1}(\t) \Big]\,.}
Another useful representation, albeit a formal one that does not converge for all $\t \in\HH$, is as an infinite sum over Eisenstein series,
\e{dgweis}{\cG_N(\t) = {N(N-1)\o 8} - \half\sum_{s=2}^\i (-1)^s c_s^{(N)}E^*_s(\t)}
For $SU(2)$, 
\e{cs2}{c_s^{(2)} =  s (1-s)(2s-1)^2}
For $SU(N)$, the $c_s^{(N)}$ obey the following recursion relation in $N$:
\e{rec}{N(N-1) c_s^{(N+1)} - (2(N^2-1)-s(1-s))c_s^{(N)} + N(N+1)c_s^{(N-1)}=0}
The recursion implies that 
\e{pols}{c_s^{(N)} = P^{(N)}(s)\, c_s^{(2)}}
where $P^{(N)}(s)$ is a polynomial of degree $2N-4$ and $P^{(2)}(s) = 1$. 

The evidence collected in \cite{Dorigoni:2021guq} for this remarkable formula is quite substantial, including finite $N$ matching to localization; matching to direct spacetime integration of Feynman integrals at low perturbative order; and large $N$ matching to expectations from type IIB string theory and S-duality. Still, it remains conjectural, and its interesting structure begs various questions. Why does $\cG_N(\t)$ have a lattice-integral representation? Do other $\cN=4$ observables have one? What is the physical meaning of the rational functions $B_N(\xi)$? Can the Laplace difference equation be derived?
 
Since $\cT_{N}(u,v;\t)$ is a correlator of protected operators, both $\cT_{N}(u,v;\t)$ and $\cG_N(\t)$ are $\sl$ invariant. Focusing on $\cG_N(\t)$, what is its spectral decomposition?
 
\ssec{Spectral decomposition}

Let us start with the $SU(2)$ case. We claim that $\cG_2(\t)$ is given by the following expression:
\es{cg2}{\cG_2(\t) = \half + {1\o 4\pi i}\ints \({\pi \o \sin \pi s}s(1-s)(2s-1)^2\) E^*_s(\t)}
In terms of spectral overlaps, we have
\es{cg2overlap}{\{\cG_2,E_s\} = {\pi\o \sin\pi s}\,c_s^{(2)}\,,\qquad (\cG_2,\phi_n)=0}
where $c_s^{(2)}$ was defined in \eqr{cs2}. The claim is that $\cG_2(\t)$ is given {\it exactly} by this expression, for all $\t\in\cF$: there is no need for regularization or resummation, a fact which is built into the spectral decomposition, convergent by definition. Note that the spectral representation \eqr{cg2} is functionally rather simpler than \eqr{dgwmain}, being a one-dimensional integral {\it without} a lattice sum on top of it. 

It is easiest to first recover the formal expression \eqr{dgweis} for $N=2$ from \eqr{cg2}. This is done by contour deformation, with the expansion coefficients $c_2^{(N)}$ encoded as residues of $(\cG_2,E_s)$ at $s\in \Z_{>1}$. This deformation also generates a constant contribution from $s=1$, where $E^*_s(\t)$ has a simple pole, cf. \eqr{Es1res}: if we deform to $\Re s > \half$, we pick up a constant term $-1/4$. This combines to give the correct constant in \eqr{dgweis}. Now, recall that \eqr{dgweis} is not convergent, but rather required Borel resummation to become well-defined. This is precisely encoded in the fact that the integrand of \eqr{cg2overlap} diverges factorially at $s\rar\pm \i$, due to the completed zeta function $\L(s)$. The spectral representation \eqr{cg2} may thus be thought of as the Borel resummation of the weak coupling expansion. 

If \eqr{cg2} is correct, then it should equal the (conjectural) expression \eqr{dgwmain} too, which is also convergent for all $\t\in\cF$. We will establish this equality in Section \ref{poincsec} as an example of a more general equality between spectral decompositions and lattice-integral representations of the same quantity $\O(\t)$, and in doing so, prove the validity of the formula \eqr{dgwmain} given only the perturbative expansions of the $k=0,1$ modes as input. 

The result for $SU(N)$, which follows by recursion \eqr{rec}, is the direct extension of the $SU(2)$ case:
\es{cgn}{\boxed{{\cG_N(\t) = {N(N-1)\o 4} + {1\o 4\pi i}\ints {\pi\o \sin\pi s}\,c_s^{(N)} E^*_s(\t)}}}
i.e.
\es{cgnoverlap}{\boxed{\{\cG_N,E_s\} = {\pi\o \sin\pi s}\,c_s^{(N)}\,,\qquad (\cG_N,\phi_n)=0}}
One again easily confirms the match to \eqr{dgweis} by performing a contour deformation. Note that the constant term can be determined even without using the explicit $c_s^{(N)}$ as follows. The constant term is determined by
\es{}{\Res_{s=1} R_s[\cG_N] &= P^{(N)}(1) \x \Res_{s=1} R_s[\cG_2]\\
&= {P^{(N)}(1)\o 2}}
So our claim is that $P^{(N)}(1) = N(N-1)/2$. This may be derived for all $N$ by setting $s=1$ in \eqr{rec}, whereby the recursion becomes
\e{recconst}{N(N-1) P^{(N+1)}(1) - 2(N^2-1)P^{(N)}(1) + N(N+1)P^{(N-1)}(1)=0}
This is solved by $P^{(N)}(1) = \a N + \b N^2$ for any $\a,\b$. Using initial data for $N=2,3$ derived from recursion fixes $\a=-\b = -1/2$, concluding the calculation.

Let us stress two striking features of this result. First, the overlap with cusp forms vanishes for all $N$. Second, the ensemble average is the constant term in \eqr{cgn}:
\e{cgnavg}{\boxed{\<\cG_N\> = {N(N-1)\o 4}}}
Note that at large $N$, we have $\<\cG_N\> \sim N^2/4$. We explicate these points in much more detail in later sections, including the systematics of the large $N$ expansion of $\cG_N(\t)$. 

\sec{Instantons and the Analytic Structure of $\cN=4$ SYM}\label{sec:instantonsN4AnalyticStructure}

We now systematically explore the consequences of spectral decomposition for $SL(2,\Z)$-invariant  observables $\O(\t)$. For clarity, we begin this section by writing formulas assuming $(\O,\phi_n)=0$, generalizing in due course. We remind the reader that $\phi_n(\t)$ has no zero mode, so all conclusions about $\O_0(y)$ are independent of this assumption. It will prove convenient to write out the Fourier modes explicitly: suppressing the cusp forms in $\O_{k>0}(y)$,
\es{Ospecexp}{\O_{0}(y) &= \<\O\> + {1\o 2\pi i}\ints \{\O, E_s\} \L(s)y^s \\
\O_{k>0}(y) &= {1\o 2\pi i}\ints \{\O,E_s\} \({\sigma_{2s-1}(k)\o k^{s-\half}}\sqrt{y} K_{s-\half}(2\pi k y)\)}
where the mode expansion was developed in \eqr{Okfourier}.

\ssec{Perturbation theory}

Our starting point is to understand how to recover weak coupling perturbation theory from this formalism. This is straightforward, and elegant. Weak coupling perturbation theory is an expansion around $y\rar\i$. For $k=0$, the perturbative expansion is therefore read off from contour deformation of \eqr{Ospecexp} in the appropriate direction. In particular, since $y\rar\i$, we must deform $y^s$ to the left. This seems in tension with the fact that the integrand may not fall off sufficiently fast at $s\rar-\i$ for this procedure to give a finite result. However, exactly this fact encodes the convergence, or lack thereof, of the perturbative expansion of $\O(\t)$. For example, if $\O(\t)$ admits an asymptotic series, then $(\O,E_s)$ will diverge factorially at $s\rar -\i$. Thus, the spectral formalism for gauge theory observables embeds the asymptotic properties of perturbative expansions in the $|s|\rar\i$ asymptotics of spectral overlaps. The virtue of the spectral decomposition lies in the existence of a {\it convergent} expression from which one can, if desired, develop the weak coupling expansion. 

For $k>0$, things work differently. We note the asymptotics 
\e{bessel}{\sqrt{y}K_{s-\half}(2\pi k y) = {1\o 2\sqrt{k}} e^{-2\pi k y}\sum_{n=0}^\i {a_n(s)\o (2\pi k y)^n}}
where
\e{bessela}{a_n(s) = {(s)_n(1-s)_n\o (-2)^n n!}.}
We highlight that the complete asymptotic expansion of $K_{s-\half}(2\pi k y)$ is a {\it single} exponential times a power series in $1/y$, for all $s$. Inserting in \eqr{Ospecexp}, 
\e{Ok}{\O_{k>0}(y) = {e^{-2\pi k y}\o 4\pi i}\int_{\Re s=\half} ds\,\{\O,E_s\} \({\sigma_{2s-1}(k) k^{-s}}\sum_{n=0}^\i {a_n(s)\o (2\pi ky)^n}\)}
Suppose we swap the sum and integral in \eqr{Ok},
\e{Ok'}{\O_{k>0}(y) \approx {e^{-2\pi k y}\o 4\pi i}\sum_{n=0}^\i y^{-n} \({1\o (-4\pi k)^n n!} \int_{\Re s=\half} ds\,\{\O,E_s\} {\sigma_{2s-1}(k) k^{-s}}(s)_n(1-s)_n\)}
{\it If} this is allowed, then it is clear that $\O_k(y)$ receives no non-perturbative, i.e. instanton-anti-instanton, contributions: such contributions would dress the $e^{-2\pi k y}$ with extra powers of $q\qb = e^{-4\pi y}$, but \eqr{Ok'} is a power series in $1/y$. The factor $\sigma_{2it}(k) k^{-\half+it}\in\mathbb{R}$ for $k\in\ZZ$ is oscillatory in $t$ but bounded in $k$. Therefore, the convergence properties of the bracketed integral are uniform in $k$. By inspection, convergence for fixed $k$ demands that the overlap decays faster than any polynomial as $t\rar\i$. 

But clearly, this cannot always hold for $\cN=4$ SYM observables, which {\it can} have \np corrections. Indeed, a short calculation (see Appendix \ref{appa}) shows that convergence of the spectral decomposition only requires
\e{OEgrowth}{\big|\{\O,E_{\half+it}\}\big| \lesssim O(e^{\pi t/2}) \qquad (t\rar\i)}
where we have suppressed power law corrections. 

Thus, the presence or absence of \np corrections to $\O_{k>0}(y)$ boils down to whether the integrals in \eqr{Ok'} converge, which in turn is determined by whether $\{\O,E_{\half+it}\}$ decays super-polynomially as $t\rar\i$. We show below that this asymptotic is elegantly determined by the Borel summability {\it of the zero mode}, $\O_0(y)$! The connection, of course, is that $\O_0(y)$ and $\O_{k>0}(y)$ are fixed (modulo cusp forms) by the same spectral overlap.

\ssec{The analytic structure of CFT spectral overlaps}\label{sec:analyticStructureCFTOverlaps}
Let us now understand how the mathematical structure of the overlaps is constrained by our physics problem. We are interested in CFT observables $\O(\t)$. These must have a good perturbative expansion: specifically, we allow only non-negative integer powers of $1/y$, and no  powers of $\log y$. This strongly constrains the form of $\{\O,E_s\}$, because we can set $k=0$, and extract the perturbative expansion by contour deformation to the real axis.\foot{All formulas for $\{\O,E_s\}$ that follow are valid even for $(\O,\phi_n)\neq 0$, since $\phi_{n,0}(y)=0$.} 

The most general such $\{\O,E_s\}$ can be written as follows:
\es{form}{\boxed{\{\O,E_s\} = {\pi \o \sin\pi s}s(1-s)\fp(s) + \fnp(s)}}
The labeling of these functions is hopefully clear. By contour deformation, $\fp(s)$ gives the complete {\it perturbative part} of $\O(\t)$, proceeding in integer powers of $1/y$.\foot{The zero mode of $\calo$ may be recovered from an inverse Mellin transform 
\e{mellindef}{\O_{0}(y) = {1\o 2\pi i}\int_{\s-i\i}^{\s+i\i} ds\, R_s[\O] y^{1-s}.}
$\s\in\R$ is taken in the so-called ``fundamental strip,'' defined as the largest open strip $s=\s+it$ in which the Mellin transform converges; since $R_s[\O]$ admits a meromorphic continuation to the entire complex plane, the fundamental strip is likewise extended to include $\s\in\R$. See e.g. the Direct Mapping Theorem of \cite{FLAJOLET19953}.} There is, in addition, the possibility of a regular part $\fnp(s)$ which, having no poles away from $s=1$, can only contribute to the {\it non-perturbative part} of the modes $\O_k(\t)$, proceeding in powers of $q\qb = e^{-4\pi y}$. Unlike the perturbative part, in the non-perturbative part of the spectral integral we cannot deform the contour to infinity and rewrite it as a sum over residues. 

The functions $\fp(s)$ and $\fnp(s)$ must obey the following properties:

{\bf 1)} Invariant under reflections, $s\rar 1-s$. 

{\bf 2)} Real for $s\in\mathbb{R}$.

{\bf 3)} Regular for $s\in\mathbb{R}$ and $s\notin [0,1]$. 

\ni These constraints follow immediately from the preceding exposition. In fact, these functions obey even stronger constraints, which we prove using number theory in Appendix \ref{appa}:

{\bf 3a)} $\fp(s)$ and $\fnp(s)$ are regular for all $s\in\CC$ away from $s=1$ (and its reflection).

{\bf 3b)} At $s=1$, 
\e{averageFromOverlap}{\lim_{s\rar 1}\,\{\O,E_{s}\} = 2\, \overline{\O}}

{\bf 3c)} On the critical line $s=\half+it$, $\{\O,E_s\}$ is finite for finite $t$. 

{\bf 4)} If $\fnp(s)=0$ (resp. $\fp(s)=0$) then $\fp(s)$ (resp. $\fnp(s)$) is entire.  

\vs

\ni To summarize, $\{\O,E_s\}$ is a meromorphic function, regular everywhere on $\CC \backslash \{s\in \mathbb{R}\}$ with poles at $s=2,3,\ldots$ and their reflections coming only from the $\sin \pi s$ factor. Determining the (continuous part of the) spectral decomposition of $\O(\t)$ boils down to finding the two real functions $\fp(s)$ and $\fnp(s)$ with these fairly rigid properties. 

Since $\{\O,E_s\}$ is defined as a Mellin transform of the zero mode, 
\e{}{\{\O,E_s\} = {1\o \L(1-s)} \mathfrak{M}\[{\O_0(y)\o y};s\]}
it is simple to read off the values of $\fp(s)$ at $s\in\Z_{>1}$ from the perturbative series: inserting \eqr{form} into \eqr{Ospecexp} and deforming the contour to develop the $1/y$ expansion gives
\e{fpasymp}{\O_0(y\rar\i) \sim \sum_{n\geq 1}(-1)^n c_n \,y^{-n} \quad\Rightarrow\quad c_n = -n(n+1)\L\Big(n+\half\Big)\fp(n+1)}
As $n\rar\i$,  
\e{}{\fp(n\rar\i) \sim -{\pi^{n+\half}\o n^2}{c_{n\rar\i}\o \G(n+\half)}\,.}
This exhibits the encoding of the asymptotics of the perturbative series $\O_0(y)$ in the $s\rar\i$ behavior of $\fp(s)$. 

As for $\fnp(s)$, since by definition it only generates non-perturbative terms at $y\rar\i$, which are necessarily of the form $y^{-\a}e^{- 4\pi n y}$ for constants $\a,n\in\Z_+$, it is just the sum of Mellin transforms
\es{}{\mM\[y^{-\a}e^{-4\pi n y};-s\] = (4\pi n)^{s+\a}\,\G(-s-\a)}
Parameterizing the most general series of non-perturbative corrections as
\e{}{\O_0(y)\supset \sum_{\a=0}^\i c_\a(\O) y^{-\a} \sum_{n=1}^\i\kappa^{(\a)}_{n}(\O) e^{-4\pi ny}\,,}
where $\kappa_n(\O)$ and $c^{(n)}_{\a}(\O)$ are constants, translates into the general formula for the \np overlap $\fnp(s)$:
\e{}{{\fnp(s) = \L(s)^{-1} \sum_{\a=0}^\i c_\a(\O) \Phi^{(\a)}_\O(-s-\a)\G(-s-\a)}}
where we have introduced a Dirichlet series $\Phi^{(\a)}_\O(s)$
\e{}{\Phi^{(\a)}_\O(s) \coloneqq  \sum_{n=1}^\i{\kappa^{(\a)}_{n}(\O) \o (4\pi n)^{s}}}
This is associated to the non-perturbative corrections $(q\qb)^n$ around each perturbative correction to $\O(\t)$.

An interesting fact is that the coefficients $c_{\a}(\O)$  and $\kappa_n^{(\a)}(\O)$ are constrained by the reflection symmetry of $\fnp(s)$, which acts as a functional equation, $\fnp(s) = \fnp(1-s)$. In other words, defining the ``completed'' function
\e{}{\Psi^{(\a)}_\O(s) \coloneqq  \L(s)^{-1} \Phi^{(\a)}_\O(-s-\a)\G(-s-\a)}
we have 
\e{spectralCrossing}{{\fnp(s) = \sum_{\a=0}^\i c_\a(\O) \Psi^{(\a)}_\O(s) = \sum_{\a=0}^\i c_\a(\O) \Psi^{(\a)}_\O(1-s)}}
A consistent set of non-perturbative contributions to $\O(\t)$ must satisfy this fascinating functional equation.\foot{One may view this as a ``bootstrap'' equation for the \np data $\kappa_n(\O)$ and $c^{(n)}_{\a}(\O)$; we do not investigate this here, but believe this deserves further study. Similarly, it would be interesting to consider the deeper meaning of these Dirichlet series for CFT observables.}

\ssec{$\sl$ Borel transform}

Suppose that $\O_0(y)$ has a factorially divergent, Borel summable expansion. Here and throughout, following the literature on resurgence, we take ``Borel summable'' to mean that the Borel transform $B[\xi]$ is non-singular in a wedge of the complex $\xi$ plane that includes $\mathbb{R}_+$.\foot{Generically, non-perturbative additions are not required for a Borel resummed quantity. However, there are ``non-generic'' cases where \np contributions are present, indeed necessary, even if the perturbative series is Borel summable. Some representative examples include ``Cheshire cat resurgence'' \cite{Kozcaz:2016wvy,Dorigoni:2017smz,Dorigoni:2019kux} as well as certain effects in string perturbation theory \cite{Grassi:2014cla, Green:2014yxa, Chester:2020vyz}. We thank Daniele Dorigoni for correspondence on the status of this question.} This is a standard situation in $\cN=4$ SYM perturbation theory at finite $N$. If $\O_0(y)$ has factorial growth, then $\fp(n\rar\i)$ is sub-factorial: 
\e{borel}{c_{n} \sim (\pi R)^{-n} n! \quad \Rightarrow \quad \fp(n) \sim - n^{-{3\o2}}R^{-n}\qquad (n\rar\i)}
The constant $R$ is the radius of convergence of the $n\gg1$ expansion of $\fp(n)$.

For resumming $\cN=4$ SYM perturbation theory, it is convenient to introduce a new, non-standard Borel transform, which we call the {\it$\sl$ Borel transform}. If $\O_0(y)$ has the expansion \eqr{fpasymp}, then we define its $\sl$ Borel transform $B[\xi]$ as 
\e{Opert}{B[\xi]\coloneqq \sum_{n=0}^\i {(-1)^{n} c_{n} \o \L(n+\half)}\,\xi^{n+1} \qquad \text{($\sl$ Borel transform)}}
This may be written in terms of the overlap using \eqr{fpasymp}: 
\e{borelb}{B[\xi] = \sum_{n=0}^\i(-1)^{n+1} n(n+1)\fp(n+1) \,\xi^{n+1}}
Its utility follows upon noticing that
\e{Lthetamellin}{\L(s) = \mathfrak{M}\[{\theta_3(y)-1\o 2};s\]}
where $\theta_3(y) = \sum_{n\in\ZZ} e^{-\pi n^2y}$ is the Jacobi theta function. The Borel resummation of the original series may then be neatly inverted:
\es{borelf}{{\widehat \O_{0}(y)} \coloneqq  y^{\half} \int_0^\i {d\xi \o \xi^{3\o2}}\({\theta_3(y\xi )-1\o 2}\){B[\xi]} }
This can be confirmed to return the original sum term-wise upon using \eqr{Lthetamellin}. The radius $R$ in \eqr{borel} is then the radius of convergence of the $\sl$ Borel transform. 

The $\sl$ Borel transform is so named because it accounts for the $\L(n+\half)$ factor which appears universally in perturbation theory for $\sl$-invariant functions. In other words, the Jacobi theta kernel resums perturbation theory while manifestly respecting S-duality. Whereas an ordinary Borel transform divides by $\G(n+1)$, the $\sl$ Borel transform also properly eliminates the Riemann zeta factor.\foot{This answers a challenge posed in \cite{Beem:2013hha}. We note that other commonly used modified Borel transforms divide by $\G(n+1)\z(n+1)$ (as in \c{Arutyunov:2016etw,Dorigoni:2021guq}) whereas here, the Riemann zeta argument is {\it twice} that of the gamma argument, by definition of the completed Riemann zeta function.}

Just as for an ordinary Borel transform, $\widehat \O_0(y)$ is analytic for $\Re(y)>0$ in the complex $y$ plane. This domain of analyticity may be extended to a wedge by considering the ``directional'' inverse Borel transform along a ray $e^{i\theta}\xi$: this yields an analytic function for $\Re (e^{i\theta}y)>0$, i.e. in the domain 
\e{}{D_\theta \coloneqq  \{ y\in \mathbb{C} ~| ~\Re(e^{i\theta}y)>0\}}
It can be shown by analytic continuation (e.g. \c{Arutyunov:2016etw}) that if $B[\xi]$ is analytic for $\text{arg}(\xi) \in [\theta_1,\theta_2]$, then $\widehat \O_0(y)$ is analytic in the {\it union} of domains $D_{\theta_1}\cup D_{\theta_2}$.

Note that the $n \rar 1-n$ transformation properties of the summand immediately imply the inversion symmetry
\e{inv}{{B[\xi]\o \sqrt{\xi}} = \sqrt{\xi}B[\xi^{-1}]}
This inversion implies the non-trivial integral relation
\e{inv2}{\int_0^\i d\xi {B[\xi]\o\xi} = \int_0^\i {d\xi\o \xi} {B[\xi]\o\xi}}
In fact, using \eqr{Lthetamellin} and \eqr{borelf}, this integral is just twice the modular --- i.e. ensemble --- average of $\O$: 
\es{Favg}{\<\O\> &= \half \int_0^\i {d\xi} {B[\xi]\o\xi}}
Finally, note that\foot{Both (\ref{Favg}) and (\ref{Bvanish}) can be seen to follow from the fact that the $\sl$ Borel transform can be written in terms of an inverse Mellin transform of the RS transform of $\calo$
\begin{equation}
  B[\xi] = \xi \, \mM^{-1}\left[{R_s[\calo]\over \Lambda(1-s)};\xi\right] \nonumber
\end{equation}
}
\e{Bvanish}{\int_0^\i {d\xi\o \xi^{3\o2}} B[\xi] = -2\Res_{s=\half} R_s[\O] = 0}
This follows from square-integrability of $\O(\t)$, together with the definition of the RS transform as an inner product involving $E_s(\t)$ \cite{zbMATH03796039}, which obeys $E_{1/2}(\t)=0$. 

An interesting implication of \eqr{Bvanish} is that $B[\xi]$ is not sign-definite for $\xi \geq 0$. The physical meaning of zeros of Borel transforms has not been properly understood in a QFT context (though see \c{Honda:2017qdb} for interesting observations in 3d $\cN=2$ SCFT). It would be nice to understand why $\sl$ Borel transforms of $\cN=4$ SYM observables must have zeros.

\ssec{The redundancy of instantons}
Having better understood the structure of overlaps $\{\O,E_s\}$ in $\cN=4$ SYM, we are in good position to prove relations among \np corrections in different $k$-instanton sectors. 

We return to the problem posed earlier, below \eqr{OEgrowth}. There is a very nice theorem about the decay of Mellin transforms along strips $s=\sigma+it$ as $t\rar\i$ (cf. Proposition 5 of \cite{FLAJOLET19953}):

\textit{Let $f(y)$ be analytic in the region}
\e{}{S_\theta = \{ y\in\CC~|~0<|y|<\i\,,~ |\text{arg}(y)|\leq \theta\}~~~\text{with}~~0<\theta<\pi}
\textit{Assume that $f(y)\sim O(y^{-\a})$ as $y\rar 0$ and $f(y)\sim O(y^{-\b})$ as $y\rar\i$ in $S_\theta$. Then
\e{}{\mathfrak{M}[f(y);\s+it] = O(e^{-\theta|t|})~~\text{as}~~|t|\rar\i}
uniformly for $\s$ in every closed subinterval of $(\a,\b)$. This extends to any subinterval of the strip in which $\mathfrak{M}[f(y);s]$ admits a meromorphic continuation.}

\ni The above theorem shows that $\{\O,E_s\}$ will decay exponentially as long as $\O_0(y)$ is analytic in a wedge of the complex $y$ plane. Happily, this follows directly from Borel summability of $\O_0(y)$: since $\O_0(y)$ is analytic in a {\it finite} wedge $\arg(y) \in [\theta_1,\theta_2]$ around the positive $y$ axis 
\es{}{\mathfrak{M}\[{\O_0(y)\o y};s\] \sim e^{-\({\pi\o 2}+\theta\)|t|}\,,\quad \theta \coloneqq  \text{min}\{|\theta_1|,|\theta_2|\}>0}
Crucially, $\theta$ is strictly positive. This holds on the critical line $\s=1/2$.\foot{Indeed, \eqr{form} implies a meromorphic continuation of $\mathfrak{M}\[{\O_0(y)\o y};s\]$ to all $s\in\mathbb{C}$, so this theorem holds for any finite $\s\in\R$.} Together with the asymptotics
\es{}{\Big |{\L(\half-it)^{-1}}\Big| \sim e^{\pi t\o 2}\log|t| \qquad (t\rar\i)}
derived in Appendix \ref{appa}, we arrive the final result
\es{overlapasymp}{\Big|\{\O,E_{\half+i t}\}\Big| \lesssim e^{- \theta t}\,,\quad \theta>0 \qquad (t\rar\i)\,.}
Thus, $\{\O,E_{\half+i t}\}$ decays exponentially as $t\rar\i$, at a rate determined by the location of the singularity of the Borel transform $B[\xi]$. This guarantees finiteness of the integrals in \eqr{Ok'}, for all $k$. Note that if $\O_0(y)$ has singularities only on the negative real axis, $y\in\RR_-$, then $\theta \rar\pi$.

We can in fact go further, by characterizing the $n\rar \i$ asymptotics of the sum \eqr{Ok'}. The best way to do this is by plugging in \eqr{form} and deforming to $\Re s > \half$, because then we can compare directly to $B[\xi]$, the Borel transform of $\O_0(y)$, in \eqr{borelb}. Dropping an overall $(-2\pi i)$, the coefficient of $y^{-n}$ is
\e{boreli1}{S_{n,k} \coloneqq  {1\o (-4\pi)^n n!} \sum_{s=1}^\i (-1)^{s} s(s-1)\fp(s)(s)_n(1-s)_n\sigma_{2s-1}(k) k^{-s-n}}
The $SL(2,\Z)$ Borel transform of $\O_k(y)$ is, in terms of $S_{n,k}$,
\e{bk}{B_k[\xi] \coloneqq  \sum_{n=0}^\i {S_{n,k}\o \L\(n+\half\)} \xi^n}
We must thus extract the $n\gg1$ asymptotics of $S_{n,k}$. This is straightforward. Set $k=1$ for the moment. We have
\es{sn1}{{S_{n,1}\o \L\(n+\half\)} &= {1\o (-4\pi)^n n!\L\(n+\half\)} \sum_{s=1}^\i (-1)^{s} s(s-1)\fp(s)(s)_n(1-s)_n\\
&\approx{1\o (-4)^n} \sum_{p=0}^\i (-1)^{p+1} (p+n+1)(p+n)\fp(p+n+1){(p+1)_{2n}\o (n!)^2}}
Here we used that $(1-s)_n= 0$ for integer $s\leq n$ and $(-x)_n  = (x-n)_n(-1)^n$, shifted the sum, and dropped irrelevant factors subleading at $n\gg1$. Taking the zero mode to be $\sl$ Borel summable with radius $R>0$ means (cf. \eqr{borel}) 
\e{}{\fp(s\gg1) \sim  R^{-s}}
So we insert this scaling into the above equation, perform the sum over $p$, and afterwards expand at large $n$. Reinstating $k$-dependence is done by using the asymptotics
\e{}{\s_{2s-1}(k)k^{-s-n}\big|_{s\gg1} \sim k^{s-n-1}}
This leads to the insertion of a $k^{p}$ in \eqr{sn1}. The sum may be performed exactly. Quoting only the $n\gg1$ result of interest, and suppressing power law prefactors,
\es{nasymp}{{S_{n,k}\o \L\(n+\half\)} &\sim {\(-{R\o (R+ k)^2}\)^n}\qquad (n\rar\i)}
Plugging into \eqr{bk} gives the desired result: {\it Borel summability of $\O_0(y)$ implies Borel summability of $\O_k(y)$ for all $k$! That is, $B_k[\xi]$ is convergent and free of singularities on $\R_+$.} Moreover, we have a very simple polynomial formula for $R_k$, the radius of convergence of ${B}_k[\xi]$, for all $k$:
\es{Rk}{\boxed{{R_k\o R} = {\(1+{k\o R}\)^2}}}
This is a powerful demonstration of the ability of $SL(2,\Z)$ to relate different instanton sectors. Other notable properties of $R_k$ include its monotonic increase with $k$, and its quadratic growth at $k\gg1$. 

This result transcends the usual philosophy of resurgence methods. In typical applications of resurgence, one must perform independent resummations in different instanton sectors. In the presence of extra symmetries, more is possible. What we are seeing here is that $\sl$-invariance is strong enough to relate all $k$-instanton sectors in a simple way, with the $k=0$ sector (and, upon including cusp forms, the $k=1$ sector) being the only independent data. In the language of the so-called ``resurgence triangle'' \cite{Aniceto:2014hoa,Dunne:2012ae}, the claim is that $\sl$-invariance moves horizontally within the triangle, determining all columns from a finite subset.

This formula may actually be extended to non-Borel summable $\O_0(y)$ as well, assuming the perturbative series still grows only factorially. By ``non-Borel summable,'' we mean that $B[\xi]$ has a singularity at $\xi=R\in\R_+$. Relative to the Borel summable case with a singularity along $\R_-$, this modifies the asymptotic \eqr{nasymp} by $R \rar -R$; in addition, while the ensuing resummation acquires \np corrections to ensure analyticity of the end result $\widehat \O_0(y)$, none of these corrections affects the {\it perturbative} expansion. Propagating this through, the same logic as above implies that non-Borel summability of $\O_0(y)$ implies non-Borel summability of $\O_k(y)$ for all $k$, where $B_k[\xi]$ is singular at $\xi=R_k$ subject to the relation \eqr{nasymp} with $(R,R_k) \rar -(R,R_k)$.

Finally, for the sake of clarity, we remind that these conclusions hold modulo possible contributions from cusp forms. We will reinstate the cusp forms in Subsection \ref{seccusp}, giving a diagnostic for their presence and generalizing \eqr{Rk} to \eqr{cuspdiag2}. Moreover, some observables have no cusp form support, in which case the above results are unconditional. 

\ssec{$\cN=4$ SYM observables as Poincar\'e sums}\label{poincsec}
The previous section derived quantitative formulas for $k$-instanton dynamics from the zero-instanton dynamics. That this should be possible followed from the fundamental determinism baked into the $\sl$ spectral decomposition. In this section we show that starting from the zero mode $\O_0(y)$, one can construct an interesting representation of the {\it full} function $\O(\t)$ --- as an $\sl$-invariant lattice sum --- which is {\it equivalent} to the spectral decomposition. 

The idea is simply to sum over $SL(2,\ZZ)$ images of $\O_0(y)$. This uniquely determines the function up to overall normalization, which is fixed by demanding that it produces the correct $\O_0(y)$.

The starting point is \eqr{borelf}, the Borel resummation of $\calo_0(y)$, which we rewrite as
\begin{equation}
  \calo_0(y) = \sqrt{y}\int_0^\infty {d\xi\over \xi^{3\over 2}}B[\xi]\sum_{n=1}^\infty e^{-\pi n^2\xi y}.
\end{equation}
We will now proceed to perform a regularized Poincar\'e sum over images. Since $\calo_0(y)$ depends only on $y$, we sum over $\Gamma_\infty\backslash PSL(2,\mathbb{Z})$, modding out by the invariant action of the modular $T$-transformation on the seed. Let us call the resulting function $\calo(\tau)$:\footnote{These Poincar\'e series are divergent, but admit a natural zeta function-like regularization. This is most clearly reflected in (\ref{eq:caloBeforeRegularization}). See \cite{Maloney:2007ud,Keller:2014xba} for a discussion of similar Poincar\'e series in the context of the partition function of pure three-dimensional quantum gravity as defined by the sum over smooth saddle points in the gravitational path integral.}
\begin{equation}
  \calo(\tau)\coloneqq \cN \sum_{\Gamma_\infty\backslash PSL(2,\mathbb{Z})}\iim(\gamma\tau)^{\half}\int_0^\infty{d\xi\over \xi^{3\over 2}}B[\xi]\sum_{n=1}^\infty e^{-\pi n^2\xi \iim(\gamma\tau)},
\end{equation}
where $\cN$ is a normalization constant that we will fix later. We now rewrite this via Poisson summation as
\begin{equation}
\begin{aligned}\label{eq:poincareSumPoissonResummed}
  \calo(\tau) &= {\cN\over 2}\sum_{\gamma\in\Gamma_\infty\backslash PSL(2,\mathbb{Z})}\left[\int_0^\infty {d\xi\over \xi^2}B[\xi]\sum_{m\in\mathbb{Z}}\exp\({-{\pi m^2\over \xi \iim(\gamma\tau)}}\) - \iim(\gamma\tau)^{\half}\int_0^\infty {d\xi\over\xi^{3\over 2}}B[\xi]\right]\,.
\end{aligned}
\end{equation}
The last term vanishes identically, cf. \eqr{Bvanish}. To proceed, we expand the exponential in (\ref{eq:poincareSumPoissonResummed}) and note that
\begin{equation}
  \sum_{\gamma\in\Gamma_\infty\backslash PSL(2,\mathbb{Z})}\sum_{m=1}^\infty \left({m^2\over\iim(\gamma(\tau))}\right)^s = \half\sum_{(m,n)\ne (0,0)}\left({|m\tau+n|^2\over y}\right)^s
\end{equation}
to arrive at
\begin{equation}\label{eq:caloBeforeRegularization}
  \calo(\tau) = {\cN\over 2}\left[\sum_{\gamma\in \Gamma_\infty\backslash PSL(2,\mathbb{Z})}\int_0^\infty {d\xi\over \xi^2}B[\xi] + \sum_{(m,n)\ne (0,0)}\int_0^\infty {d\xi\over \xi ^2}B[\xi]\exp\({-{\pi |m\tau+n|^2\over \xi y}}\)\right].
\end{equation}
The last formal equality we need is that $\sum_{\gamma\in\Gamma_\infty\backslash PSL(2,\mathbb{Z})} = E_0(\tau) = 1$, a regularization inherited from the meromorphic continuation of the Eisenstein series. We can then combine the two terms and make use of the inversion property (\ref{inv}) satisfied by the $SL(2,\mathbb{Z})$ Borel transform to arrive at
\begin{equation}
  \calo(\tau) = {\cN\over 2} \sum_{(m,n)\in\mathbb{Z}^2}\int_0^\infty {d\xi}{B[\xi]\over \xi}\exp\({-{\pi \xi |m\tau+n|^2\over y}}\)\,.
\end{equation}
This is the final result up to the determination of the normalization constant $\cN$. One may borrow the analysis of e.g. \cite{Dorigoni:2021guq}, done there for the specific observable $\calo = \cG_N$ but applicable more generally, to find $\cN = 1/2$. Thus we have
\es{Flattice}{\boxed{\O(\t) = {1\o4}\sum_{(m,n)\in\ZZ^2}  \int_0^\i d\xi {B[\xi]\o \xi} \exp\({-\pi \xi {|m+n\t|^2\o y}}\).}}

To summarize, we have constructed the observable $\O(\t)$ with zero mode $\O_0(y)$ as a Poincar\'e sum of $\O_0(y)$. This is not sensitive to any details of $B[\xi]$. The resulting formula is necessarily equivalent to its spectral decomposition. In other words, one can construct any modular function $\O(\t)\in L^2(\cF)$ uniquely modulo cusp forms from its zero mode $\O_0(y)$ in (at least) two equivalent ways: first, as a Poincar\'e sum over $\O_0(y)$, with the resulting lattice-integral expression \eqr{Flattice}; and second, as a spectral decomposition \eqr{Ospecexp}. Of course, as we've been stressing throughout this work, in the absence of cusp forms the $k>0$ sectors do not contain any new information.

This construction also makes clear that the resulting function is orthogonal to cusp forms, because any Poincar\'e sum over a zero mode has vanishing cusp form overlap: the ``unfolding trick'' reduces the overlap to a $y$ integral over the zero mode of the cusp forms, $\phi_{n,0}=0$, cf. \eqr{cuspzero}.

We observe that the constant $(m,n)=(0,0)$ term is half of the ensemble average of $\O(\t)$, cf. \eqr{Favg}.

The formula also applies to non-Borel summable $\O_0(y)$, with suitable tweaking of $B[\xi]$. Singularities in $B[\xi]$ for $\xi\in\R_+$ lead to \np terms in $\O_0(y)$, which arise from taking $B[\xi] \rar \tilde B[\xi]$ for some appropriate $\tilde B[\xi]$ whose resummation {\it is} analytic. A typical approach is ``median resummation,'' which does so by defining the resummation $\widehat \O_0(y)$ by integrating $\Re B[\xi]$ instead of $B[\xi]$ --- see e.g. \c{Dorigoni:2014hea}. Since the derivation of \eqr{Flattice} only used the definition \eqr{borelf}, it exists even for non-Borel summable cases upon substituting $\tilde B[\xi]$ instead.

\ssec{Integrated correlator II: Derivation}
Everything we have established in the previous subsections is beautifully on display for the integrated correlator $\cG_N(\t)$, whose form we are able to efficiently explain and derive. 

In terms of the overlap functions in \eqr{form}, let us write
\e{eq548}{\{\cG_N,E_s\} = {\pi\o \sin\pi s}s(1-s)f_{{\rm p},N}(s) + f_{{\rm np},N}(s)}
For $SU(2)$, from \eqr{cs2} and \eqr{cg2overlap} we have 
\e{}{f_{{\rm p},2}(s) = (2s-1)^2\,,\quad f_{{\rm np},2}(s) = 0}
with the $SU(N)$ results determined by recursion \eqr{rec}. This is extremely simple. In fact, it is the {\it simplest} possible non-trivial overlap consistent with the necessary constraints! Setting $\fnp(s)=0$ to eliminate \np terms, we recall from Section \ref{sec:analyticStructureCFTOverlaps} that $\fp(s)$ must then be an entire, real, even function of $s-\half$. The simplest entire functions are polynomials. If $\fp(s)$ is constant, the corresponding $\sl$ Borel transform is $B[\xi] \propto 2\xi^2/(1+\xi)^3$, which is sign definite for $\xi> 0$, contradicting \eqr{Bvanish}. The next simplest possibility is the degree-two monomial, which is the case realized by $\cG_2(\t)$.

The lattice-integral formula for $\cG_N(\t)$ was given in \eqr{dgwmain}. Comparing to \eqr{Flattice}, we see immediately that 
\e{BDGW}{{B[\xi]\o 2\xi} = B_N(\xi)}
The kernel of the lattice-integral is simply the $\sl$ Borel transform of the zero mode. (One can check for sanity, e.g. by integrating numerically, that the $\sl$ Borel resummation of $\cG_{N,0}(y)$ is equal to the Borel resummations in \c{Dorigoni:2021guq} that were performed with respect to more conventional kernels.)

It is satisfying that this derivation elucidates the meaning of $B_N(\xi)$, which had previously been opaque. We also understand why it must be independent of the instanton index $k$, and why it obeys the inversion symmetry
\e{}{\sqrt{\xi} B_N(\xi) = {B_N(\xi^{-1})\o \sqrt{\xi}} }
This was seen to be crucial for the $\sl$ invariance of the lattice sum in \cite{Dorigoni:2021guq}; we now understand inversion symmetry as a consequence of $\sl$ invariance. Moreover, the integral of $B_N(\xi)$ is simply its ensemble average, as implied by \eqr{Favg}:
\e{}{\< \cG_N\> = \int_0^\i d\xi B_N(\xi) = {N(N-1)\o 4}}
where the second equality uses the result previously extracted from the spectral decomposition, cf. \eqr{cgnavg}. In addition, due to \eqr{Bvanish}
\e{}{\int_0^\i {d\xi\o \sqrt{\xi}} B_N[\xi] = 0}
These two equalities were observed in \cite{Dorigoni:2021guq}. 

In Subsection \ref{poincsec} we showed that the lattice-integral representation \eqr{Flattice} is {\it derived} from the zero mode $\O_0(y)$. Applying this to $\O = \cG_N$ thus furnishes a proof of the formula \eqr{dgwmain}, taking only the perturbative expansion of the zero mode $\cG_{N,0}(y)$ as input. Moreover, \eqr{Flattice} is manifestly equivalent to the spectral decomposition. To make the proof truly complete, it remains to show that $(\cG_N,\phi_n)=0$. This follows from showing that the one-instanton mode $\cG_{N,1}(y)$ is fully reproduced by the lattice-integral representation \eqr{dgwmain}, so in total, one needs both $\cG_{N,0}(y)$ and $\cG_{N,1}(y)$ as input. 

Finally, let us check the radius of convergence formula \eqr{Rk}, which should hold exactly for $\cG_N(\t)$. At $k=0$, one quickly deduces that $R=1$, so \eqr{Rk} predicts that
\e{Rkgn}{R_k = (1+k)^2}
To test this, it is convenient to use equation (3.22) of \cite{Dorigoni:2021guq} at $N=2$,
\es{}{\cG_{2, k} (y) &= e^{-2\pi k y} \mathop{\sum_{p,q \,\in\,\Z_+}}_{p q =k}  z^2 (p+q) [ z \left(11 p^2+2 p q+11 q^2\right) (p-q)^2+2 z^2 (p+q)^2
   (p-q)^4 \\
   & +9 p^2 -12 p q+9 q^2]  - \frac{\sqrt{\pi}}{2}  z^{3\o2} e^{z (p+q)^2} \left[4 z^3
   \left(p^2-q^2\right)^4+24 z^2 \left(p^2+q^2\right) \left(p^2-q^2\right)^2 \right. \\ 
 &\left.  +\, 3 z \left(9 p^4 -2 p^2 q^2+9 q^4\right) + 3 \left(p^2+q^2\right)\right]
   \text{erfc}\left(\sqrt{z} (p+q)\right) }
where $z\coloneqq \pi y$. The leading $y\gg1$ behavior comes from the complementary error function alone, which admits an expansion
\e{}{\pi (p+q)\sqrt{z}\,e^{z (p+q)^2}\text{erfc}\left(\sqrt{z} (p+q)\right) =\sum_{n=0}^\i \(-{1\o \pi(p+q)^2}\)^{n} \G\Big(n+\half\Big) y^{-n}}
Therefore, applying our previous definitions for the $\sl$ Borel transform,
\e{}{R_k = \underset{p,q}{\text{max}}  \big[(p+q)^2\big]~~~\text{s.t.}~~~ {p,q \in\Z_+}\,,~{p q =k}}
The maximum is given by $(p,q) = (k,1)$, thus proving \eqr{Rkgn}. The result for general $N$ follows from recursion.

\sssec{Comment on integrated $\la 22pp\ra$ correlator}

One may instead consider the supersymmetric integrated $\< 22pp\>$ correlator, where ``$p$'' denotes the half-BPS superconformal primary scalar $\O_p$ of dimension $\D=p$ in the $[0\,p\,0]$ representation of $SU(4)_R$. Let us call this quantity $\cG_{22pp}(\t)$, slightly abusing earlier notation and leaving the $N$-dependence implicit. $\cG_{22pp}(\t)$ may also be computed via localization. Adding sources for integrated $\O_p$ to $Z_{S^4}(\t;m)$,
\e{}{Z_{S^4}(\t,\t_p;m) \coloneqq  \int d^{N-1} a \(\prod_{i<j} {a_{ij}^2 H^2(a_{ij})\o H(a_{ij}-m)H(a_{ij}+m)}\) |Z_{\rm inst}(\t,\t_p)|^2 \exp\({-2\sum\limits_{p=2}^\i \pi^{p\o2} \Im(\t_p')\sum_i (a_i)^p}\)}
where $\t'_2 \coloneqq \t$, one then differentiates with respect to sources to obtain
\e{22ppdef}{\cG_{22pp}(\t) = \p_{\t_p'}\p_{\tb_p'} \p_{m}^2 \log Z_{S^4}(\t,\t_p;m)\big|_{m=\t_p'=0}}
This is slightly schematic due to operator mixing issues on $S^4$; see \c{Gerchkovitz:2016gxx,Binder:2019jwn} for an exact expression.

What is the spectral decomposition of $\cG_{22pp}(\t)$? In particular, we are interested in whether the $p$-dependence affects the structural simplicity seen for $p=2$. We address this by examining $y\rar\i$ perturbation theory, setting $Z_{\rm inst}(\t,\t_p)=1$. Compared to the $p=2$ case, the derivatives of $\log Z_{S^4}(\t,\t_p;m)$ simply bring down $p-2$ extra powers of the eigenvalues $a_i$. These powers multiply $p$-dependent coefficients. These coefficients do not contain factorials, nor do the manipulations leading to $\cG_{22pp}(\t)$, including the mixing matrix, lead to any. Therefore, generalizing the computations for $\cG_N(\t)$, we learn that {\it the perturbative expansion of $\cG_{22pp}(\t)$ is Borel summable.} Applying the logic of resurgence for generic observables would imply that $\fnp(s)=0$ for $\cG_{22pp}(\t)$ as well. 

As for the cusp forms, we have not tried to formulate a rigorous argument as to their absence; this would involve the structure of $Z_{\rm inst}(\t,\t_p)$, whose explicit form is unknown. However, the structural uniformity in $p$ of the localization integral suggests (though certainly does not prove) the following stronger statement: $\cG_{22pp}(\t)$ is Borel summable, and obeys $\fnp(s)= (\cG_{22pp},\phi_n)=0$ for all $N$ and $p$. 

\ssec{Including cusp forms}\label{seccusp}

We now re-introduce cusp forms $\phi_n(\t)$ to the preceding analysis. The upshot is equation \eqr{cuspdiag2} for the radius $R_k$ that generalizes \eqr{Rk} by including cusp forms, which follows rather intriguingly from a deep number-theoretic property of their Fourier expansions.

Let us first explain why determining the cusp form contribution to $\O_{k>0}(y)$ is inherently much harder, for both physics and math reasons, than determining the continuous, Eisenstein contribution.

Maass cusp forms exhibit arithmetic chaos \cite{sarnak,hejrack}. The fundamental data characterizing the cusp forms --- namely, the spectral parameters $t_n$ and the Fourier coefficients $a_k^{(n)}$ --- are not known explicitly, but are known to obey chaotic distributions. For example, the $t_n$ exhibit Poissonian statistics at $n\gg1$, while the $a_k^{(n)}$ for prime $k$ are conjecturally asymptotically equidistributed with respect to a semi-circle distribution, at either large $k$ or large $n$.\foot{The word ``arithmetic'' refers to the arithmeticity of $SL(2,\Z)$, or to the existence of an infinite number of mutually commuting generators, known as Hecke operators, of which each cusp form is an eigenvalue. This symmetry leads to ``milder'' forms of chaos, e.g. Poissonian statistics for the $t_n$, than is seen in other contexts.} There is good ``experimental'' evidence for these claims \cite{hejrack,1993MaCom..61..245H,Steil:1994ue}. We will utilize one such property more explicitly below. 

At a technical level, this means that it is infeasible to study the sum over $\phi_n(\t)$ analytically. One could instead imagine approximating the discrete sum by treating $(\O,\phi_n)$ distributionally, then performing contour deformations in the complex $s_n$ plane away from the critical line $s_n=\half+it_n$. But besides being rather radical (and not, to our knowledge, something done in the mathematics literature), since the overlap $(\O,\phi_n)$ is not determined as a Mellin transform, unlike the Eisenstein sector, its asymptotics as $n\rar\i$ are anyway not obviously constrained by analyticity of the Fourier modes of $\O(\t)$. 

Nevertheless, we {\it can} incorporate the presence of cusp forms.

First, note that the relation \eqr{Rk} gives a one-way diagnostic of the presence of cusp forms. If \eqr{Rk} does not hold for some $k$, then $(\O,\phi_n)\neq 0$ for some $n$: 
\e{cuspdiag}{{R_k\o R} \neq {\(1+{k\o R}\)^2}\quad \Rightarrow \quad (\O,\phi_n)\neq 0}
What about the converse? It is not obvious that cusp forms must modify $R_k$ at all. For example, their contribution could (as far as we can see) be convergent or even purely non-perturbative, in which case they give a vanishing contribution to the Borel resummation, and hence to $R_k$.\foot{This happens if, for example, $(\O,\phi_n)\neq 0$ for a finite subset of $\phi_n(\t)$.} However, let us consider below the interesting situations, where the cusp forms give a factorially divergent contribution to perturbation theory, so that their contributions are not swamped by everything else. Then in fact, \eqr{cuspdiag} holds in both directions, as we now show. 

The contribution of $\phi_n(\t)$ to the $y^{-m}$ term of $\O_k(y)$ is given by\foot{The $L^2$-normalized cusp forms $\phi_n(\t)$ being used throughout this paper have Fourier coefficients  $a_k^{(n)}\widetilde a_1^{(n)}$, where $\widetilde a_1^{(n)} := 1/\sqrt{(\nu_n,\nu_n)}$ is simply a norm factor. Recall that $a_1^{(n)}=1$, cf. \eqr{a1nhecke}.}
\e{Okcusp}{\O_k(y)\big|_{y^{-m}} \supset {1\o (-4\pi k)^m m!} \sum_{n=1}^\i a_k^{(n)}\widetilde a_1^{(n)}{(\O,\phi_n)}\(\half+i t_n\)_m\(\half-i t_n\)_m}
The full $y^{-m}$ term is given by the sum of \eqr{Okcusp} and \eqr{Ok'}. Let us use $R_1^{(\phi)}$ to denote the radius of convergence of the $\sl$ Borel transform of the cusp contribution at $k=1$:
\e{}{\sum_{n=1}^\i \widetilde a_1^{(n)}{(\O,\phi_n)}\(\half+i t_n\)_m\(\half-i t_n\)_m \,\sim\, (m!)^2 4^m  \Big(R_1^{(\phi)}\Big)^{-m} \qquad (m\rar\i)\,.} 
Likewise, use $R_1^{(E)}$ to denote the radius of convergence of the $\sl$ Borel transform of the Eisenstein contribution at $k=1$, determined previously in \eqr{Rk}. The total radius of convergence of $B_k[\xi]$ at $k=1$ is the minimum of the convergence radii in the cusp and Eisenstein sectors,
\e{}{R_1 = \text{min}\(R_1^{(E)},R_1^{(\phi)}\)}
We now ask what is the cusp contribution at $k>1$. Associated to it will be a radius of convergence $R^{(\phi)}_{k}$, such that $R_k$, the total radius of convergence of $B_k[\xi]$, is
\e{}{R_k = \text{min} \(R_k^{(E)},R^{(\phi)}_{k}\)}

The Fourier coefficients $a_k^{(n)}$ are believed to obey the {\it Ramanujan-Petersson conjecture}: for all $n$,
\e{}{|a_k^{(n)}| \leq 2\qquad (k\text{ prime})\,.}
This implies that the sole $k$-dependence of $R^{(\phi)}_{k}$ comes from the $k^{-m}$ term in the Fourier expansion of the Bessel function, because it is not possible for the sum over cusp forms to generate any parametric $k$-dependence. This is in distinction to the Eisenstein series where $a_k(s) = \s_{2s-1}(k) k^{-s} \rar k^{s-1}$ at $k\gg1$, and hence the integral over $s$ generated the $k$-dependence seen earlier. In fact, instead of invoking Ramanujan-Petersson we can use the best {\it proven} bound, of Kim and Sarnak \c{kimsarn}: for all $n$,
\e{}{|a_{k}^{(n)}| \leq k^{7/64} + k^{-7/64}\qquad (k\text{ prime})\,.}
This scaling is still insufficient to change the asymptotics because it does not depend on $n$. Thus, \eqr{Okcusp} implies
\e{}{R^{(\phi)}_{k} = k R_1^{(\phi)} \qquad (k\text{ prime})\,.}
Comparing $R^{(\phi)}_{k}$ with $R^{(E)}_k$, we see that at sufficiently large prime $k=k_*$, the former must become smaller ---  for any values of the $k=1$ radii --- because it grows linearly, rather than quadratically, in $k$. Hence, 
\e{}{R_{k> k_*} = k R_1^{(\phi)}\qquad (k\text{ prime})\,,}
where $k_*$ is defined as the (integral part of the) largest root of the quadratic equation $k_* R_1^{(\phi)} = {R_{k_*}^{(E)}}$, with $R_{k_*}^{(E)}$ defined in \eqr{Rk}. 

Actually, we can drop the prime condition by using Hecke relations (e.g. \cite{Iwaniec2002SpectralMO}). The Hecke operators $T_k$ act on cusp forms as
\e{}{T_k\, \phi_n = a_k^{(n)} \phi_n}
and obey $T_jT_k = \sum_{d|(j,k)} T_{j k/d^2}$. This determines the $a_k^{(n)}$ for non-prime $k$ as a polynomial in the $a_k^{(n)}$ for prime $k$, for every $n$. Therefore, all $a_k^{(n)}$ obey a polynomial bound in $k$ that is independent of $n$. 

Altogether, when cusp forms contribute to the Borel transform we have the following result: 
\es{cuspdiag2}{{R_k\o R} = \min\( \(1+{k\o R}\)^2\,,~{k\o R}\,R_1^{(\phi)} \)}
    where $R_1^{(\phi)}$ is infinite if $(\O,\phi_n)=0$ for all $n$. This also implies that \eqr{cuspdiag} holds in both directions, providing a rigorous, albeit mildly conditional, two-way diagnostic of the presence of cusp forms. More conceptually, this subsection demonstrates a simple point: {\it instantons in $\cN=4$ SYM exhibit arithmetic chaos, encoded in the cusp forms.} If it happens that $(\O,\phi_n)=0$, as is the case for the integrated correlator $\mathcal{G}_N(\t)$, one could rightfully call $\O(\t)$ non-chaotic in this sense.

\ssec{On the strong coupling expansion}
To conclude this section, let us ask what the spectral decomposition implies about the {\it strong} coupling expansion, $y\rar 0$. To be clear, S-duality implies that strong coupling can be mapped to weak coupling. The purpose of this short coda is to derive some precise consequences of this.

To develop the $y\rar 0$ expansion of $\O_0(y)$, we again start from the first equation of \eqr{Ospecexp}, but now deforming to the right, $\Re s >\half$. This yields
\e{strong}{\O_0(y) 
 = \<\O\> +\sum_{s=2}^\i (-1)^s s(s-1) \fp(s)\L(s) y^{s} \qquad (y\rar0)}
There are a few notable features here. First, there is a universal factor $\varphi(s)$ relating the coefficients of the weak and strong coupling expansions of $\O_0(y)$, seen by comparing \eqr{fpasymp} and \eqr{strong}. In \cite{Dorigoni:2021guq}, this was regarded as a striking observation for $\cG_N(\t)$. We are showing that it is a simple consequence of $SL(2,\ZZ)$-invariance for any square-integrable observable. Note that $\varphi(s\gg1) \sim 1/\sqrt{s}$, so the asymptotics at $y\rar 0$ and $y\rar\i$ are essentially identical. Second, there is never an $O(y)$ term, which follows from $\Res_{s=1}R_{1-s}[\O]= \Res_{s=1}\(\varphi(s)^{-1}R_s[\O]\)=0$. Finally, the ensemble average equals the zero mode at $y=0$:
\e{avgstrong}{ \<\O\> = \O_0(y= 0)}
In other words, dialing $\O_0(y)$ from weak to strong coupling adds a factor of the ensemble average. This is a non-trivial statement, notwithstanding S-duality: we are taking the strong coupling limit of the {\it zero mode}, which is not $SL(2,\Z)$ invariant, not of $\O(\t)$, which is $SL(2,\Z)$ invariant. We will see a large $N$ version of this statement in Section \ref{sec:averageAndSugra}, with powerful implications. 

\sec{Large $N$: Generalities}\label{sec:largeN}

We will be concerned with two types of large $N$ limits. 

The first limit is the 't Hooft limit, defined as $N\rar\i$, $g\rar 0$ with $\l\coloneqq  g^2 N$ fixed. In terms of our variable $y=\Im\t$, 
\e{}{N\rar\i, ~~\l = {4\pi N\o y}~\text{fixed} \qquad \text{('t Hooft)}}
In the 't Hooft limit, the perturbative $1/N$ expansion organizes into a genus expansion. Instantons are non-perturbatively suppressed as
\e{nzmode}{\O_k(y) \propto e^{-2\pi k y} \sim e^{-8\pi^2k  {N \o\l}}}
so the genus expansion applies to the zero mode, 
\e{genus}{\O_0(y) = \sum_{g=0}^\i N^{2-2g}\, \O_0^{(g)}(\l)}
There are also other non-perturbative corrections in $N$ which correct the zero mode itself at each order in the genus expansion, to be derived below.

The second limit is $N\rar\i$ with $g$ fixed, sometimes called the ``very strongly coupled limit'' \c{Azeyanagi:2013fla}:
\e{vsclim}{N\rar\i, ~~y~\text{fixed}\qquad \text{(very strongly coupled)}}
The genus expansion \eqr{genus} does not hold in the very strongly coupled limit, but taking $y$ fixed and then sending $y\rar\i$ recovers the leading $\l\gg1$ behavior in the 't Hooft limit.

Before delving into details, let us state one of the main overarching points here. As first explained in the previous section on perturbation theory at finite $N$, to develop a perturbative expansion in some parameter from the spectral decomposition, one first expands the spectral overlap $\{\O,E_s\}$, and then performs a contour deformation to pick up poles. The presence or absence of {\it non-perturbative} corrections to the resulting expansion is encoded in the $|s|\rar\i$ asymptotics of the overlap. In the large $N$ expansions to follow, this approach will reveal various corrections. The power of $\sl$ is that perturbative and \np physics are intertwined into one and the same automorphic object. 

We first need to understand how the spectral overlaps are constrained. Let us present, then explain, the result. At large $N$, the spectral overlap $\{\O, E_s\}$ is of the following form:
\es{nspec}{\boxed{\{\O, E_s\} = \sum_{g=0}^\i N^{2-2g}\({\pi \o \sin\pi s} s(1-s)\(N^{-s} \fp^{(g)}(1-s) + (s \rar 1-s)\) + \sum_{m=0}^\i N^{-3-{m\o 2}} \fnp^{(g,m)}(s)\)}}
where $\sl$-invariance requires the reflection symmetry $\fnp^{(g,m)}(s) = \fnp^{(g,m)}(1-s)$. 

These two sets of terms are the large $N$ expansions of the perturbative and \np parts of \eqr{form}, respectively. Let us now explain \eqr{nspec} in slightly more detail. The reader interested only in its consequences may skip to the next section. 

The first, perturbative term in \eqr{nspec} can be understood by considering the 't Hooft limit, but applies to any large $N$ limit since $\{\O, E_s\}$ is independent of $y$. The key observation is that any function $f(\l)$ is produced by an overlap with $N^{-s}$ scaling: recalling that $\{\O, E_s\}$ is $\L(s)^{-1}$ times the RS transform, we have
\es{Rb}{R_{1-s}[f(\l)] &= \int_0^\i dy\, y^{-1-s} f(\l) \\&= (4\pi N)^{-s} \int_0^\i d\l \,\l^{s-1} f(\l) \\ &= (4\pi N)^{-s} \,\mM[f(\l);s]}
where $\l$ is the Mellin integration variable in the final line. Therefore, the genus expansion \eqr{genus} implies that $\fp^{(g)}(1-s)$ contains information of the 't Hooft expansion at genus $g$, for {\it any} $\l$. The details of the $\l\ll1$ and $\l\gg1$ expansions --- including possible \np corrections in $\l$ --- are encoded in the polar structure of $\fp^{(g)}(1-s)$ for $s\in\R$. The presence of the reflected solution in \eqr{nspec} follows from the $\sl$-invariance of the spectral decomposition, i.e. the reflection symmetry of $\{\O,E_s\}$. Note that since we are expanding at $N\rar\i$ first, the polar structure of $\fp^{(g)}(s)$ can be non-trivial, unlike $\fp(s)$ at finite $N$. 

The second, non-perturbative term in \eqr{nspec} generates all exponential terms $\sim e^{-4\pi ny}$, as discussed in Section \ref{sec:analyticStructureCFTOverlaps}. These are \np in the 't Hooft limit; consistent with the previous paragraph, their $N$-scaling is independent of $s$. It may not be immediately obvious why, or how, $\fnp^{(g,m)}(s)$ generates exponentials. After all, to obtain the contribution of $\fnp^{(g,m)}(s)$, we simply integrate along the critical line $s=\half+it$; why is this non-perturbative at $y\rar\i$? The answer can be nicely understood from basic properties of the Rankin-Selberg method as adapted by Zagier \cite{zbMATH03796039}. At a fixed order in the $1/N$ expansion, $\O(\t)$ must be $\sl$ invariant, although it need not be square-integrable. However, consistency with the genus expansion in the 't Hooft limit means that the leading power of $y$ at fixed order in $1/N$ is bounded above. As shown in \cite{zbMATH03796039}, if an $\sl$-invariant function $F(\t)$ with bounded power-law growth at $y\rar\i$ has a zero mode
\e{}{F_0(y) = \sum_{i=1}^m {c_i\over n_i!}y^{\alpha_i}\log^{n_i}y + \text{(exponential)}}
for some finite $m$, with $n_i\in\Z_{\geq 0}$ and $c_i, \alpha_i\in\mathbb{C}$ with bounded $\Re(\a_i)$, then $F(\t)$ can be written as
\begin{equation}
  F(\tau) \coloneqq \sum_{i\,|\,\alpha_i\geq \half}\left. c_i{\partial^{n_i}\over\partial s^{n_i}}E_s(\tau)\right|_{s=\alpha_i} + F_{\rm spec}(\tau)
\end{equation}
The Eisenstein series subtract the powers of $y^{\alpha_i\geq \half}$ in a modular-invariant way to restore square-integrability, such that $F_{\rm spec}(\t)\in L^2(\cF)$ and hence admits a spectral decomposition. Then since the zero mode of the Eisenstein series has no exponential terms in $y$, any  exponential terms in $F_0(y)$ must come solely from $F_{\rm spec}(\t)$.

We pause to briefly note a small subtlety. Taking $F_0(y)$ to be free of logs, as in perturbative CFT applications, we have 
\e{}{F_{{\rm spec},0}(y) = \sum_{i\,|\,\alpha_i< \half} c_iy^{\a_i} - \sum_{i\,|\,\alpha_i\geq \half}c_i\varphi(\alpha_i) y^{1-\alpha_i} + \text{(exponential)}}
If the powers of $y$ do not exactly cancel (and they need not), then $F_{{\rm spec},0}(y)$ has, in addition to the exponentials, spurious powers of $y$ that are not present in $F_0(y)$ (and vice-versa). In order to obtain these powers from the spectral integral $F_{\rm spec}(\t)$, one should shift the spectral contour, picking up residues of spurious poles of $\{F_{\rm spec},E_s\}$ along the way.\foot{If there is a finite number of such spurious poles, this shift incurs no residue at infinity. All examples we know are of this type. This is very neatly exhibited by a certain class of solutions to inhomogeneous Laplace equations appearing in perturbative string theory and $\cN=4$ SYM correlators, as we recall in detail in Appendix \ref{app:inhomogeneous}.} The remaining spectral integral has the form of an inverse Mellin transform. 

Finally, we comment on the range of the sum over $m$ in \eqr{nspec}. Here, unlike elsewhere in this work, we use some mild input from holography: in particular, that the first non-perturbative terms in the AdS$_5 \x S^5$ effective action appear at 14-derivative order, via $D^6R^4$ invariants and their superpartners. This may be seen as a fact about type IIB string theory, or as following on general grounds from on-shell constraints in theories of maximal supergravity combined with the flat space limit of AdS$_5 \x S^5$ \c{Wang:2015jna,Wang:2015aua}. 

Summarizing and returning to our problem: the function $\fnp^{(g,m)}(s)$ contributes non-perturbative terms $(q\qb)^n$, plus a possible set of spurious poles. Moreover, since $\fnp^{(g,m)}(s)$ generates exponentials in $y$, 
\e{fnpgrowth}{\L(s)\fnp^{(g,m)}(s) \sim \G(|s|) \qquad (s\rar\pm \i)}
for every $(g,m)$.
 
One more comment before moving on. Although above we discussed non-perturbative contributions to the spectral overlap $\{\calo,E_s\}$ between a CFT observable and the Eisenstein series, essentially identical considerations apply to the overlap with the Maass cusp forms, $(\calo,\phi_n)$. In particular, the cusp forms do not enter the AdS$_5 \x S^5$ effective action until 14-derivative order, and so $(\calo,\phi_n)$ is suppressed by a factor of $1/N^3$ compared to the leading perturbative effect $f_p(s)$. Moreover, like $\fnp(s)$, the large $N$ expansion of $(\calo,\phi_n)$ must proceed in half-integer powers of $N$:
\begin{equation}
  (\calo,\phi_n) = \sum_{g=0}^\infty N^{2-2g}\sum_{m=0}^\infty N^{-3-{m\over 2}}(\calo,\phi_n)^{(g,m)}.
\end{equation}
Indeed, in the concrete example of large $N$ expansion of the integrated correlator $\cF_N(\t)$ in the very strongly coupled limit in Section \ref{subsec:F4}, we will see that these two contributions to the spectral overlap ($\fnp(s)$ and $(\calo,\phi_n)$) are intertwined, as they descend from a common modular invariant, the solution to the inhomogeneous Laplace equation described in Appendix \ref{app:inhomogeneous}.
 
We now consider the two large $N$ limits. It will prove beneficial to treat the 't Hooft limit first, where consistency with the genus expansion will strongly constrain the allowed poles of the overlap. For brevity, when referring to the \np part of \eqr{nspec} we sometimes refer to $\fnp(s)$ without the $(g,m)$ superscript, bearing in mind its large $N$ expansion. 

\sec{'t Hooft Limit}\label{sectH}

The strategy to analyze the 't Hooft limit is clear: we substitute $y=4\pi N/\l$ in the spectral decomposition \eqr{Ospecexp} and analyze it at fixed $\l$, taking further perturbative $\l\ll1$ or $\l\gg1$ limits afterwards. We introduce a useful notation 
\e{}{\tl \coloneqq  {\l\o 4\pi}}
We safely ignore the cusp forms, which are suppressed as $\sim e^{-N}$ in the 't Hooft limit. Importantly, while we will generalize in due course (see Section \ref{subsec:fnp}), we begin by setting 
\e{}{\fnp(s)=0}
This turns off the \np terms $\sim (q\qb)^n$. We employ this strategy because it leads to very strong constraints, and will allow us to cleanly identify signatures of \np physics in the large $N$ limit. 

In the 't Hooft limit, the zero mode of the completed Eisenstein series is
\e{E0}{E^*_{s,0}(y) = \L(s) N^s\tl^{-s} + (s \rar 1-s)}
Plugging \eqr{E0} and the overlap \eqr{nspec} into the spectral decomposition \eqr{Ospecexp} gives
\e{O0largen}{\O_0(\l) = \<\O\> + {1\o 2\pi i}\ints {\pi \o \sin\pi s} s(1-s) \sum_{g=0}^\i N^{2-2g} \(\L(s) \tl^{-s} + \L(1-s) N^{1-2s}  \tl^{s-1}\) \fp^{(g)}(1-s)}
The two terms under the integral are of a very different nature:

\bul The first term is manifestly consistent with the 't Hooft limit, and can be expanded at weak or strong coupling, or evaluated at finite $\tl$. At $\tl\ll1$, we deform to the left, while at $\tl\gg1$, we deform to the right. 

\bul The second term has an explicit $N^{1-2s}$ factor. Because we are taking the large $N$ limit, we are forced to deform to the right, for any $\tl$.\foot{Bearing in mind earlier remarks, we must ask whether the coefficient has a factorial divergence at $s\rar\i$, thus generating potential \np corrections in $N$. We analyze this question in the next subsection.} This generates two constraints on $\fp^{(g)}(1-s)$:

\begin{enumerate}[label=\textbf{\arabic*})]
\item {\bf \textit{Consistency with the genus expansion:}} For $\Re s >\half$, the poles of the integrand must lie at $s=\half+m$ with $m\in \Z_+$. In particular, the $\sin\pi s$ poles must be cancelled by zeros,
\e{fpzeros}{{\fp^{(g)}(s)=0\,,\quad s\in\Z_{-}}}
On the other hand, simple poles for half-integer $s$ are allowed,
\e{}{\Res_{s=\half-m}\[\fp^{(g)}\(s\)\] \neq 0\,,\quad m\in \Z_{+}}
In addition, $\L(1-s)$ has a pole at $s=1$, which contributes a constant term. Taking this all into account, we have, for fixed $\tl$,
\es{O0exp}{\O_0(\l) = &\(\<\O\> - \half \sum_{g=0}N^{1-2g} \fp^{(g)}(0)\) - \sum_{g=0}^\i\sum_{m=1}^\i  N^{2-2g-2m}  \varphi\Big(\half+m\Big)  \mathsf{R}_m^{(g)}\tl^{m-\half}\\ &+  {1\o 2\pi i}\sum_{g=0}^\i N^{2-2g}\ints  {\pi \o \sin\pi s} s(1-s)  \L(s)\tl^{-s} \fp^{(g)}(1-s) }
where we define a ``residue function''
\es{alphadef}{\mathsf{R}_m^{(g)} &\coloneqq  \Res_{s=\half+m}\Big[{\pi\o \sin \pi s} s(1-s)\L(s)\fp^{(g)}(1-s)\Big]}
We can now deform this as needed to develop perturbative expansions. 

\item {\bf \textit{Consistency with the $\tl\ll1$ expansion:}} These residues at $s=\half+m$ generate allowed powers of $1/N$, but multiplied by {\it half-integer} powers of $\tl$, plainly visible above. These terms are present for any fixed $\tl$. They must cancel contributions from the remaining integral when we go to weak coupling, $\tl\ll1$, where only integer powers of $\tl$ are allowed. Developing the $\tl\ll1$ expansion by deforming \eqr{O0exp} to the left, the offending terms are cancelled if and only if the following condition on the residues is met:
\e{cancel}{\varphi\Big(\half+m\Big)\,\mathsf{R}_m^{(g)}= \mathsf{R}_{-m}^{(g+m)}}
This intriguing condition relates perturbative data of different genera. 

\end{enumerate}

After the dust settles, we have the following weak coupling expansion
 \es{O0weak'}{{\O_0(\l\ll 1) \approx \sum_{g=0}^\i N^{2-2g}  \sum_{m=1}^\i(-1)^{m+1} m(m+1) \L\Big(\half+m\Big)\fp^{(g)}(1+m)\,\tl^{m}}}
In terms of the residue function,
  \es{O0weak}{\boxed{\O_0(\l\ll 1) \approx \sum_{g=0}^\i N^{2-2g}  \sum_{m=1}^\i \mathsf{R}_{-m-\half}^{(g)} \,\tl^{m}}}
Note that the absence of $\log\tl$ terms, or any other non-integer powers of $\tl$, requires that $\fp^{(g)}(s)$ is regular for all $\Re s>\half$ away from the half-integer poles appearing in the cancellation condition \eqr{cancel}. We have thus completely determined the structure of allowed poles and zeros of $\fp^{(g)}(s)$. 

The payoff of this work happens at strong coupling, $\tl\gg1$. Deforming the second line of \eqr{O0exp} to the right now, and applying the previous constraints on the polar structure of $\fp^{(g)}(s)$, yields the strong coupling expansion

\begin{empheq}[box=\fbox]{equation}\label{O0strong}
\begin{split}
\O_0(\l\gg1) &\approx \mathsf{C}(N)- \sum_{g=0}^\i\sum_{m=1}^\i N^{2-2g}\(N^{-2m}\varphi\Big(\half+m\Big) \mathsf{R}_m^{(g)} \tl^{m-\half}+\mathsf{R}^{(g)}_m \tl^{-{1\o2}-m} \)
\end{split}
\end{empheq}

The first term is a constant,
\e{}{\mathsf{C}(N) \coloneqq  \<\O\> - \half \sum_{g=0}^\i N^{1-2g} \fp^{(g)}(0)}

The next group of terms are the terms described earlier, but in the strong coupling expansion they do not cancel with anything. They may at first seem peculiar --- a finite set of {\it positive} powers of $\l$ at each even power of $1/N$ --- but they are precisely what is needed for the renormalization of the $1/N$ expansion at strong coupling. As explained in \c{Aharony:2016dwx}, they are, from the gravity point of view, the string theory regularization of loop-level divergences in AdS$_5 \x S^5$ supergravity, where $g+m=L$ with $L$ the loop level. Consistent with this, they are $1/N^2$ suppressed compared to the leading order.\foot{At $N^0$, for example, $(g,m)=(0,1)$ gives a $\sqrt{\l}$ term, which is precisely the regularization of one-loop supergravity by $R^4$ (and its superpartners) in AdS$_5 \x S^5$. At $1/N^2$, $(g,m) = (0,2)$ is the regularization of two-loop supergravity by $D^4R^4$, while $(1,1)$ is the regularization of one-loop supergravity corrected by the tree-level $R^4$ term. This pattern continues.} 

What we are mainly interested in is the last group of terms, containing the conventional genus sum over perturbative series in $1/\tl$. Only half-integer powers of $\tl$ appear, a fact which we have derived solely from consistency conditions on the weak coupling and genus expansions. This expression contains much physics, which we analyze in stages, beginning now with the implications for \np effects. 

\ssec{Non-perturbative effects implied by $\sl$}\label{subsubsec:nonpertSL2Z}

In this subsection we derive a surprising result: as a direct consequence of S-duality, the convergence of a weak coupling expansion at $\l\ll1$ directly implies the existence of \np corrections in both $\l\gg1$ and in $N\gg1$, with the precise \np scales set by the {\it weak} coupling radius of convergence. 

\sssec*{Non-perturbative corrections in $\l$}
Let us here reproduce the $\l\ll1$ expansion \eqr{O0weak} and the $\l\gg1$ expansion \eqr{O0strong} at fixed genus $g$, dropping terms terms in $\O_0^{(g)}(\l\gg1)$ that do not contribute asymptotically and writing the residue function in terms of $\fp^{(g)}(s)$:
\es{thsum}{\O_0^{(g)}(\l\ll1)&= -\sum_{m=1}^\i(-1)^{m} m(m+1) \L\Big(\half+m\Big)\fp^{(g)}(1+m)\,\tl^{m}\\
\O_0^{(g)}(\l\gg1) &=-\sum_{m=1}^\i(-1)^{m}\pi  \Big(m-\half\Big) \Big(m+{1\o2}\Big)\L\Big(\half+m\Big)\Res_{s=\half-m}\[ \fp^{(g)}(s)\] \tl^{-{1\o2}-m}}
This makes clear that the $\l\ll1$ and $\l\gg1$ convergence are determined by the $|m|\rar\i$ asymptotics of $\fp^{(g)}(m)$ in opposite directions along $\R$.

Suppose that the $\l\ll1$ expansion is convergent, with radius of convergence $|\l|\leq \l_*$. This is the typical situation in planar $\cN=4$ SYM. Planar combinatorics determine $\l_*=\pi^2$ for generic $\O(\t)$ \cite{cmp/1103901558,Beisert:2006ez}, but for the sake of generality we will leave $\l_*$ explicit. At $m\rar +\i$, 
\e{}{\L\Big(\half+m\Big) \approx \G(m+1)\frac{\pi ^{-m-\frac{1}{2}}}{\sqrt{m}}\qquad (m\rar\i)}
This implies that\foot{Here and henceforth in this subsection, we use the symbol $\sim$ to denote the asymptotic behavior up to multiplicative power law corrections and beyond.}
\e{}{\fp^{(g)}(m+1) \sim \G(m+1)^{-1}\({4\pi^2\o \l_*}\)^{m}\qquad (m\rar\i)}
In particular, $\fp^{(g)}(x)$ decays factorially at large positive argument. We now make a technical assumption, borne out by examples, that the $|x|\rar\i$ limit of $\fp^{(g)}(x)$ commutes with $x\rar-x$.\foot{This is weaker than, but included in, the natural condition that $\fp^{(g)}(s)$ admit a uniform asymptotic expansion for $s\in\CC$.} This in turn implies
\e{fpoddres}{\Res_{s=\half-m}\[\fp^{(g)}\(s\)\] \sim \G\Big(\!-m+\half\Big)^{-1}\({4\pi^2\o \l_*}\)^{-m}\qquad (m\rar\i)}
We plug these into \eqr{thsum}. For large positive integer $m$, 
\es{}{\G(m+1)\G\Big(\!-m+\half\Big)^{-1}&\sim (-1)^m4^{-m}\G(2m)\qquad (m\rar\i)}
Inserting into the strong coupling expansion \eqr{thsum}, we see that the $\l\gg1$ expansion diverges double-factorially without alternating sign! This is a hallmark of Borel non-summability. A summand scaling asymptotically as 
\e{}{ \({4\pi^2 \l\o \l_*}\)^{-m} \G(2m)}
requires non-perturbative corrections in powers of $e^{-{2\pi\o \sqrt{\l_*}}\sqrt{\l}}$. 
Therefore, the full expression for the zero mode at $\l\gg1$, perturbatively in $1/N$, is
\e{nplargeth}{\boxed{\O_0(\l\gg1) = \sum_{g=0}^\i N^{2-2g} \sum_{n=0}^\i e^{-{2\pi n\o \sqrt{\l_*}}\sqrt{\l}}\,\O_{0,n}^{(g)}(\l)}}
where $\O_{0,n}^{(g)}(\l)$ dresses the $n$'th exponential correction by a perturbative series in $1/\l$. The purely perturbative $n=0$ piece is given explicitly in \eqr{O0strong}, while the \np $n>0$ terms are obtained by applying resurgence to $\O_{0,0}^{(g)}(\l)$. 

It is remarkable that the numerical factor in the exponent is determined by $\l_*$, the {\it weak coupling} radius of convergence.\foot{In principle $\l_*$ could depend on $g$, though we are not aware of any such examples. If this happens, one should take $\l_* \rar \l_{*,g}$ in this formula.} For the canonical radius $\l_*=\pi^2$, the \np corrections are in powers of $e^{-2\sqrt{\l}}$. 

\sssec*{Non-perturbative corrections in $N$}

The exact same argument that led to \eqr{nplargeth} also applies to the second term in \eqr{O0largen}. The contour deformation develops a formal series in positive integer powers of $\l/N^2$, which diverges double-factorially at $s\rar\i$ with the same leading-order asymptotics. The previous analysis carries over, but with the substitution $\tl \rar N^2/\tl$. This yields \np corrections to the 't Hooft limit, for fixed $\tl$, in powers of $\exp({-{8\pi^2\o \sqrt{\l_*}}{N\o \sqrt{\l}}})$. 

Thus, for an $SL(2,\Z)$-invariant observable $\O(\t)$ with $\fnp(s)=0$ whose $\l\ll1$ expansion has radius of convergence $\l_*$, the 't Hooft expansion in $1/N$ for fixed $\l$ receives non-perturbative corrections exponentially small in $N$, of the following form:
\es{thfull}{\boxed{\O_0(\l) = \sum_{g=0}^\i N^{2-2g} \(\O_0^{(g)}(\l) + \O_{0,{\rm np}}^{(g)}(N,\l)\)}}
where 
\e{}{{\O_{0,{\rm np}}^{(g)}(N,\l) = \sum_{n=0}^\i e^{-{8\pi^2 n\o \sqrt{\l_*}}{N\o\sqrt{\l}}}\, \O_{0,{\rm np}|n}^{(g)}\({N^2\o \l}\)}}
$\O_{0,{\rm np}|n}^{(g)}\({N^2\o \l}\)$ dresses the $n$'th exponential correction by a perturbative expansion in ${\l\o N^2} \ll 1$. The purely perturbative piece $\O_{0,{\rm np}|0}^{(g)}\({N^2\o \l}\)$ was given in the first line of \eqr{O0exp},
\e{O00hat}{\O_{0,{\rm np}|0}^{(g)}\({N^2\o \l}\) = - {1\o N} \sum_{m=1}^\i   \varphi\Big(\half+m\Big) \mathsf{R}_m^{(g)}\({4\pi N^2\o\l}\)^{\half-m}}
We will compute $\O_{0,{\rm np}}^{(g)}(N,\l)$ for $\O(\t) = \cG_N(\t)$ by resurgence in the next subsection.\foot{This is a slight abuse of notation since $\O_{0,{\rm np}|0}^{(g)}\({N^2\o \l}\) $ is itself perturbative in $N$. This is the series on which we perform resurgence to determine the $n>1$, genuinely non-perturbative terms. Note that compared to the computation of $\O_{0,n}^{(g)}(\l)$, the two perturbative series to which we apply resurgence differ by a factor of $\varphi(s)$. This will slightly modify the details of the non-perturbative corrections, as we will see applied to $\mathcal{G}_N(\t)$ in Subsection \ref{subsubsec:npCorrectionsIntegratedCorrelator}.} Let us stress once again that $\l_*$, the radius of convergence at weak 't Hooft coupling, controls the strength of these non-perturbative corrections.

\sssec*{Nonzero modes}

We can also include nonzero modes, i.e. instantons, suppressed exponentially as \eqr{nzmode} in the 't Hooft limit. In each $k$ sector, one should expect a structure like \eqr{thfull} --- the sum of a perturbative (in $N$) piece and non-perturbative corrections of the same general form as above --- with an overall exponential suppression. A mechanical calculation generalizing the previous subsections, now using the Bessel asymptotics appearing in the nonzero modes, yields
\e{Okth}{\O_{k>0}(y) = e^{-8\pi^2 k {N\o\l}}\sum_{r=1}^\i N^{{3\o2}-r}\O_{k,r}(\l)} 
where
\e{}{\O_{k,r}(\l) \coloneqq  \sum_{g=0}^\i\sum_{m=1}^\i\sum_{n=0}^\i  f_{g,m,n}(\l;k)\,\delta_{2g+m+n,r}}
with
\e{}{f_{g,m,n}(\l;k) \coloneqq  -{\mathsf{R}^{(g)}_m}\,{\sigma_{2m}(k)k^{-\half-m}\o \L\Big(\half+m\Big)}{a_n\(\half+m\)\o \({8\pi^2 k}\)^n}\l^n}
where $a_n\(\half+m\)$ is the Bessel function coefficient defined in \eqr{bessela}. This represents the perturbative $1/N$ expansion of $\O_{k>0}(y)$ at fixed $\l$. Only odd half-integer powers appear, starting at $\sqrt{N}$. At a fixed order in $1/N$, there is a finite number of terms. \eqr{Okth} may receive further possible non-perturbative contributions in $N$, depending on whether the asymptotics of $\O_{k,r}(\l)$ at $r\rar\i$ are Borel summable. We will not study this here.

\ssec{Comments + String theory interpretation}\label{secstringnp}

The $1/N$ expansion of CFT observables is expected to be asymptotic. Our analysis shows that for $\sl$-invariant observables in $\cN=4$ SYM or other SCFTs, not only is it asymptotic, but \np corrections in $N$ are necessary. This is clearly true if $\fnp(s)\neq 0$, which by definition introduces \ii terms controlled by the scale $q\qb \sim \exp({-16\pi^2 {N\o \l}})$. If, on the other hand, $\fnp(s)=0$, the analysis above shows that there are {\it still} \np corrections, instead controlled by the scale $ \sim \exp({-{8\pi^2\o \sqrt{\l_*}}{N\o \sqrt{\l}}})$. It would be valuable to confirm the growth condition assumed in Subsection \ref{subsubsec:nonpertSL2Z}, on which these conclusions rely, in full generality.

Let us explain the physical nature of the \np scales. We take $\l_*=\pi^2$ here to reduce clutter. In the 't Hooft limit, the Eisenstein zero mode generates two types of terms in the integrand of \eqr{O0largen}. One term generates functions of $\l$ at every genus $g$, including terms at $\l\gg1$ suppressed by integer powers of the square of a \np scale\foot{We have chosen to write the corrections as the square of a \np scale in the spirit of \c{Alday:2007mf} (where the \np scale in $\Gamma_{\rm cusp}$ is interpretable as a mass gap of the $O(6)$ sigma model) and of various other applications of resurgence to quantum systems (see e.g. \c{Marino:2020ggm} for a recent discussion).}
\e{npscalel}{\Lambda_{\l} \coloneqq  \exp\(-\sqrt{\l}\)}
The other term generates functions of $N^2/\l$. This may be thought of as the worldsheet coupling in what we call the {\it S-dual 't Hooft limit}:
\es{}{N\rar\i\,,~~ \ls= {16\pi^2 N\o g^2}~\text{fixed}\qquad (\text{S-dual 't Hooft})}
In terms of $\l$, 
\e{}{\ls = {16\pi^2 N^2\o\l}}
Given a sequence of $\cN=4$ SYM theories parameterized by $N$, each of which enjoys an $SL(2,\Z)$ invariance, one takes the S-dual 't Hooft limit by approaching large Yang-Mills coupling as $N\rar\i$. From this point of view, the \np corrections in $N$ that we have derived are simply the S-dual of the \np corrections \eqr{nplargeth}, now suppresed by the square of a \np scale
\e{npscalels}{\Lambda_{\ls} \coloneqq  \exp\(-\sqrt{\ls}\)}
Since $\ls\propto N^2\gg1$ for any fixed $\l$, they are non-perturbative in $N$ in the ordinary `t Hooft limit. 

We can now understand the type IIB string theory duals of the various \np corrections. The AdS/CFT dictionary includes the basic entry
\e{}{\sqrt{\l} = {1\o \a'} = 2\pi T_{\rm F1}\,,\quad \sqrt{\ls} = {1\o \a' g_s} = 2\pi T_{\rm D1}}
where $T_{\rm F1}$ and $T_{\rm D1}$ are fundamental string and D-string tensions, respectively, in AdS units. Thus,

\begin{itemize}

\item Terms controlled by $\L_\l$ are fundamental string worldsheet instantons, with endpoints on the AdS$_5 \x S^5$ boundary. In AdS units, the \np scale is $\L_\l = e^{-2\pi T_{\rm F1}}$.

\item Terms controlled by $\L_{\ls}$ are D-string worldsheet instantons, with endpoints on the AdS$_5 \x S^5$ boundary. In AdS units, the \np scale is $\L_{\ls} = e^{-2\pi T_{\rm D1}}$.

\item Terms controlled by $q\qb = e^{-16\pi^2/g_s}$ are spacetime instanton-anti-instantons. 

\end{itemize}

\ni To summarize the main result in bulk terms, we have holographically argued for the non-Borel summability of string perturbation theory on AdS$_5 \x S^5$: the $g_s\ll1$ expansion must be completed by spacetime instantons, D-string instantons, or both. This generalizes previous observations about string perturbation theory in flat space \c{Gross:1988ib,Shenker:1990uf}. We also point out previous studies of \np corrections to sphere free energies in ABJM theory in the 't Hooft regime \c{Drukker:2011zy,Marino:2012zq}.\foot{Since the dual AdS$_4 \x \mathbb{CP}^3$ background solves type IIA, there are no D-strings.}

\ssec{Integrated correlator III: 't Hooft limit and D-string instantons from resurgence}\label{subsubsec:npCorrectionsIntegratedCorrelator}

Let us pause the formalism and apply everything so far to $\cG_N(\t)$. Recall that $f_{{\rm np},N}(s)=0$.

In terms of the coefficients $c_s^{(N)}$ defined in \eqr{dgweis}, the relations \eqr{cgn} and \eqr{form} imply
\e{}{f_{{\rm p},N}(s)  = {1\o s(1-s)}c_s^{(N)}}
The large $N$ expansion of the coefficients $c_s^{(N)}$ was performed in \cite{Dorigoni:2021guq}. Translating to current notation,\foot{In \cite{Dorigoni:2021guq}, the ``reflected'' solution $\fp^{(g)}(1-s)$ appearing in \eqr{nspec} was absent from the large $N$ expansion of the coefficients $c_s^{(N)}$. This was fine for their purposes because the $c_s^{(N)}$ were only used in the integer-index expansion \eqr{dgweis}, and $\fp^{(g)}(1-s)$ vanishes on all integers $s\in\Z_{>1}$, as one can confirm with the expressions \eqr{fpdgw}.}
\es{fpdgw}{\fp^{(0)}(s) &=  {2^{2s-2} (2s-1)^2  \Gamma \left(s-\frac{1}{2}\right) \o \sqrt{\pi }s \Gamma   (s+2)\G(s)}\\
\fp^{(1)}(s) &= \frac{2^{2 s-6} (2s-1)^2 (s-6) \Gamma \left(s-\frac{3}{2}\right)}{3
   \sqrt{\pi } s \Gamma (s-1)\G(s)}\\
   \fp^{(2)}(s) &= \frac{2^{2s-11} (2s-1)^2 (5 s^2 - 47 s + 30) \Gamma
   \left(s-\frac{5}{2}\right)}{45 \sqrt{\pi } s\Gamma (s-4)\G(s)}\\
     \fp^{(3)}(s) &=\frac{2^{2s-15} (2s-1)^2 \left(35 s^4-602 s^3+2749 s^2-4582 s+1680\right)
   \Gamma \left(s-\frac{7}{2}\right)}{2835 \sqrt{\pi } s\Gamma (s-6)\G(s)}}
These obey the relations
\es{}{\fp^{(g)}(0) &= -\half\delta_{g,0}\\\fp^{(g)}(s)&=0\,, \qquad s\in\Z_-}
The latter confirms \eqr{fpzeros}, while the former implies that the constant term in \eqr{O0exp} is 
\e{}{\<\O\> -\half \sum_{g=0}^\i N^{1-2g} \fp^{(g)}(0) = \<\O\> + {N\o4} = {N^2\o4}}
This is correct \cite{Dorigoni:2021guq} (see also \eqr{eq:GNLargeN}).

In the't Hooft limit, let us write\foot{The parameter $\mathfrak{g}$ differs from our previous definition of the genus in \eqr{O0largen} --- here we group terms in powers of $1/N$, rather than associating the ``renormalization terms'' to a given genus (but see the comment around \eqr{renormGN} relating the two) --- so as to allow simpler comparison to \cite{Dorigoni:2021guq}.}
\e{}{\cG_N(\l) = \sum_{\mathfrak{g}=0}^\i N^{2-2\mathfrak{g}} \cG_N^{(\mathfrak{g})}(\l)}
The order $N^2$ term is (now writing in terms of $\l=4\pi \tl$)
\e{g0spec}{\cG^{(0)}(\l) = {1\o4} + {1\o 2\pi i}\int_{\Re s=\half} ds {\pi\o \sin\pi s}s(1-s)\L(1-s)\({\l\o 4\pi}\)^{s-1} \fp^{(0)}(s)}
Again, we emphasize a main point of our treatment: {\it the spectral decomposition gives the full result for any $\l$.} In \cite{Dorigoni:2021guq}, it was shown that the (median) Borel resummation of the $\l\gg1$ limit at $\mathfrak{g}=0$ coincides with the following integral representation obtained by resumming the convergent $\l\ll1$ expansion:
\e{g0dgw}{\cG^{(0)}(\l) = {\l\o 4\pi^2}\int_0^\i dw \,w^3 \,{{}_1F_2\({5\o2};2,4;-{w^2\l\o \pi^2}\)\o \sinh^2 w}}
Computing numerically for various values of $\l$, one can confirm that \eqr{g0spec} and \eqr{g0dgw} are equal for finite $\l$. 

Let us next verify that contour deformation of \eqr{g0spec} produces the correct physics at $\l\gg1$, both perturbatively and non-perturbatively. At $\mathfrak{g}=0$, \cite{Dorigoni:2021guq} finds 
\es{cg0th}{\cG^{(0)}_N(\l) = {1\o4} + \sum_{n=1}^\i b_n^{(0)} \l^{-n-\half}\,,}
where
\e{}{b^{(0)}_n = \frac{2^{2-2 n} \Gamma \left(n-\frac{3}{2}\right) \Gamma \left(n+\frac{3}{2}\right) \Gamma (2 n+1)\zeta (2 n+1) }{\pi  \Gamma (n)^2}}
At $\mathfrak{g}=1$, \cite{Dorigoni:2021guq} finds
\e{cg1th}{\cG^{(1)}_N(\l) = -{\sqrt{\l}\o 16} + \sum_{n=1}^\i b_n^{(1)} \l^{-n-\half}\,,}
where
\e{}{b^{(1)}_n = -{n^2(2n+11)\G(n+\half)\G(n+{3\o2})^2\z(2n+1)\o 24\pi^{3\o2}\G(n+2)}}
One easily confirms the match to \eqr{g0spec} and its genus one counterpart. Notice the term linear in $\sqrt{\l}$ in \eqr{cg1th}. From our $\sl$-based point of view, this should be thought of as a renormalization term for genus zero --- in other words, while it is of order $N^0$, its origin is the first term in \eqr{O0strong} at $g=0$, with $\mathsf{R}_m^{(0)}$ determined in terms of $\fp^{(0)}(s)$ by \eqr{alphadef}. One can indeed confirm the prediction of \eqr{O0strong} for the relative coefficient of this renormalization term to the $\l^{-3/2}$ term in \eqr{cg0th}, which is
\e{renormGN}{\varphi\({3\o2}\)^{-1}(4\pi)^2 = -16 b_1^{(0)} = 48\z(3)\,,}
giving a nice confirmation of this point of view. 

Beyond perturbation theory, $\cG_N^{(\mathfrak{g})}(\l)$ also contains \np terms at $\l\gg1$ precisely as predicted by our general treatment. From the $\l\ll1$ expansion of $\cG^{(\mathfrak{g})}(\l)$, one finds a convergent expansion with $\l_*=\pi^2$. Therefore, we predict a non-Borel summable $\l\gg1$ expansion with \np corrections in powers of $\sim e^{-2\sqrt{\l}}$ (dressed by perturbative series in $1/\l$). We can see this easily enough from the spectral integrand of \eqr{g0spec}. At $\mathfrak{g}=0$,
\es{}{\text{\tt integrand of \eqr{g0spec}}&= \({\pi^{{3\o2}-s}\o \sin\pi s}(1-s){ (2s-1)^2} \)\l^{s-1}{\L(1-s)\Gamma \left(s-\frac{1}{2}\right)\o \Gamma   (s+2)\G(s)}}
At $s\rar+\i$, this behaves sub-factorially, but at $s\rar-\i$, it behaves double-factorially, $\sim \l^s (2|s|)!$, with a relative alternating sign. At $\mathfrak{g}=0$, the \np terms were derived in \cite{Dorigoni:2021guq} to be
\e{}{\cG^{(0)}_{N,0}(\l)\Big|_{\rm np} \propto  \sum_{\ell=1}^\infty a_\ell \, (2\sqrt{\l})^{ 1-\ell}\, {\mbox Li}_{\ell-1}(e^{-2\sqrt{\l}})}
where $a_\ell\in\mathbb{Q}$. This expansion proceeds in powers of $e^{-2\sqrt{\l}}$, confirming our general analysis. Similar expressions were derived in \cite{Dorigoni:2021guq} at $\mathfrak{g}>0$.

\sssec*{D-string instantons from resurgence}

We now derive the non-perturbative, D-string instanton corrections to $\mathcal{G}_N(\t)$ in the 't Hooft limit at fixed $\lambda$. As previously argued at the end of Section \ref{subsubsec:nonpertSL2Z}, the computation of the non-perturbative corrections to the 't Hooft limit (i.e. in $1/N$) at finite $\lambda$ are related to those at $\lambda \rar\i$ via resurgence of asymptotic perturbative expansions that differ by simple replacements. In particular, examining the second and third terms in (\ref{O0strong}), we see that we replace $\tilde\lambda \to N^2/\tilde\lambda$ and $\Lambda(m+\half)\to\Lambda(m)$ in the perturbative expansion. For convenience we introduce the rescaled S-dual 't Hooft coupling
\e{}{\tls \coloneqq  {\ls\o 4\pi}= {N^2\o \tl}}

As in Section \ref{subsubsec:nonpertSL2Z}, we denote\footnote{Really we should be writing $\left(\cG_N\right)^{(\mathfrak{g})}_{0,\rm{np}|n}$, but we are suppressing the subscript $N$ in an effort to keep the notation manageable.} by $\cG_{0,{\rm np}|n}^{(\mathfrak{g})}(\tls)$ the part of the $N^{2-2\mathfrak{g}}$ term that admits a purely perturbative expansion in $\tls^{-1}$ around $e^{-2n\sqrt{\ls}}$. We will focus on the non-perturbative corrections to the $\mathfrak{g}=0$ part in order to illustrate the general idea. At $\mathfrak{g}=0$, the leading term is given by
\begin{equation}\label{eq:hatGNAsymptoticSeries}
  \cG_{0,{\rm np}|0}^{(0)}(\tls) = {1\over N}\sum_{n=1}^\infty \widehat B_n^{(0)}\tls^{\half-n},
\end{equation}
where
\begin{equation}
  \widehat B_n^{(0)} = {\Lambda(n)\over\Lambda(n+\half)}{b_n^{(0)}\over (4\pi)^{\half+n}} = {4^{-2n}\pi^{-1-n}\Gamma(n-{3\over 2})(2n+1)\over \Gamma(n)}\Gamma(2n+1)\zeta(2n).
\end{equation}
This series is manifestly double-factorially divergent. To proceed, we follow \cite{Dorigoni:2021guq} in their analysis of the resurgence of the strong-coupling expansion, and define the following modified Borel transform
\begin{equation}
  \cB\left[\cG_{0,{\rm np}|0}^{(0)}\right](\xi) \coloneqq \sum_{n=1}^\infty {\widehat B_n^{(0)}\over\Gamma(2n+1)\zeta(2n)}\xi^{2n} = {2w^2(4w^2-3)\over\sqrt{\pi}\sqrt{1-w^2}},\quad w \coloneqq {\xi\over 4\sqrt{\pi}}.
\end{equation}
We note the presence of the branch cut on the positive real $\xi$ axis, which signals that the original asymptotic series (\ref{eq:hatGNAsymptoticSeries}) was not Borel summable. 

The Borel summation one would have naively liked to define is given by\footnote{To see this, note that $\int_0^\infty {d\xi\over 4\sinh^2({t\xi\over 2})}\xi^{2n} = \zeta(2n)\Gamma(2n+1)t^{-2n-1}$.}
\begin{equation}
  \widehat\cG_{0,{\rm np}|0}^{(0)} \stackrel{?}{=} {t^2\over N}\int_0^\infty {d\xi\over 4\sinh^2({t\xi\over 2})}\cB\left[\cG_{0,{\rm np}|0}^{(0)}\right](\xi),\quad t\coloneqq \sqrt{\tls},
\end{equation}
but this is not well-defined due to the branch cut. Thus there are ambiguities in the Borel resummation, reflecting the need for new non-perturbative terms. We will proceed by applying median resummation as in \cite{Dorigoni:2021guq}. In order to apply the median resummation, we will need to consider the directional Borel resummation
\begin{equation}
  \left(\widehat\cG_{0,{\rm np}|0}^{(0)}\right)_{\theta} = {t^2\over N}\int_0^{e^{i\theta}\infty} {d\xi\over 4 \sinh^2({t\xi\over 2})}\cB\left[\cG_{0,{\rm np}|0}^{(0)}\right](\xi).
\end{equation}
This defines an analytic function in the wedge $\rre(e^{i\theta}t)>0$ of the complex $t$ plane with the same perturbative expansion as (\ref{eq:hatGNAsymptoticSeries}). The necessary non-perturbative corrections are captured by the discontinuity, given by the difference between the lateral resummations
\begin{equation}
  \left(\widehat\cG_{0,{\rm np}|0}^{(0)}\right)_{\rm disc} = \left[\left(\widehat\cG_{0,{\rm np}|0}^{(0)}\right)_{+}-\left(\widehat\cG_{0,{\rm np}|0}^{(0)}\right)_{-}\right]= {t^2\over N}\int_0^{\infty} {d\xi\over 4 \sinh^2({t\xi\over 2})}\Discc \cB\left[\cG_{0,{\rm np}|0}^{(0)}\right](\xi),
\end{equation} 
where for a CFT observable $\O$ with Borel transform $\cB[\O](\xi)$, the lateral resummations are defined in terms of the directional resummation as
\begin{equation}
  \widehat\calo_{\pm} \coloneqq \lim_{\theta\to 0^{\pm}}\widehat\calo_\theta
\end{equation}
and
\begin{equation}
  \Discc \cB[\O](\xi) = \cB[\O](\xi+i0) - \cB[\O](\xi-i0).
\end{equation}
The median resummation is obtained by adding the resummation discontinuity to one of the lateral resummations
\begin{equation}\label{eq:medianResummation}
  \widehat\calo_{\rm med} \coloneqq \widehat\calo_{\pm } \mp \half \widehat\calo_{\rm disc}
\end{equation}

The non-perturbative corrections are then captured by the following integral
\begin{equation}
\begin{aligned}
  \left(\widehat\cG_{0,{\rm np}|0}^{(0)}\right)_{\rm disc} &= {2t^2\over N}\int_0^{\infty} {dw\over \sinh^2({2\sqrt{\pi}t w})}\Discc\left(w^2(4w^2-3)\over\sqrt{1-w^2}\right)\\
  &= {4it^2\over N} \int_1^\infty {dw\over \sinh^2(2\sqrt{\pi}t w)}{w^2(4w^2-3)\over\sqrt{w^2-1}}.
\end{aligned}
\end{equation}
To parse this, we expand the $\sinh^2$ factor in the denominator and shift the integration variable to get 
\begin{equation}
\begin{aligned}\label{eq:discontinuityResummation}
  \left(\widehat\cG_{0,{\rm np}|0}^{(0)}\right)_{\rm disc} &= {16 i t^2\over N}\sum_{k=1}^\infty k e^{-4\sqrt{\pi}t k}\int_0^\infty du \, e^{-4\sqrt{\pi}t k u} {(u+1)^2\left(4(u+1)^2-3\right)\over\sqrt{u(u+2)}}\\
  &= {2i\over\pi \sqrt{\ls}N}\sum_{k=1}^\infty {2k\sqrt{\ls}(3+k^2\ls)K_0(2k\sqrt{\ls}) + (6+5k^2 \ls)K_1(2k\sqrt{\ls})\over k^2}.
\end{aligned}
\end{equation}
where we re-inserted $t=\sqrt{\tls}$. This captures the perturbative expansions around each non-perturbative contribution $e^{-2k\sqrt{\ls}}$ as predicted by the analysis of Section \ref{subsubsec:nonpertSL2Z}, since $K_n(x) \sim \sqrt{\pi\over 2 x}e^{-x}\left(1+O(x^{-1})\right)$ as $x\gg1$.

The result \eqr{eq:discontinuityResummation} gives the non-perturbative corrections we are after. They are dual to AdS$_5 \x S^5$ D-string instanton corrections to $\cG_N(\t)$. We can anyway proceed to compute the full median resummation by adding this discontinuity to one of the lateral resummations as in (\ref{eq:medianResummation}) to obtain the following manifestly real result
\begin{equation}
\begin{aligned}
 \left(\widehat\cG_{0,{\rm np}|0}^{(0)}\right)_{\rm med} &= {t^2\over N}\int_0^\infty {d\xi\over 4 \sinh^2({t\xi\over 2})}\rre \cB\left[\cG_{0,{\rm np}|0}^{(0)}\right](\xi)\\
  &= {2t^2\over N} \int_0^1 {dw\over \sinh^2(2\sqrt{\pi} t w)}{w^2(4w^2-3)\over \sqrt{1-w^2}}.
\end{aligned}
\end{equation}
The latter integral can actually be done in closed form in terms of special functions. One finds 
\begin{equation}\label{eq:medianResummation2}
\begin{aligned}
  \left(\widehat\cG_{0,{\rm np}|0}^{(0)}\right)_{\rm med} =& \, {1\o 3N} \sum_{k=1}^\infty \bigg[{1\over 2k}\bigg(-{k\sqrt{\ls}}\({8k^2\o\pi}\ls + 9 I_1(2k\sqrt{\ls})\)+6(3+k^2\ls)I_2(2k\sqrt{\ls}) \\
  & \, +9k\sqrt{\ls} L_1(2k\sqrt{\ls}) -6(3+k^2\ls)L_2(2k\sqrt{\ls})\bigg)\bigg],
\end{aligned}
\end{equation}
where $I_n(x)$ is the modified Bessel function, and $L_n(x)$ is the modified Struve function, defined as the solution to the inhomogeneous differential equation 
\e{}{x^2 L_n''(x) + x L_n'(x) -(x^2+n^2)L_n(x) = {4\({x\o2}\)^{n+1}\o \sqrt{\pi} \G\(n+\half\)}.}
The median resummation (\ref{eq:medianResummation2}) reproduces the perturbative expansion (\ref{eq:hatGNAsymptoticSeries}) and includes non-perturbative corrections implied by the non-Borel summability of the perturbative series.

\ssec{The general result: Restoring instanton-anti-instanton effects}\label{subsec:fnp}

Let us now consider a general spectral overlap \eqr{nspec}, in which $\fnp(s) \neq 0$. This reintroduces instanton-anti-instanton pairs in the $1/N$ expansion. How does this affect the physics in the 't Hooft limit?

The strong coupling expansion when $\fnp(s)=0$ is given in \eqr{O0strong}. There is something striking about that equation: only half-integer powers of $1/\l$ appear! Therefore, for $\sl$-invariant observables $\O(\t)$,
\vs
\ni {\it Integer powers of $1/\l$ are allowed in the strong 't Hooft coupling expansion if and only if $\fnp(s)\neq 0$.}
\vs
\ni This is a simple diagnostic of \np physics.\foot{In fact, this must also be true at {\it finite} $N$, else the large $N$ limit would be trivial. We restate and apply this to $\cN=4$ SYM anomalous dimensions in Section \ref{secfiniteN}.} It is straightforward to unveil the mechanism for this by retracing our steps, and to derive modified expansions. Now including $\fnp(s)$, the zero mode in the 't Hooft limit becomes
\es{O0largen2}{\O_0(\l) = \<\O\> &+\sum_{g=0}^\i N^{2-2g}\Bigg[ {1\o 2\pi i}\ints {\pi \o \sin\pi s} s(1-s)  \(\L(s) \tl^{-s} + \L(1-s) N^{1-2s}  \tl^{s-1}\) \fp^{(g)}(1-s)\\ &+ \sum_{m=0}^\i N^{-3-{m\o2}} {1\o 2\pi i} \ints \L(1-s)N^{1-s} \tl^{s-1}  \fnp^{(g,m)}(s)\Bigg]}
With $\fnp(s)\neq 0$ the condition \eqr{fpzeros} is no longer required for consistency with the genus expansion: any integer powers of $1/N$ generated by the poles of the $\sin\pi s$ factor can now be cancelled by residues of $\fnp(s)$.\foot{These are what we referred to as ``spurious'' poles earlier.} Equating integer powers of $N$ and $\tl$ yields the linear relation
\e{fpint}{\Res_{t=s}\[\fnp^{(g,m)}(t)\] = (-1)^{s+1} s(1-s) \fp^{(g)}(1-s)\,,\quad m=2(s-3)\,,\quad s\in \Z_{\geq 3}}
When $\fnp^{(g,m)}(s)=0$, we recover \eqr{fpzeros}. 

Other similar effects of $\fnp^{(g,m)}(s)$ are also straightforward to obtain by contour deformation of \eqr{O0largen2}, following the logic outlined at the beginning of Section \ref{sectH}. In particular, \eqr{cancel} is modified to include a contribution from $\fnp^{(g,m)}(s)$; and spurious poles of $\fnp^{(g,m)}(s)$ can contribute new ``renormalization terms'' --- generalizing those in \eqr{O0strong} --- which have both integer and half-integer powers of $\tl$.

After all is said and done, the most general strong coupling expansion may be written as follows: 
\es{O0strong2}{\boxed{\O(\l\gg1) \approx \mathsf{C}(N) - \sum_{g=0}^\i N^{2-2g}\sum_{m=0}^\i\( \mathsf{a}_m^{(g)} \l^{-{3+m\o 2}} + N^{-2-2m}\,\mathsf{b}_{\a_m}^{(g)}\l^{\a_m}\)}}

The first term is a constant, now of the form
\e{eq:mathsfCN}{\mathsf{C}(N) = \<\O\> - \half\sum_{g=0}^\i N^{2-2g}\(N^{-1}\fp^{(g)}(0) + N^{-3}\sum_{m=0}^\i N^{-{m\o 2}} \fnp^{(g,m)}(0)\)}
Plugging in for the ensemble average $\<\O\>$ leads immediately to some very interesting conclusions, which we defer to Section \ref{sec:averageAndSugra}. 

The second term is the perturbative strong coupling expansion, now including both half-integer and integer powers, the latter being nonzero iff $\fnp^{(g,m)}(s)\neq 0$, cf. \eqr{fpint}. In terms of the residue function defined in \eqr{alphadef},
\begin{equation}\label{acoeffs}
  \mathsf{a}_m^{(g)} = (4\pi)^{3+m\over 2}\mathsf{R}^{(g)}_{{m\over 2} + 1}.
\end{equation}
In writing \eqr{O0strong2}, we have used the result \eqr{alphadef} that $\mathsf{R}_{1/2}^{(g)}=0$ (i.e. there is no $s=1$ pole).

The third term is the sum of $1/N$ renormalization effects described earlier. $\a_m$ is an index, which is summed over in \eqr{O0strong2}, taking values in a set of non-negative half-integers bounded above by $m$. The allowed values of $\a_m$ may be enumerated by combining \eqr{O0largen2} with consistency of the genus expansion.\foot{One can further constrain this by feeding in known facts about the type IIB string effective action at low orders in $\a'$, combined with the structure of the AdS$_5 \x S^5$ compactification. Let us also remark that $\alpha_m=0$ should be understood as signifying powers of $\log\l$. That these may be present in $\cN=4$ SYM observables is suggested by the existence of non-analytic threshold terms in 10D flat space scattering amplitudes, appearing there as powers of $\log(s\a')$  (e.g. \c{Green:2008bf, Alday:2018pdi}). The significance of the reorganization in \eqr{O0strong2} of the usual $1/N$ expansion --- which collects all terms of a given power in $1/N$, as opposed to the expansion \eqr{O0strong2} which falls out from S-duality considerations --- would be nice to explore further.} The coefficients $\mathsf{b}_{\a_m}^{(g)}$ can likewise be written in terms of $\fp(s)$ and residues of $\fnp^{(g,m)}(s)$, but this is not terribly enlightening: unlike when $\fnp(s)=0$, where the renormalization terms in \eqr{O0strong} were related by $\sl$ to the weak coupling expansion and hence determined in terms of $\fp(s)$, the residues of spurious poles of $\fnp^{(g,m)}(s)$ are {\it a priori} unconstrained.

How does $\fnp^{(g,m)}(s)\neq 0$ change the \np worldsheet instanton physics derived in Subsection \ref{subsubsec:nonpertSL2Z}? On the one hand, the perturbative $\l\gg1$ and $\ls\gg1$ series still must diverge double-factorially. Indeed, the earlier argument implies sign-definite double-factorial growth for the integer and half-integer series independently. On the other, that argument does not tell us these series' relative sign. Therefore, Borel summability may or may not hold, and $F$- and $D$-string worldsheet instantons may or may not be required. A specific Borel summable example is the planar ``octagon'', $\mathbb{O}(z,\zb)$ \c{Coronado:2018ypq, Coronado:2018ypq2, Kostov:2019stn, Kostov:2019auq, Bargheer:2019kxb, Belitsky:2019fan, Bargheer:2019exp, Belitsky:2020qrm}. This is related to a certain BPS four-point function, $A(z,\zb) \sim [\mathbb{O}(z,\zb)]^2$, in a limit of large BPS charge; we refer the reader to the references for definitions. Analysis of the $\l\gg1$ expansion of the planar octagon based on the determinant formula of \c{Kostov:2019auq} finds integer powers of $1/\l$, and therefore $\fnp(s)\neq 0$, but the expansion is Borel summable \c{Belitsky:2020qrm}, with asymptotically alternating signs.\foot{Interestingly, the planar octagon receives non-perturbative corrections in $\l$ despite its Borel summability; our analysis herein implies that the AdS$_5 \x S^5$ fundamental string worldsheet instanton effects, colorfully depicted in \c{Bargheer:2019exp}, should be accompanied by equally colorful D-string worldsheet instantons. It would be nice to apply the present techniques to expose the rich non-perturbative physics of the octagon.}

\sec{Very Strongly Coupled Limit}\label{sec:VSC}

We have seen in the previous section that consistency with the 't Hooft expansion places stringent constraints on the analytic structure of the perturbative part of the spectral overlap of large $N$ $\cN=4$ SYM observables. Here we will see how these constraints carry over to the structure of the expansion of these observables in the \emph{very strongly coupled} (VSC) limit of large $N$ and fixed complexified coupling, cf. \eqr{vsclim}.

Mimicking Section \ref{sectH}, we will start by assuming for the moment (though we will generalize) that there are no instanton-anti-instanton contributions to $\calo(\t)$, in particular that $\fnp(s)=0$. We will also assume $(\calo,\phi_n) = 0$. In this case, recall that the spectral decomposition is given by the following in the large $N$ limit: 
\begin{equation}\label{eq:VSCfnp=0}
  \calo(\tau) = \langle\calo\rangle + \sum_{g=0}^\infty  N^{2-2g} {1\over 2\pi i} \int_{\rre s = \half} ds\,{\pi\over\sin\pi s}s(1-s)N^{-s}\fp^{(g)}(1-s)\Lambda(s)E_s(\tau).
\end{equation}
Upon taking into account the conditions on the spectral overlap when $\fnp(s)=0$ that were derived in Section \ref{sectH}, contour deformation leads to the following large $N$ expansion in the VSC limit: 
\begin{equation}\label{eq:VSCfnp=0Expansion}
  \calo(\tau) = \left(\langle\calo\rangle - \half \sum_{g=0}^\infty N^{1-2g}\fp^{(g)}(0)\right) - \sum_{g=0}^\infty \sum_{m=1}^\infty N^{{3\over 2}-2g-m}\mathsf{R}_m^{(g)}E_{\half+m}(\tau),
\end{equation}
where the residue function $\mathsf{R}^{(g)}_m$ is defined in (\ref{alphadef}). We pause here to emphasize that the large $N$ limit and the spectral decomposition do not commute --- since the spectral decomposition gives the exact CFT observable at any $N$, it is important that one performs the spectral decomposition first, and takes the large $N$ limit afterwards. We see in (\ref{eq:VSCfnp=0Expansion}) that order-by-order in $1/N$, the observable $\calo(\t)$ is written in terms of Eisenstein series with order greater than $\half$, which are not themselves square-integrable and do not admit a non-trivial spectral decomposition. As a corollary, the modular (ensemble) average and the large $N$ limit do not commute: at each order in $1/N$, the modular average diverges.

In the case that $\fnp(s)$ and $(\calo,\phi_n)$ are non-vanishing, the spectral decomposition \eqr{eq:VSCfnp=0} is augmented by new terms to become
\begin{equation}\label{eq:VSCGeneral}
\begin{aligned}
  \calo(\tau) &= \langle\calo\rangle + \sum_{g=0}^\infty  N^{2-2g} {1\over 2\pi i} \int_{\rre s = \half} ds\,{\pi\over\sin\pi s}s(1-s)N^{-s}\fp^{(g)}(1-s)\Lambda(s)E_s(\tau)\\
    & \, + \sum_{g=0}^\infty \sum_{m=0}^\infty N^{-1-2g-{m\over 2}}\left({1\over 4\pi i}\int_{\rre s = \half}ds\, \fnp^{(g,m)}(s)\Lambda(s)E_s(\tau)+\sum_{n=1}^\infty (\calo,\phi_n)^{(g,m)}\phi_n(\tau)\right).
\end{aligned}
\end{equation} 
As discussed in Section \ref{subsec:fnp}, in the presence of non-perturbative effects the constraint that $\fp^{(g)}(s)=0$ for $s\in\Z_-$ is relaxed; instead, $\fp^{(g)}(s)$ and $\fnp^{(g,m)}(s)$ are related for $s\in\Z_-$ as in (\ref{fpint}). This allows integer powers of $1/N$ to appear in the large $N$ expansion in the VSC limit, leading to
\begin{empheq}[box=\fbox]{equation}\label{eq:VSCGeneralExpansion}
\begin{aligned}
  \calo(\tau) =& \, \left(\langle\calo\rangle -\half\sum_{g=0}^\infty N^{1-2g}\fp^{(g)}(0)  \right) - \sum_{g=0}^\infty \sum_{n=2}^\infty N^{{3\over 2}-2g-{n\over 2}}\mathsf{R}^{(g)}_{n\over 2}E_{n+1\over 2}(\tau)\\
  & \, + \sum_{g=0}^\infty \sum_{m=0}^\infty N^{-1-2g-{m\over 2}}\left({1\over 4\pi i}\int_{\rre s = \half}ds\, \fnp^{(g,m)}(s)\Lambda(s)E_s(\tau)+\sum_{n=1}^\infty (\calo,\phi_n)^{(g,m)}\phi_n(\tau)\right).
\end{aligned}
\end{empheq}
The non-perturbative spectral overlaps $\fnp^{(g,m)}(s)$ exhibit growth in the $|s|\to\infty$ limit that forbids deforming the spectral contour to infinity as is done with $\fp(s)$; this is ultimately why they lead to non-perturbative physics in the weak-coupling limit. They may, however, have a finite number of spurious poles at positive integer values of $s$ in the right half $s$-plane, with residues that are related to the perturbative part via (\ref{fpint}), which is required for consistency of the weak-coupling expansion. 

The terms on the second line of \eqr{eq:VSCGeneral} can, in general, generate modular invariants that are much more complicated than the Eisenstein series; an example which appears at integer powers of $1/N$ in the integrated correlator of \cite{Chester:2020dja,Chester:2020vyz} that we study in Section \ref{subsec:F4} is described in detail in Appendix \ref{app:inhomogeneous}.

\subsection{Example: $\cG_N(\t)$}
Here we apply this formalism to study the integrated correlator $\cG_N(\t)$. Since this observable is free of non-perturbative effects, we need only the perturbative overlaps, which are given up to genus three in (\ref{fpdgw}). This data is enough to compute $\cG_N(\t)$ up to order $N^{-{11\over 2}}$ in the VSC limit. Computing the residues that appear in (\ref{eq:VSCfnp=0Expansion}), we find explicitly
\begin{equation}\label{eq:GNLargeN}
\begin{aligned}
  \mathcal{G}_N(\tau) =& \, {N^2\over 4} -{3\over 2^{4}}N^{\half}\widetilde E_{3\o2}(\tau)  + {45\over 2^8}N^{-\half}\widetilde E_{5\o2}(\tau) +3 N^{-{3\over 2}}\left({1575\over 2^{15}}\widetilde E_{7\o2}(\tau)- {13\over 2^{13}}\widetilde E_{3\o2}(\tau)\right) \\
  & \, + 225N^{-{5\over 2}}\left({441\over 2^{18}}\widetilde E_{9\o2}(\tau) - {5\over 2^{16}}\widetilde E_{5\o2}(\tau)\right) + 63 N^{-{7\over 2}}\left({1575\over 2^{15}}\widetilde E_{11\o2}(\tau) - {44625\over 2^{25}}\widetilde E_{7\o2}(\tau) + {73\over 2^{22}}\widetilde E_{3\o2}(\tau)\right)\\
  & \, + 945 N^{-{9\over 2}} \left({31216185\over 2^{31}}\widetilde E_{13\o2}(\tau) - {41895\over 2^{26}}\widetilde E_{9\o2}(\tau) + {1639\over 2^{27}}\widetilde E_{5\o2}(\tau)\right)\\
  & \, + 33 N^{-{11\over 2}} \left({1220198104125\over 2^{38}}\widetilde E_{15\o2}(\tau) - {12033511875\over 2^{36}}\widetilde E_{11\o2}(\tau) + {61486425\over 2^{34}}\widetilde E_{7\o2}(\tau) - {109447\over 2^{32}}\widetilde E_{3\o2}(\tau)\right) \\&+ O(N^{-{13\over 2}}),
\end{aligned}
\end{equation}
where
\begin{equation}\label{eq:ETildeDef}
  \widetilde E_s(\tau) \coloneqq \chi(s) E_s(\tau)
\end{equation}
for
\begin{equation}\label{eq:chiDefinition}
  \chi(s) \coloneqq {2\Lambda(s)\over\Gamma(s)}.
\end{equation}
This precisely agrees with the large $N$ expansion of this integrated correlator previously found by other methods in \cite{Chester:2019jas,Dorigoni:2021guq}. The structure of the large $N$ expansion of $\cG_N(\t)$ cleanly reifies the prior analysis and confirms the prediction (\ref{eq:VSCfnp=0Expansion}).

\ssec{Example with non-perturbative effects: $\cF_N(\t)$}\label{subsec:F4}
We now explore the spectral decomposition of the other integrated correlator of \cite{Chester:2020dja,Chester:2020vyz} as a more non-trivial working example in which (as we will see) both $\fnp(s)$ and the cusp form overlap are nonzero. This observable is computed by integrating the $\O_{\mathbf{20'}}$ four-point function over Euclidean space weighted with the following supersymmetric measure \cite{Chester:2020dja}
\begin{equation}
  \cF_N(\tau) \coloneqq -{32 c^2 \over \pi}\int_0^\infty dr\int_0^\pi d\theta\, r^3\sin^2\theta\left({1+u+v\over u^2}\right)\bar D_{1,1,1,1}(u,v)\cT_N(u,v;\tau) - 48\zeta(3)c,
\end{equation}
where $c= {N^2-1\over 4}$, the cross ratios $u,v$ are given in terms of $r,\theta$ as in (\ref{uvrtheta}) and $\bar D_{1,1,1,1}$ is a scalar box integral given by 
\begin{equation}
  \bar D_{1,1,1,1}(u,v) = {1\over z-\bar z}\left(\log(z\bar z)\log{1-z\over 1-\bar z}+2\LLi_2(z) -2\LLi_2(\bar z)\right),
\end{equation}
where $u=z\zb$ and $v=(1-z)(1-\zb)$. 

Like $\cG_N(\t)$, this integrated correlator is a protected observable that can be obtained by taking derivatives of the sphere free energy in the $\cN=2$-preserving mass-deformed theory,
\begin{equation}
  \cF_N(\t) =\left.\partial_{m}^4 \log Z_{S^4}(\t;m)\right|_{m=0}
\end{equation}
The large $N$ expansion of this integrated correlator at fixed coupling was worked out in \cite{Chester:2020vyz} as\footnote{In \cite{Chester:2020vyz}, this result was phrased in terms of the function $\cE_{r;s_1,s_2}(\tau)$, which is related to the function that appears below and in Appendix \ref{app:inhomogeneous} as
\begin{equation}
  \cE_{r;s_1,s_2}(\tau) = 4\zeta(2s_1)\zeta(2s_2)F_{r;s_1,s_2}(\tau) = \pi^{s_1+s_2}\widetilde F_{r;s_1,s_2}(\tau).
\end{equation}} 
\begin{equation}
\begin{aligned}\label{eq:d4m4LogZ}
  \cF_N(\tau) =& \, 6N^2 + 6 N^{\half}\widetilde E_{3\o2}(\tau) + C_0 - {9\over 2}N^{-\half}\widetilde E_{5\o2}(\tau) - {27\over 2^3}N^{-1}\widetilde{F}_{3;{3\over 2},{3\over 2}}(\tau)\\
  & \, + N^{-3/2}\left({117\over 2^8}\widetilde E_{3\o2}(\tau) - {3375\over 2^{10}}\widetilde E_{7\o2}(\tau)\right) + N^{-2}\left(C_1 + {14175\over 704}\widetilde{F}_{6;{5\over 2},{3\over 2}}(\tau) - {1215\over 88}\widetilde{F}_{4;{5\over 2},{3\over 2}}(\tau)\right)\\
  & \, + N^{-5/2}\left({675\over 2^{10}}\widetilde E_{5\o2}(\tau) - {33075\over 2^{12}}\widetilde E_{9\o2}(\tau)\right) + N^{-3}\bigg[\alpha_3 \widetilde{F}_{3;{3\over 2},{3\over 2}}(\tau)  \\
  & \, + \sum_{r=5,7,9}\left(\alpha_r \widetilde{F}_{r;{3\over 2},{3\over 2}}(\tau) + \beta_r\widetilde{F}_{r;{5\over 2},{5\over 2}}(\tau) + \gamma_r\widetilde{F}_{r;{7\over 2},{3\over 2}}(\tau)\right)\bigg]+O(N^{-{7\over 2}})
\end{aligned}
\end{equation}
where $\widetilde{F}$ is the solution to the inhomogeneous Laplace equation studied in Appendix \ref{app:inhomogeneous} rescaled as follows
\begin{equation}
  \widetilde{F}_{r;s_1,s_2}(\tau) = \chi(s_1)\chi(s_2)F_{r;s_1,s_2}(\tau),
\end{equation}
$\widetilde{E}_s(\t)$ is given in (\ref{eq:ETildeDef}) and $\alpha_r,\beta_r,\gamma_r,C_i$ are constants written down in \cite{Chester:2020vyz}. 

From the structure of the large $N$ expansion of $\cF_N(\t)$ in (\ref{eq:d4m4LogZ}) some novel features are already apparent. The first is that, compared to the large $N$ expansion of $\cG_N(\t)$ given in (\ref{eq:GNLargeN}), we see that there are integer powers of $1/N$ in addition to the half-integer powers. The reason for this is as anticipated by the previous discussion: the non-perturbative contribution to the spectral overlap, $\fnp(s)$, is non-vanishing! Thus $\fp^{(g)}(s)$ is no longer required to vanish on the negative integers, and integer powers of $1/N$ can appear. While, similarly to $\cG_N(\t)$, the coefficients of the half-integer powers of $1/N$ are the familiar Eisenstein series of half-integer order, we see that the coefficients of the integer powers of $1/N$ involve the solutions to the inhomogeneous Laplace equation studied in detail in Appendix \ref{app:inhomogeneous}. In particular, this implies that the overlap of $\cF_N(\t)$ with the Maass cusp forms is nonzero. To summarize, from the large $N$ expansion (\ref{eq:d4m4LogZ}) we conclude that
\begin{equation}
\begin{aligned}\label{eq:FNCuspOverlap}
  \fnp(s) & \ne 0 \\
  (\cF_N,\phi_n) &\ne 0, \quad n = 1,2, \ldots
\end{aligned}
\end{equation}

\ni Since the large $N$ overlaps are simply the expansion of the finite-$N$ overlaps viewed as functions of $N$, (\ref{eq:FNCuspOverlap}) holds at finite values of $N$ as well. Remarkably, because the cusp forms enter via the solutions to the inhomogeneous solutions to the Laplace equation, the cusp form overlap $(\cF_N,\phi_n)$ is actually computable, at least in a $1/N$ expansion, in terms of certain $L$-functions associated with the cusp forms.

The VSC limit of the spectral decomposition of $\cF_N$ has the following general structure
\begin{equation}
\begin{aligned}\label{eq:FNSpectral}
  \cF_N(\tau) =& \, \langle{\cF_N}\rangle + \sum_{g=0}^\infty N^{2-2g}\Bigg[{1\over 2\pi i }\int_{\rre s = \half}ds\, {\pi\over\sin(\pi s)}s(1-s)N^{-s}\fp^{(g)}(1-s)E_s^*(\tau) \\
  & \,+ \sum_{m=0}^\infty N^{-3-{m\over 2}}\left({1\over 4\pi i }\int_{\rre s=\half} ds\, \fnp^{(g,m)}(s)E_s^*(\tau) + \sum_{n=1}^\infty (\cF_N,\phi_n)^{(g,m)} \phi_n(\tau)\right)\Bigg]
\end{aligned}
\end{equation}
We will now provide each element of this decomposition by matching to \eqr{eq:d4m4LogZ}. At first it is not clear how the inhomogeneous functions in \eqr{eq:d4m4LogZ} fit into this framework. The key is to think of these functions themselves in the spectral, rather than Fourier, decomposition.

We begin by discussing the perturbative part $\fp^{(g)}(s)$. The novel feature compared to $\cG_N(\t)$ is that $\fp^{(g)}(s)$ need no longer vanish for $s\in\Z_-$. Indeed, the spectral decomposition of the inhomogeneous solutions have Eisenstein series in their spectral expansion (see Appendix \ref{app:inhomogeneous}):
\begin{equation}\label{eq:integerEisensteinsInF}
\begin{aligned}
  F_{r;s_1,s_2}(\tau) &= {E_{s_1+s_2}(\tau)\over \mu(s_1+s_2)-\mu(r+1)} + {\varphi(s_2)E_{1-s_2+s_1}(\tau)\over\mu(1-s_2+s_1)-\mu(r+1)} + (\text{spectral})\\
  F_{r;s,s}(\tau) &= {E_{2s}(\tau)\over\mu(2s)-\mu(r+1)} - {2\varphi(s)\widehat E_1(\tau)\over\mu(r+1)} + (\text{constant}) + (\text{spectral}),
\end{aligned}
\end{equation}
where recall that $\mu(s) = s(1-s)$. The function $\widehat E_1(\tau)$, discussed in further detail in Appendix \ref{app:inhomogeneous}, is extracted from the $s\rar1$ behavior of $E_s(\t)$, 
\begin{equation}
  \widehat E_1(\tau) \coloneqq \lim_{s\to 1}\left(E_s(\tau) - {3\over \pi(s-1)}\right)
\end{equation}
The integer-order Eisenstein series that appear via the spectral decomposition of $F_{r;s_1,s_2}$ (cf. (\ref{eq:integerEisensteinsInF})) at integer powers of $1/N$ in (\ref{eq:d4m4LogZ}) are realized as the residues of ${\pi s(1-s)\over\sin(\pi s)}\fp^{(g)}(1-s)E_s^*(\tau)$ at $s\in\Z_+$. It will be convenient to split up $\fp^{(g)}(s)$ as follows 
\begin{equation}
  \fp^{(g)}(s) = f^{(g)}_{\rm p,{\mathbb{Z}}+\half}(s) + f^{(g)}_{\rm p,\mathbb{Z}}(s) ,
\end{equation}
where the product of $f^{(g)}_{\rm p,\mathbb{Z}}(1-s)$ and $f^{(g)}_{{\rm p},{\mathbb{Z}}+\half}(1-s)$ with $\sin(\pi s)^{-1}$ contributes residues at positive integer and half-integer values of $s$, respectively, upon contour deformation. In particular, $f^{(g)}_{{\rm p},{\mathbb{Z}}+\half}(1-s)$ must vanish for $s\in\Z_+$.

To get a feeling for how this works, let us work out the genus-zero contributions to the perturbative part of the spectral decomposition as explicitly as possible. The half-integer powers of $1/N$ in (\ref{eq:d4m4LogZ}) are consistent with 
\begin{equation}
  f^{(0)}_{\rm p,{\mathbb{Z}}+\half}(1-s) = {4^{2-s}(2s-1)(2s+3)\Gamma({3\over 2}-s)\over\sqrt{\pi}s\Gamma(2-s)\Gamma(3-s)} 
\end{equation}
so that
\begin{equation}
\begin{aligned}
  & \, {1\over 2\pi i} \int_{\rre s = \half}ds \, {\pi \over \sin(\pi s)}s(1-s)N^{2-s}f_{\rm p,{\mathbb{Z}}+\half}^{(0)}(1-s) E_s^*(\tau)\\
  = & \,  -10 N + N^{1/2} \,6 \widetilde E_{3/2}(\tau) - N^{-1/2}\,{9\over 2}\widetilde E_{5/2}(\tau) - N^{-3/2}{3375\over 2^{10}}\widetilde E_{7/2}(\tau) - N^{-5/2}{33075\over 2^{12}}\widetilde E_{9/2}(\tau) + O(N^{-7/2}).
\end{aligned}
\end{equation}
We note the presence of the $O(N)$ term which is not present in (\ref{eq:d4m4LogZ}) and thus must be cancelled by something. Similarly, although we have not been able to guess its functional dependence on $s$ explicitly, the genus-zero contribution $f^{(0)}_{\rm p, \mathbb{Z}}(s)$ must give the following
\begin{equation}
\begin{aligned}
  &\,{1\over 2\pi i} \int_{\rre s = \half}ds \, {\pi \over \sin(\pi s)}s(1-s)N^{2-s}f_{\rm p,{\mathbb{Z}}}^{(0)}(1-s) E_s^*(\tau)\\
  = & \, -N{f_{\rm p,\mathbb{Z}}^{(0)}(0)\over 2} -N^{-1}{27\chi({3\over 2})^2\over 8}E_3(\tau) - N^{-2}{135 \chi({3\over 2})\chi({5\over 2})\over 128}E_4(\tau)- N^{-3}\left({6885 \chi({5\over 2})^2\over 4096}+{42525 \chi({7\over 2})\chi({3\over 2})\over 16384}\right)E_5(\tau)\\& + O(N^{-4}).
\end{aligned}
\end{equation}

The non-perturbative contributions to the spectral overlap in (\ref{eq:d4m4LogZ}) are actually quite straightforward to work out given our knowledge of the spectral decomposition of the solutions to the inhomogeneous Laplace equation as in Appendix \ref{app:inhomogeneous}. For example, from (\ref{eq:d4m4LogZ}) we can read off 
\begin{equation}\label{eq:FNfnp}
\begin{aligned}
  \Lambda(s)f_{\rm np}^{(0,0)}(s) &= -{27\over 2^3}{\widetilde K_{1-s,{3\over 2}, {3\over 2}}\over \mu(s)+12}\\
  \Lambda(s)f_{\rm np}^{(0,2)}(s) &= \left({14175\over 704(\mu(s) + 42)} - {1215\over 88( \mu(s)+20)}\right)\widetilde K_{1-s,{5\over 2},{3\over 2}},
\end{aligned}
\end{equation}
where $\widetilde K_{s,s_1,s_2}$ is given in (\ref{eq:KtildeEis3}).

Similarly, the overlap of $\cF_N(\t)$ with the Maass cusp forms can be written in terms of the Clebsch-Gordan coefficient given in equation (\ref{eq:CGEis2Cusp}) describing the triple product between two Eisenstein series and a cusp form. In particular, we have 
\begin{equation}\label{eq:FNcusp}
\begin{aligned}
  (\cF_N,\phi_n)^{(0,0)} &= -{27\over 2^3}{\widetilde K^n_{{3\over 2} {3\over 2}}\over \mu_n + 12} \\
  (\cF_N,\phi_n)^{(0,2)}&=  \left({14175\over 704(\mu_n + 42)} - {1215\over 88( \mu_n+20)}\right)\widetilde K^n_{{5\over 2},{3\over 2}},
\end{aligned}
\end{equation}
where $\widetilde K^n_{s_1,s_2}$ is given in (\ref{eq:KtildeEis2Cusp}). 

\sec{Interlude: Non-perturbative Effects in Anomalous Dimensions at Finite $N$}\label{secfiniteN}

A main point of Section \ref{subsubsec:nonpertSL2Z} was to demonstrate how strongly the \np corrections at large $N$ are constrained by $\sl$ symmetry. Among other things, we discovered that integer powers of $1/\l$ in the strong 't Hooft coupling expansion --- equivalently, integer powers of $1/N$ in the VSC expansion --- necessarily imply instanton-anti-instanton corrections in powers of $q\qb$, order-by-order in $1/N$. Somewhat obviously, that result implies the existence of \np corrections at {\it finite} $N$:
\vs
\ni {\it If $\O(\t)$ contains integer powers of $1/\l$ in the strong 't Hooft coupling expansion, then $\O(\t)$ receives non-perturbative, instanton-anti-instanton corrections at finite $N$.}
\vs
\ni The logic is simply that in the spectral formulation, the $q\qb$ corrections at large $N$ appear  via nonzero $\fnp^{(g,m)}(s)$ in the spectral overlap, and the $\fnp^{(g,m)}(s)$ are the large $N$ expansion of the finite $N$ function $\fnp(s)$. One can thus rigorously deduce the existence of $q\qb$ corrections at {\it finite} $N$ from data at {\it large} $N$.\foot{One may imagine that the corrections could kick in only above some $N_*$, or vanish for a sporadic set of integers $N$. On the other hand, the corrections cannot vanish for all integer $N$ unless the large $N$ limit exhibits oscillations in $N$; this seems pathological, but can at any rate be assessed on a case-by-case basis.}

We now apply this to the spectrum of conformal dimensions of gauge-invariant local operators:
\e{specdef}{\O(\t) = \text{Spec}(\mathbb{D})(\t)}
where $\mathbb{D}$ is the dilatation operator of $\cN=4$ SYM. The italicized result above may be used to prove that {\it unprotected operator dimensions receive non-perturbative, instanton-anti-instanton corrections at finite $N$.} 

Let us assemble some ingredients. Local operators, eigenstates of $\mathbb{D}$ in radial quantization, are specified by their quantum numbers under the maximal bosonic subgroup of the $\cN=4$ superconformal algebra $PSU(2,2|4)$: namely, a conformal dimension $\D(\t)$, Lorentz spins $(j_1,j_2)$, and an $SU(4)_R$ representation with Dynkin labels $[p_1\,p_2\,p_3]$, with $p_i\in\Z_{\geq 0}$. As BPS-protected dimensions are constant in $\t$, we may restrict our discussion to unprotected operators.\foot{For more details on $\cN=4$ superconformal representation theory, see e.g. \c{Beem:2016wfs, Cordova:2016emh}. For modern bootstrap results for $\cN=4$ SCFTs at finite $N$, see e.g. \c{Beem:2013qxa, Beem:2013sza, Alday:2013opa, Beem:2016wfs, Bissi:2020jve, Chester:2021aun}.}

By S-duality, the spectrum of the dilatation operator is $\sl$ invariant:
\e{}{\text{Spec}(\mathbb{D})(\g\t) = \text{Spec}(\mathbb{D})(\t)\,,\quad \g\in SL(2,\Z)}
Introducing a shorthand for the $SU(2,2) \x SU(4)_R$ quantum numbers
\e{}{\mathsf{Q} \coloneqq  \{(j_1,j_2);[p_1\,p_2\,p_3]\}}
we introduce the grading
\e{}{\mathbb{D} =\, \bigoplus_\mathsf{Q}\, \mathbb{D}_\mathsf{Q}}
acting on the Hilbert space of states on $S^3$. In a sector of fixed charge $\mathsf{Q}$, the spectrum may be ordered by increasing conformal dimension,
\e{}{\text{Spec}\(\mathbb{D}_\mathsf{Q}\) = \{\D_\mathsf{Q}^{(1)},\D_\mathsf{Q}^{(2)},\D_\mathsf{Q}^{(3)},\ldots\}\,,\quad \D_\mathsf{Q}^{(1)} < \D_\mathsf{Q}^{(2)} < \D_\mathsf{Q}^{(3)} < \ldots}
where these inequalities hold for all $\t\in\cF$.\foot{ In writing this, we have employed the standard assumption that any accidental degeneracies which arise at some $\t$ are resolved, such that level crossings are avoided. Strictly speaking, this remains an expectation rather than a bulletproof fact about $\cN=4$ SYM. However, there are arguments for it \c{Korchemsky:2015cyx}, and it is compatible with (but does not assume) general lore about the structure of Regge trajectories in CFTs. Our discussion may be suitably generalized to account for possible complexities.} Each $\D_\mathsf{Q}^{(i)}(\t)$ admits a spectral decomposition.

Let us first prove the desired result for the lightest scalar singlet, generalizing to the rest afterwards. (We use the 't Hooft limit, but the VSC limit is equally applicable.) To simplify notation, we write, following \c{Beem:2013qxa, Beem:2016wfs},
\e{}{\D_0(\t) \coloneqq \D^{(1)}_{\{(0,0);[000]\}}(\t)}
In the 't Hooft limit, 
\e{}{\D_0(\t) = 2 + \g_0^{(0)}(\l) + {\g_0^{(1)}(\l)\o N^2} + O\({1\o N^4}\)}
At $\l\ll1$, $\D_0(\t)$ may be identified with the dimension of the Konishi operator, $\D_K(\t)$ . As the coupling increases, $\D_0(\t)$ transitions to the dimension of the scalar singlet double-trace operator $[\O_{\mathbf{20'}}\O_{\mathbf{20'}}]$ of leading twist --- in particular, $\g_0^{(0)}(\l\gg1) = 2$ (up to non-perturbative $1/\l$ corrections). Thus, to deduce \np corrections of $\D_0(\t)$, it suffices to identify the $1/\l$ asymptotics of the double-trace anomalous dimension. Taking
\e{}{\D_0(\t) = 4 + {\g_0^{(1)}(\l)\o N^2} + O\({1\o N^4}\)\qquad (\l\gg1)\,,}
the anomalous dimension $\g_0^{(1)}(\l)$ obeys \c{DHoker:1999kzh}
\e{g0th}{\g_0^{(1)}(\l) = -16 +{\g^{(1)}_{(0,0)}\o \l^{3/2}} + {\g^{(1)}_{(1,0)} \o \l^{5/2}} + {\g^{(1)}_{(0,1)}\o \l^3} + \ldots\qquad (\l\gg1)}
The notation \eqr{g0th} signals the fact that $\g_0^{(1)}(\l)$ may be efficiently extracted from the connected four-point function $\la \O_{\mathbf{20'}}\O_{\mathbf{20'}}\O_{\mathbf{20'}}\O_{\mathbf{20'}}\ra$, which admits a $1/\l$ expansion in crossing-symmetric polynomials (dual to AdS$_5 \x S^5$ quartic vertices) labeled by two integers $(a,b)\geq 0$; this is simply the original analysis of \c{Heemskerk:2009pn}, with an $\cN=4$ superconformal dressing.\foot{A lightning review is as follows. After processing $\cN=4$ superconformal Ward identities and stripping R-symmetry polarizations, the unprotected part of $\la \O_{\mathbf{20'}}\O_{\mathbf{20'}}\O_{\mathbf{20'}}\O_{\mathbf{20'}}\ra$ is determined by an undressed scalar four-point function. In Mellin space, this so-called ``reduced'' correlator admits an expansion \c{Rastelli:2016nze, Rastelli:2017udc}
\e{}{M(s,t) = M_{\rm sugra}(s,t) + \sum_{a,b=0}\l^{-3/2-a-3b/2} M_{(a,b)}(s,t)}
where $M_{(a,b)}(s,t) = \s_2^a \s_3^b + \ldots$, where $\s_n \coloneqq  s^n+t^n+u^n$ and $\ldots$ represents lower powers of $s,t,u$. The coefficients $\g_{(a,b)}^{(1)}$ follow from the conformal block expansion of $M(s,t)$. Note that the terms $M_{(a,b)}$ are holographically generated by scalar superpartners of the $D^{2k}R^4$ contact terms in the quartic gravitational effective action of classical string theory on AdS$_5 \x S^5$, with the relation $2k=4a+6b$. See  \c{Goncalves:2014ffa,Aprile:2017xsp,Aprile:2018efk,Caron-Huot:2018kta,Drummond:2019odu,Chester:2019pvm,Aprile:2020mus,Alday:2021vfb} for a tranche of computations of double-trace anomalous dimensions at strong coupling.} In particular, the coefficient $\g_{(0,1)}^{(1)}$, fixed by (a scalar superpartner of) the non-vanishing $D^6R^4$ vertex in AdS$_5 \x S^5$, is nonzero. By the logic at the start of this Section, this means that $\fnp(s)\neq 0$ for $\D_0(\t)$ at finite $N$, and therefore $\D_0(\t)$ receives $q\qb$ corrections, concluding the proof. 

The above language is intentionally agnostic about any Lagrangian identification of the operators themselves: we should not ask, non-perturbatively, what an operator is ``made of''. However, at infinitesimal coupling, we {\it do} know what operators these dimensions are describing: the departure from the free fixed point turns on infinitesimal anomalous dimensions for the operators of free $\cN=4$ SYM. We noted above that $\D_0(\t)$ is identified with $\D_K(\t)$ in a neighborhood of weak coupling. Accordingly, we note what our result implies for the Konishi dimension {\it per se}: since the $q\qb$ corrections are \np near $y\rar\i$, precisely the region where $\D_0(\t) = \D_K(\t)$, the Konishi dimension receives non-perturbative $q\qb$ corrections to its weak coupling expansion. For context and self-containedness, let us assemble all previously-known perturbative data about $\D_K(\t)$. We expand the Fourier modes $\D_{K,k}(y)$ in powers of $q\qb$ as
\e{}{\D_{K,k}(y) = \sum_{n=0}^\i\D_{K,k}^{(n)}(y) \,(q\qb)^n}
For the zero mode, the perturbative expansion is known\foot{Recall that $y=4\pi/g^2$. Note that the analytic expansion of $\D_{K,0}(y)$ in the planar limit is known through 11 loops \c{Marboe:2018ugv}.} through four loops \c{Fiamberti:2008sh,Bajnok:2008qj,Velizhanin:2009gv}
\e{Kweak}{\D_{K,0}^{(0)}(y) = 2 + {3 N\o \pi} y^{-1} - {3N^2\o \pi^2} y^{-2} + {21N^3\o 4\pi^3}y^{-3} + {N^4\o 4\pi^4}y^{-4}\(-39+9\z(3) - 45\z(5)\(\half+{6\o N^2}\)\) +\ldots}
For the nonzero modes, direct instanton computations \cite{Alday:2016tll} give the leading 1-instanton term 
\e{}{\D_{K,1}^{(0)}(y) =  -{27\kappa_N\o 5\pi^2(N^2-1)} y^{-2}  + \ldots\,, \qquad\text{where}~~ \kappa_N \coloneqq  {2\G(N-\half)\o \sqrt{\pi} \G(N-1)}} 
for finite $N$, and the leading $k$-instanton term
\e{}{\D_{K,k}^{(0)}(y) \sim -{54\pi^{-5/2}\o  N^{3/2}} k^{-7/2}\s_{2}(k) \,y^{-2}+\ldots  \qquad  (N\rar\i)}
for large $N$. Prior to this work, the status of $\D_{K,k}^{(n>0)}(y)$ was unknown. The result that we have proven here is that, for some range of finite $N$,
\e{Knp}{\D_{K,k}^{(n\geq0)}(y) \neq 0\,.}
Computing these explicitly is an attractive target for $\cN=4$ SYM instanton calculus.

This result for $\D_0(\t)$ generalizes to the complete spectrum of unprotected operator dimensions $\D_{\mathsf{Q}}^{(i)}(\t)$. The point is that the unprotected spectrum of planar $\cN=4$ SYM at $\l\gg1$ consists {\it solely} of multi-trace composites of BPS single-trace operators. So a version of the above argument suffices to establish the existence of $q\qb$ corrections in general. We demonstrate this generalization with the entire leading even-spin Regge trajectory in the $SU(4)_R$-singlet sector,
\es{}{\D_{\ell}(\t) \coloneqq \D^{(1)}_{\{\({\ell\o2},{\ell\o2}\);[000]\}}\,,\quad \ell\in2\,\Z_{\geq 0}}
The $\l\gg1$ operator representatives are the $SU(4)_R$-singlet double-trace operators $[\O_{\mathbf{20'}}\O_{\mathbf{20'}}]_{0,\ell}$, of twist four and spin $\ell$, so
\e{}{\D_\ell(\t) = 4 +\ell+ {\g_\ell^{(1)}(\l)\o N^2} + O\({1\o N^4}\)\qquad (\l\gg1)\,.}
These are the leading-twist members of the larger class $[\O_{\mathbf{20'}}\O_{\mathbf{20'}}]_{n,\ell}$, of twist $4+2n$ and spin $\ell$. The $M_{(a,b)}(s,t)$ term of the four-point correlator contributes to bounded even spins $\ell \leq 2a+3b$, and all $n$. Thus, for fixed $\ell$, the same leading integer power of $1/\l$ turns on for all $n$, while for fixed $n$, an integer power of $1/\l$ turns on at sufficiently high spin $\ell$ \c{Heemskerk:2009pn}. Therefore, applying our earlier logic, $\D_\ell(\t)$ receives non-perturbative $q\qb$ corrections for all $\ell$. And in analogy to the $\ell=0$ case, the $\l\ll1$ representatives are the ``twist-two'' single-trace superconformal primary singlets, one for each spin $\ell$ --- therefore, all twist-two anomalous dimensions receive $q\qb$ corrections.\foot{The four-loop anomalous dimension for  twist-two operators was recently computed in \c{Kniehl:2021ysp}.}

We may unify the above findings in terms of the leading even-spin Regge trajectory, with Reggeon spin $j(\nu)$. The dimension of the trajectory, $\mathbf{\D}(\nu) \coloneqq \Delta(j(\nu))$, obeys
\e{}{\mathbf{\D}(\nu) = 2\pm i\nu}
where $\nu$ depends on $y$. All even-spin twist-two operators live at points $j(\nu)\in 2\Z_{\geq 0}$ on this trajectory. Our conclusion is that the entire trajectory receives $q\qb$ corrections. Decomposing $\mathbf{\D}(\nu)$ in Fourier modes $\mathbf{\D}_{k}(\nu)$, 
\e{}{\mathbf{\D}_{k}(\nu) = \sum_{n=0}^\i\mathbf{\D}_{k}^{(n)}(\nu) \,(q\qb)^n\,,\quad \mathbf{\D}_{k}^{(n\geq 0)}(\nu) \neq 0}

The extension to double-trace operators $[\O_p\O_p]$ of other half-BPS operators $\O_p$ in the $[0\,p\,0]$ representation for $p>2$, or to $K$-trace operators, proceeds along identical lines using bulk $2K$-point contact interactions. These mix with the $[\O_{\mathbf{20'}}\O_{\mathbf{20'}}]_{n,\ell}$ operators for $n>0$ analyzed above \c{Aprile:2018efk}, which complicates the explicit form of the eigenfunctions of $\mathbb{D}$, but the method of proof is the same, and should lead likewise to $q\qb$ corrections for the subleading Regge trajectories.\foot{These conclusions apply to {\it all} unprotected operator dimensions modulo the (to us) unlikely logical possibility that for some privileged subclass of unprotected multi-trace operators, integer powers do not appear at any order in $1/\l$.}

\sec{AdS$_5 \x S^5$ Supergravity is Ensemble-Averaged String Theory}\label{sec:averageAndSugra}

This section may be read independently of those before.

Our starting point is equation \eqr{O0strong2}, the most general perturbative $\l\gg1$ expansion consistent with $\sl$ invariance of an observable $\O(\t)$. Let us reproduce it here: 
%
\es{O0strong3'}{\O(\l\gg1) \approx \mathsf{C}(N) - \sum_{g=0}^\i N^{2-2g}\sum_{m=0}^\i\( \mathsf{a}_m^{(g)} \l^{-{3+m\o 2}} + N^{-2-2m}\,\mathsf{b}_{\a_m}^{(g)}\l^{\a_m}\)}
The coefficients $\mathsf{a}_m^{(g)}$ and $\mathsf{b}_{\a_m}^{(g)}$, independent of $N$ and $\l$, take explicit forms in terms of residues of the perturbative and non-perturbative pieces --- denoted $\fp^{(g)}(s)$  and $\fnp^{(g,m)}(s)$ , respectively ---  of the spectral overlap $\{\O,E_s\}$ at genus $g$, written in \eqr{nspec}. The $\mathsf{a}_m^{(g)}$ terms generate the usual $1/\l$ expansion, while the $\mathsf{b}_{\a_m}^{(g)}$ terms are ``renormalization terms'' that cut off divergences in the $1/N$ expansion at large $\l$ \c{Aharony:2016dwx} (see Subsection \ref{subsec:fnp} for details). Their explicit forms are not important here. However, the function $\mathsf{C}(N)$, constant in $\l$, {\it is} important:
\es{const}{\mathsf{C}(N) = \<\O\> - \half\sum_{g=0}^\i N^{2-2g}\(N^{-1}\fp^{(g)}(0) + N^{-3}\sum_{m=0}^\i N^{-{m\o 2}} \fnp^{(g,m)}(0)\)}
$\<\O\>$ is the ensemble average with respect to the Zamolodchikov measure. Let us simplify this expression. By definition, 
\es{Olargen}{\<\O\> &= \Res_{s=1}\big[\L(1-s)\{\O, E_s\}\big] \\
&= \half\sum_{g=0}^\i N^{2-2g}\(\fp^{(g)}(1) + N^{-1}\fp^{(g)}(0) + N^{-3}\sum_{m=0}^\i N^{-{m\o 2}} \fnp^{(g,m)}(0)\)}
where we used the general form of the spectral overlap \eqr{nspec}. This simplifies the constant $\mathsf{C}(N)$ considerably, yielding the expansion 
\es{O0strong3}{\O(\l\gg1) &= \sum_{g=0}^\i N^{2-2g}\(\half \fp^{(g)}(1) + \sum_{m=0}^\i\( \mathsf{a}_m^{(g)} \l^{-{3+m\o 2}} + N^{-2-2m}\,\mathsf{b}_{\a_m}^{(g)}\l^{\a_m}\)\)}
In particular, the leading-order contribution to the average \eqr{Olargen} at genus $g$,
\es{}{\llangle\O^{(g)}\rrangle &\coloneqq  \lim_{N\rar\i} N^{2g-2} \<\O^{(g)}\> = \half \fp^{(g)}(1)\,,}
furnishes the entire constant term in \eqr{O0strong3} at every genus:
\es{O0strong4}{\boxed{{\O(\l\gg1) = \sum_{g=0}^\i N^{2-2g}\(\llangle\O^{(g)}\rrangle+ \sum_{m=0}^\i\( \mathsf{a}_m^{(g)} \l^{-{3+m\o 2}} + N^{-2-2m}\,\mathsf{b}_{\a_m}^{(g)}\l^{\a_m}\)\)}}}

This is a remarkable expression. It is the usual sum over genera, expanded in $1/\l$ (and renormalized) at every genus, with one key property: {\it the strong coupling limit is simply the ensemble average!} Focusing on leading order in large $N$, we therefore conclude that
\es{infavg}{\O(\l\rar\i) = \<\O\>}
Since the $1/\l$ expansion is, by the AdS/CFT correspondence, dual to the low-energy expansion of classical type IIB string theory around the supergravity limit, this is equivalently stated as
\es{sugraavg}{\O_{\rm sugra} = \<\O\>}
In other words, {\it an observable in classical type IIB supergravity is simply the classical limit of its $\sl$ ensemble average over type IIB string moduli space.} 

Let us exhibit this equivalence for the integrated correlator $\cG_N(\t)$. At large $N$ and large $\l$ \c{Binder:2019jwn},
\e{}{\cG_N(\l\gg1) \approx {N^2\o 4}\(1+O(\l^{-3/2})\)\qquad (N\gg1, \l \gg1)}
On the other hand, \eqr{cgnavg} implies
\e{}{\<\cG_N\> \approx {N^2\o 4} \qquad (N\gg1)}
So indeed, $\cG_{N,\,{\rm sugra}} = \<\cG_N\>$ at large $N$.

The extension of this correspondence to all orders in $1/N$ is manifest from \eqr{O0strong4}. The constant term at genus $g$ is the large $N$ limit of the genus-$g$ average:
\e{infavg2}{\O^{(g)}(\l\rar\i) = \llangle\O^{(g)}\rrangle}
Holographically, the $\l\rar\i$ limit of $\O(\t)$ at genus $g$ is dual to the $(g-g_*)$-loop supergravity result, where we define $g_*$ to be the leading non-trivial genus at which $\O(\t)$ is nonzero. So \eqr{sugraavg} extends to all genera as
\e{infavg3}{\O_{\text{sugra}}^{((g-g_*)\text{-loop})} = \llangle\O^{(g)}\rrangle}
Since the quantity $\llangle\O^{(g)}\rrangle$ is finite, the supergravity computation at each order in $G_N\sim 1/N^2$ is to be understood as regularized by the string scale cutoff. As noted earlier, this is exactly the role of the $\mathsf{b}_{\a_m}^{(g)}$ terms, dual to local counterterms in AdS$_5 \x S^5$ implementing the string theory regularization of gravitational UV divergences.\foot{We note in passing that $\sl$ invariance appears to select a certain renormalization scheme --- that is, the freedom to add finite counterterms at each order $1/N$ --- such that the renormalized $(g-g_*)$-loop bulk computation matches the specific quantity $\llangle \O^{(g)}\rrangle$ defined by conformal manifold integration.}

It is important to remember that the large $N$ limit and the ensemble average do not commute: one must take the ensemble average of $\O(\t)$ first. For one, the order-by-order average of the $1/N$ expansion is formally divergent: the modular functions appearing in the $1/N$ expansion of $\O(\t)$ are not square-integrable, as discussed around \eqr{eq:VSCfnp=0Expansion}; indeed, even integrating them once against the hyperbolic measure gives a divergence. What is more, attempting to bypass this issue by applying standard regularization techniques to these formally divergent quantities yields incorrect results. This is clearly on display for the integrated correlator $\cG_N(\t)$, studied earlier. One may write it formally as a sum of integer-index Eisenstein series $E_s^*(\t)$, as in \eqr{dgweis}. In the usual regularization (oft-employed in string theory) \`a la Zagier \cite{zbMATH03796039},  $\<E_s^*\> \rar 0$. But regularizing \eqr{dgweis} this way gives half of the correct result, $\<\cG_N\>= N(N-1)/4$. On a more physical level, non-commutativity is also visible in the fact that $\<\O\>$ generically contains a term suppressed by $1/N$, but the genus expansion proceeds in powers of $1/N^2$.\foot{This term is the ``S-dual'' of the leading, $O(N^2)$ term. Its origin is not obvious from the bulk point of view. Perhaps it should be understood using D-branes.}

To recapitulate, we have discovered that AdS$_5 \x S^5$ supergravity is, simultaneously, the classical limit of ensemble averaged type IIB string theory, and the low-energy limit of classical type IIB string theory, with the analogous statements for $\cN=4$ SYM implied by holography. 

The traditional holographic paradigm is left intact --- individual $\cN=4$ SYM theories are dual to full type IIB string theory on AdS$_5 \x S^5$, sans ensemble averaging --- while nevertheless, as seen in other holographic dualities in lower dimensions, an ensemble average {\it does} generate a simple bulk dual, which in this case is AdS$_5 \x S^5$ supergravity. The ensemble average here is literal, a well-defined integration over a conformal manifold $\cM$, equivalently, a string moduli space. And indeed, the resulting theory is {\it extremal}: having specified an $\cN=4$ superconformal gauge theory, the large $N$ averaged CFT has the largest possible spectral gap consistent with the symmetries, given by the infinite-coupling limit of the microscopic planar CFT.

This result also provides new strategies for {\it deriving} supergravity observables from $\cN=4$ SYM. Instead of computing $\O(\t)$ in the planar limit as a function of $\l$ and taking $\l\rar\i$, one could compute the average $\<\O\>$ as a function of $N$ and take $N\rar\i$. It is conceivable that the latter is computationally simpler, requiring coarse-grained information at every $N$, whereas the former, traditional approach requires complete control over strong 't Hooft coupling dynamics. Intriguingly, it is also true that the supergravity/strong coupling result is very closely related to the {\it weak} coupling expansion. In particular, the weak coupling expansion of $\O(\l) - \O(0)$, where $\O(0)$ is the free result, is given in \eqr{O0weak'}. Writing it in simpler notation as
 \e{}{\O(\l) - \O(0)  = \sum_{g=g_*}^\i N^{2-2g}  \sum_{m=1}^\i \mathsf{c}^{(g)}_m(\O) \l^m \qquad (\l\ll1)}
 and comparing to \eqr{O0strong3}, one observes that the supergravity result is simply the analytic continuation of the perturbative expansion coefficients:
\es{}{\O_{\rm sugra} = -\lim_{m\rar 0} \mathsf{c}^{(g_*)}_m(\O)}
This is a tantalizing, and more mysterious, sense in which the strong coupling dynamics are directly encoded in the weak coupling expansion, beyond the usual tenets of integrability or resurgence. 

What is the broader meaning of this for holography? It is natural to posit that this paradigm extends beyond the $\cN=4$ SYM/AdS$_5 \x S^5$ duality to other string/M-theory compactifications. The general point of view is that in the large $N$ limit, automorphic averages over U-duality symmetries of string theory, i.e. ensemble averages over generalized S-dualities in CFT, localize onto extremal points in the moduli space. We defer further comments on this and other implications for AdS/CFT to Section \ref{sec:remarks}. 

\sssec*{Comment on modularity of string states}\label{modstring}

One reading of \eqr{O0strong3} is that any planar observable $\O(\t)$ which diverges as $\l\rar\i$ cannot be $\sl$ invariant. From the CFT point of view, this includes anomalous dimensions of all operators which become single-trace at large $N$, e.g. twist-two operators. These dimensions are not non-perturbatively well-defined for all $\t$, so one should not --- and as we verify from a symmetry perspective, {\it must} not --- treat them as $\sl$ invariant. From the bulk point of view, one may ask how the $\sl$ symmetry of string theory acts on stringy states. They are not $\sl$ invariant, but one would like to characterize the representation of $\sl$, e.g. whether it is finite- or infinite-dimensional. The latter seems more likely, perhaps with the size of the representation controlled by $\l$. It is also unclear (to us) whether {\it all} states whose masses scale with $\a'$ mix under $\sl$, or whether (say) ``short string'' states with $\D\sim \l^{1/4}$ and ``semiclassical'' states \c{Tseytlin:2003ac} with $\D\sim \sqrt{\l}$ are distinguished by the action of $\sl$. 

\sec{Statistics of the $\sl$ Ensemble}\label{sec:SL2ZStatistics}

Motivated by the large $N$ equivalence between averages and supergravity, we turn to the statistics of the $\sl$ ensemble. This will allow us to quantify the extent to which supergravity approximates an individual member of the $\sl$ ensemble at large $N$.

\subsection{Higher moments of CFT observables}

We have seen that the spectral decomposition of S-duality-invariant observables $\O(\t)$ in $\cN = 4$ SYM cleanly distinguishes the ensemble average over the conformal manifold with respect to the Zamolodchikov measure.\foot{From now on, every reference to the ensemble average assumes the Zamolodchikov measure unless otherwise noted.} In particular, the spectral decomposition $\calo(\t)$ encodes the deviation from the ensemble average $\langle \calo\rangle$, which we call $\calo_{\rm spec}(\tau)$:
\begin{equation}
  \calo(\tau) = \langle \calo \rangle + \calo_{\rm spec}(\tau).
\end{equation} 
A natural question is the extent to which more intricate statistics of observables in the $SL(2,\mathbb{Z})$ ensemble are captured by their spectral overlaps.

For example, the variance of an observable in the $SL(2,\mathbb{Z})$ ensemble,
\begin{equation}
  \cV(\calo) \coloneqq \langle \calo^2 \rangle - \langle\calo\rangle^2\,,
\end{equation}
is given by the modular (ensemble) average of $\calo^2_{\rm spec}(\t)$:
\begin{equation}
  \cV(\calo) = \langle \calo_{\rm spec}^2\rangle =  \vvol(\cF)^{-1}\int_{\cF} {dxdy\over y^2} \, \calo_{\rm spec}^2(\tau) = \Res_{s=1}R_s[\calo_{\rm spec}^2].
\end{equation}
Although the RS transform on the right-hand side of this equation could be computed directly by integrating the triple products of eigenfunctions using the Clebsch-Gordan coefficients of \cite{zbMATH03796039, watson2008rankin, Benjamin:2021ygh} (and reviewed in Appendix \ref{app:inhomogeneous}), there is a shortcut. The ensemble average of $\calo_{\rm spec}^2(\t)$ can be written in terms of the Petersson inner product of $\calo_{\rm spec}(\t)$ with itself,
\begin{equation}
  \langle\calo_{\rm spec}^2\rangle = \vvol(\cF)^{-1}(\calo_{\rm spec},\calo_{\rm spec})\,,
\end{equation}
which can be evaluated using Parseval's identity, leading to the following elegant formula for the variance in terms of the spectral overlaps of $\calo(\t)$:
\begin{equation}\label{eq:spectralVariance}
  \boxed{\cV(\calo) = \vvol(\cF)^{-1}\left({1\over 4\pi i}\int_{\rre s = \half} ds\, |(\calo,E_s)|^2 + \sum_{n=1}^\infty (\calo,\phi_n)^2\right)}
\end{equation}
Thus the second moment of $\calo_{\rm spec}(\t)$ integrated over the fundamental domain $\cF$ is equal to the second moment of the spectral overlaps. Given the convergence properties of the spectral integral in the decomposition of $\calo(\t)$ itself, this manifestly converges. We note that the variance (\ref{eq:spectralVariance}) is non-vanishing for any CFT observable with a non-trivial spectral decomposition --- which is to say, for any CFT observable that varies non-trivially on the conformal manifold.

For example, the variance of the integrated correlator $\cG_N(\t)$ in the $SU(2)$ theory is easily computed from its spectral decomposition as
\begin{equation}
  \cV(\cG_2) = {\vvol(\cF)^{-1}\over 4\pi i}\int_{\rre s = \half}ds\, \left({\pi\over \sin(\pi s)}s(1-s)(2s-1)^2\right)^2 \Lambda(s)\Lambda(1-s) \approx 0.0214690
\end{equation}
Recall that the average of $\cG_2(\t)$ over the conformal manifold is given by $\langle \cG_2\rangle = \half$.

The higher moments of $\calo_{\rm spec}(\t)$ are not simply given by the corresponding higher moments of the spectral overlaps. For example, the fourth moment $\langle\calo_{\rm spec}^4\rangle$ is proportional to the inner product $(\calo_{\rm spec}^2,\calo_{\rm spec}^2)$, whose computation via the Parseval identity would involve the projections $(\calo_{\rm spec}^2,E_s)$ and $(\calo_{\rm spec}^2,\phi_n)$. The latter can be evaluated by integrating the triple products of Appendix \ref{app:inhomogeneous} weighted by spectral overlaps, but in particular they are not simply given by the square of the overlaps $(\calo,E_s)$ and $(\calo,\phi_n)$. For example, even if $(\calo,\phi_n) = 0$, it is not necessarily true that $(\calo_{\rm spec}^2,\phi_n)=0$. This is due to the nonzero triple product $K^n_{s_1,s_2}$ given in equation (\ref{eq:CGEis2Cusp}), which reflects the fact that the product of Eisenstein series has cusp forms in its spectral decomposition. These overlaps are written explicitly in (\ref{eq:OspecSquaredOverlaps}). 

\subsection{The variance at large $N$}
Here we will study the contribution of $|(\O,E_s)|^2$ to the variance in the large $N$ expansion. One expects on general principles that the variance $\cV(\O)$ is suppressed compared to the squared average $\langle \calo\rangle^2$ at large $N$, and we will see explicitly that this is indeed the case. 

We will start by considering the perturbative part, i.e. setting $\fnp(s) = 0$. Since the variance involves the second moment of the full spectral overlap $(\calo,E_s)$ (rather than the rescaled overlap $\{\O,E_s\}$), it will be convenient to define
\begin{equation}
  \mf_{\rm p}^{(g)}(s) \coloneqq {\pi\over\sin\pi s}s(1-s)\Lambda(s)\fp^{(g)}(s)
\end{equation}
so that the large $N$ expansion of the spectral overlap can be written as
\begin{equation}
  (\calo,E_s) = \sum_{g=0}^\infty N^{2-2g}\left(N^{s-1}\mf_{\rm p}^{(g)}(s) + \varphi(s)^{-1}N^{-s}\mf_{\rm p}^{(g)}(1-s)\right).
\end{equation}
On the critical line $\rre s = \half$, we then have the following expansion of the squared spectral overlap:
\begin{equation}
  |(\calo,E_s)|^2 = \sum_{g_1,g_2=0}^\infty N^{4-2(g_1+g_2)}\left[N^{-1}\mf_{\rm p}^{(g_1)}(1-s)\mf_{\rm p}^{(g_2)}(s) + N^{-2s}\varphi(s)^{-1} \mf_{\rm p}^{(g_1)}(1-s)\mf_{\rm p}^{(g_2)}(1-s) + (s\to 1-s)\right].
\end{equation}
Assuming that $\O(\t)$ starts at $g=0$ (the generalization to $g_*>0$ is obvious), the leading contributions to the variance in the $1/N$ expansion are thus given by
\begin{equation}\label{eq:overlapSquaredExpansion}
\begin{aligned}
  \int_{\rre s = \half} ds \, |(\calo,E_s)|^2 &= \, 2N^3 \int_{\rre s = \half}ds\, \mf_{\rm p}^{(0)}(s)\mf_{\rm p}^{(0)}(1-s)\\
  & \quad+ 2\int_{\rre s = \half}ds\, N^{4-2s}\varphi(s)^{-1}\mf_{\rm p}^{(0)}(1-s)\mf_{\rm p}^{(0)}(1-s) + \ldots,
\end{aligned}
\end{equation}
where the $\ldots$ denote higher-genus contributions whose $1/N$ suppression is manifest. The first term is of order $N^3$, with manifestly positive coefficient. The spectral contour for the second term must be deformed to the right. The poles of the integrand for $\Re s > \half$, in order of increasing $\Re s$, include a simple pole at $s=1$, a double pole at $s={3\over 2}$ (double because $\mf_{\rm p}^{(0)}(1-s)$ appears squared), and others with $\Re s>{3\o2}$. The simple pole at $s=1$ gives an order $N^2$ term, while the double pole at $s={3\o2}$ gives order $N\log N$ and order $N$ terms. Using the definition of $\mf_{\rm p}^{(g)}(1-s)$ and the relation
\e{}{\Res_{s=1}\[\L(s)\L(1-s)\] = {\pi \o 12}}
one finds
\es{varlargeN}{\cV(\O) = N^3\({\vvol(\cF)^{-1}\over 2\pi i}\int_{\rre s = \half}ds\, \mf_{\rm p}^{(0)}(s)\,\mf_{\rm p}^{(0)}(1-s) \)- {N^2\o 4}\(\fp^{(0)}(0)\)^2 + O(N\log N)}
As we note in the following subsection, the $1/N$ expansion is asymptotically divergent. 

Since the average is of order $N^2$, we thus conclude that the variance is parametrically suppressed at large $N$:
\begin{equation}\label{eq:varianceSuppression}
  {\cV(\calo)\over\langle\calo\rangle^2}  \sim {1\o N}~.
\end{equation}
An identical suppression holds if the leading contribution to $\O(\t)$ enters at higher genus $g_*>0$. In that case, one would have $\cV(\calo)\sim N^{3-4g_*}$ and $\langle\calo\rangle^2 \sim N^{4-4g_*}$.

\begin{figure}[t]
\centering
\includegraphics[scale=1]{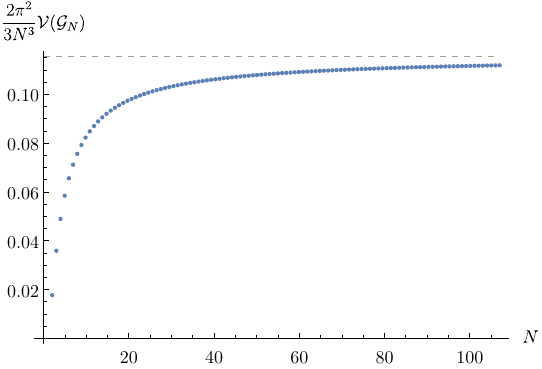}
\caption{A plot of ${2\pi^2\over 3N^3}\cV(\cG_N)$ as a function of $N$ for $N\leq 107$. This matches well onto large $N$ asymptotics \eqr{eq:VGNLargeN}.}
\label{figVGN}
\end{figure}

\begin{figure}[t]
\centering
{
\subfloat{\includegraphics[scale=0.26]{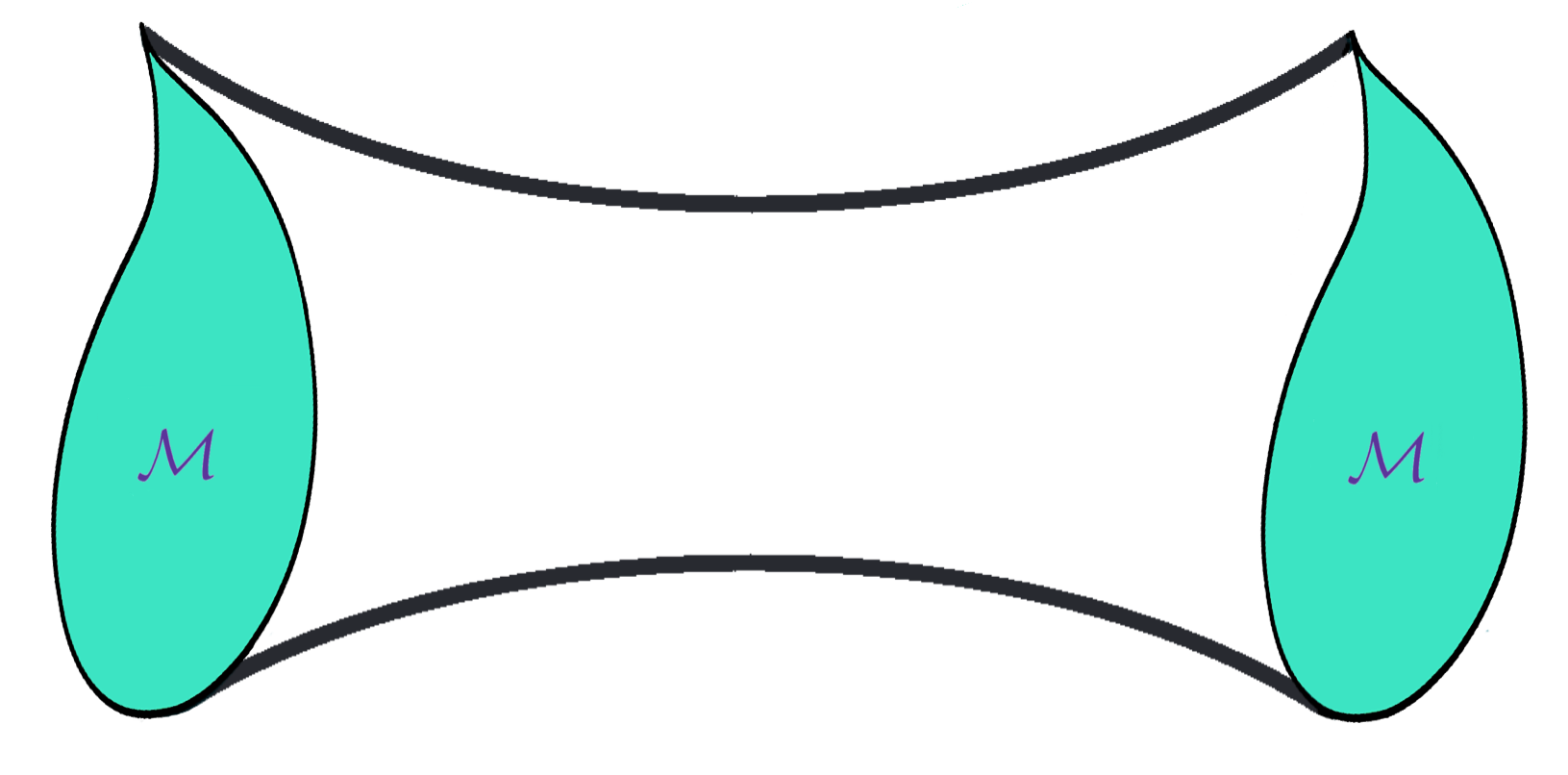}}
\qquad\qquad\qquad
\subfloat{\includegraphics[scale=0.40]{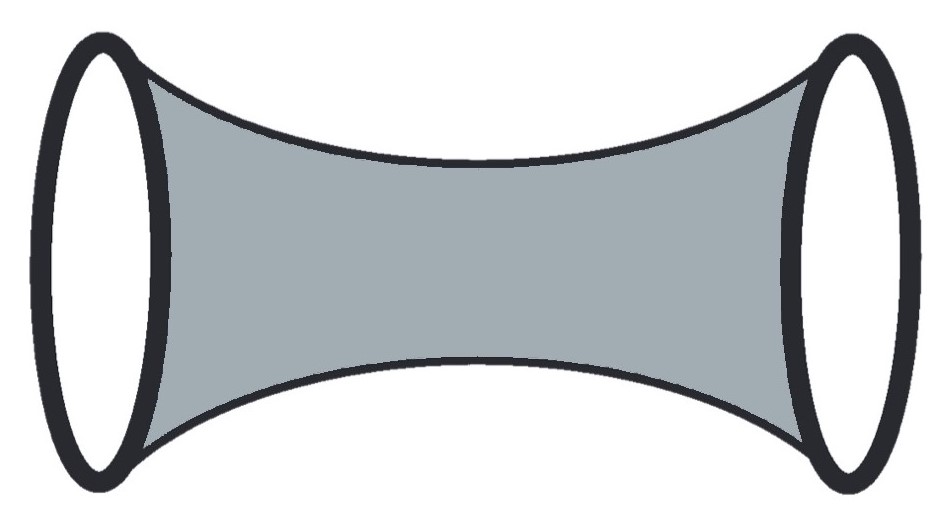}}
}
\caption{Coupling-dependent observables $\O(\t)$ have nonzero variance in the $\sl$ ensemble. This correlation may be represented by a ``wormhole'' in an abstract space containing two copies of the $\cN=4$ SYM conformal manifold $\cM$ (left). At large $N$, this invites a geometric re-interpretation, as a spacetime wormhole in AdS$_5 \x S^5$ with strongly coupled planar $\cN=4$ SYM living on the conformal boundaries (right).
}
\label{figworm}
\end{figure}

As an example, let's again study the variance of the integrated correlator $\cG_N(\t)$ (which recall has the special property that $\fnp(s) = 0$), now at large $N$. The genus-zero contribution to the variance is given to leading order in $1/N$ by 
\begin{equation}
\begin{aligned}
  \cV(\cG_N) &\approx 
   {3N^3\over 2\pi^2} \int_{-\infty}^\infty dt\, {16\pi t^2 \csch^2(\pi t)\zeta(-2it)\zeta(2it)\over 4t^2+9} + \ldots,
\end{aligned}
\end{equation}
where we used $ \mf_{\rm p}^{(0)}(s)$ from (\ref{fpdgw}) and the $\ldots$ denote both contributions that are subleading at genus zero (i.e. the terms on the second line of (\ref{eq:overlapSquaredExpansion})) and the parametrically suppressed contributions from higher genera. This can be evaluated to give
\begin{equation}\label{eq:VGNLargeN}
  {2\pi^2\over 3N^3}\cV(\cG_N) \approx 0.115551\,.
\end{equation} 
We can check this against the exact brute force evaluation of $\cV(\cG_N)$ at large but finite values of $N$. In Figure \ref{figVGN} we plot the result of direct numerical integration of $|(\cG_N,E_s)|^2$ for $N\leq 107$. The result nicely confirms \eqr{eq:VGNLargeN}.

The $1/N$ suppression of the variance is a quantitative measure of the sense in which (super)gravity is self-averaging:  in any  typical member of the $SL(2,\Z)$ ensemble at large $N$, observables $\O(\t)$ are well-approximated by their supergravity values, $\O_{\rm sugra}$. This may be compared to recent discussions of ``apparent ensemble averaging'' in holographic duality --- where moduli are either not present, or fixed and identified between the bulk and boundary --- in which gravity is said to reliably compute those quantities which are self-averaging in the large $N$ random matrix theory that ostensibly governs black hole microstates (i.e. it computes large $N$ limits of smooth functions of $N$ which well-approximate the actual, discrete sequences at large integer $N$) \cite{Schlenker:2022dyo}. The relation between gravity and averaging derived in our work, where the averaging is over couplings, is different. Both appear to be true for $\cN=4$ SYM/AdS$_5 \x S^5$ holography.

One might ask how the re-introduction of non-perturbative $q\qb$ physics would modify the discussion here. From (\ref{eq:VSCGeneral}) we see that the non-perturbative contributions to the spectral overlap $\fnp(s)$ are suppressed by a factor of $N^{-3}$ compared to the leading perturbative effects, as required by consistency with the AdS$_5\times S^5$ effective action (as we describe in Section \ref{sec:largeN}). They would thus give contributions subleading in $1/N$ compared to (\ref{eq:overlapSquaredExpansion}), and would not affect the suppression (\ref{eq:varianceSuppression}). The same argument applies to cusp form contributions $(\O,\phi_n)\neq 0$. Therefore, the $1/N$ suppression derived above is robust.

\subsection{Non-perturbative corrections to the variance}
We now show that, if the $\lambda\ll 1$ expansion of $\O(\t)$ converges, the large $N$ variance receives non-perturbative corrections in $N$, with strength controlled by the radius of convergence of the $\l\ll1$ expansion, precisely analogously to the non-perturbative effects in the 't Hooft limit treated in Section \ref{subsubsec:nonpertSL2Z}. In particular, we will see that these corrections appear at every genus.

The basic logic is the same as that described in Section \ref{subsubsec:nonpertSL2Z} in the 't Hooft limit. Let us again start by suppressing $q\qb$ contributions, setting $\fnp(s) = 0$. It is convenient to grade the perturbative contributions to the variance by the total genus $g = g_1 + g_2$, so that
\begin{equation}
  \cV(\calo) = \sum_{g=0}^\infty N^{4-2g}\cV^{(g)}(\calo)
\end{equation}
where 
\es{eq:genusGVariance}{ \cV^{(g)}(\calo) = {{\vol(\cF)^{-1}}\o 2\pi i}\sum_{g_1=0}^g\Bigg[{1\over N}&\int_{\rre s = \half}ds\, \mf_{\rm p}^{(g_1)}(s)\mf_{\rm p}^{(g-g_1)}(1-s) \\+&\ints N^{-2s}\varphi(s)^{-1} \mf_{\rm p}^{(g_1)}(1-s)\mf_{\rm p}^{(g-g_1)}(1-s)\Bigg]} 
As derived in the previous subsection, the first term in (\ref{eq:genusGVariance}) enters at a fixed order in $1/N$, while contour deformation of the second term to $\Re s >\half$ furnishes an expansion in powers of $1/N$. This sum is badly divergent. To see this, recall our previous result (\ref{fpoddres}) (subject to an assumption) on the factorial growth of the residues of the perturbative part of the spectral overlap at negative odd half-integer values of $s$, and that $\mf_{\rm p}(s)$ carries an additional factor of $\Lambda(s)$ compared to $\fp(s)$. Together these imply that the integrand of the second line of (\ref{eq:genusGVariance}) diverges quadruple-factorially. When the genera are equal, $2g_1 = g$, this is manifestly sign-definite and hence non-Borel summable. In particular, this is true of the leading contribution in $1/N$ at $g=0$ (or, more generally, at $g=g_*$).\footnote{Beyond the leading genus, sign-definiteness is guaranteed only if the $s\to\infty$ asymptotics of $\fp^{(g)}(1-s)$ have the same sign for all $g$. This is borne out in the examples we study in this paper, but we do not have a proof of this.} Assuming (\ref{fpoddres}), we then have the following asymptotic growth of the terms in the perturbative expansion in (\ref{eq:genusGVariance}), up to factors sub-exponential in $s$: 
\begin{equation}
\begin{aligned}
  N^{-2s}\varphi(s)^{-1}\mf_{\rm p}^{(g_1)}(1-s)\mf_{\rm p}^{(g-g_1)}(1-s)  &\sim \Gamma(4s)\left({8\pi^{3\over 2}\sqrt{N}\over\sqrt{\l_*}}\right)^{-4s}\qquad (s\to \infty).
  \end{aligned}
\end{equation}
Recall that $\l_*$ is the radius of convergence of the $\l\ll1$ expansion of $\calo(\t)$. Thus we conclude that the non-Borel summability of the original perturbative series (\ref{eq:genusGVariance}) implies the existence of non-perturbative corrections at large $N$ in powers of $\exp({-{8\pi^{3/ 2}\sqrt{N}\over \sqrt{\l_*}}})$. For the radius of convergence $\l_* = \pi^2$ that is canonical in planar $\cN = 4$ SYM, these non-perturbative effects proceed in powers of $\Lambda_N^2$ where 
\begin{equation}\label{eq:LambdaN}
  \Lambda_N \coloneqq e^{-{4\sqrt{\pi N}}}.
\end{equation}
Combining perturbative and non-perturbative effects, and using the canonical $\l_*=\pi^2$ for simplicity, the variance of a CFT observable in $\cN = 4$ SYM thus has the following structure at large $N$
\es{}{  \cV(\calo) = \sum_{g=0}^\infty N^{4-2g}\left(\cV_{\rm p}^{(g)}(\calo) + \cV_{\rm np}^{(g)}(\calo)\right)}
where $\cV_{\rm p}^{(g)}(\calo)$ is the $1/N$ expansion derived by contour deformation of (\ref{eq:genusGVariance}) and\footnote{This is slightly schematic. Note that the combination $\mf_{\rm p}^{(g_1)}(1-s)\mf_{\rm p}^{(g-g_1)}(1-s)$ in (\ref{eq:genusGVariance}) generically has double poles, so the perturbative expansion splits into two series, one with powers $N^{-1-2m}$ and another with powers $N^{-1-2m}\log N$. Thus there are really two towers of non-perturbative corrections due to non-Borel summability of these perturbative series.}
\begin{equation}
  \cV_{\rm np}^{(g)}(\calo) = \sum_{n=1}^\infty e^{-8n\sqrt{\pi N}}\cV_{{\rm np}| n}^{(g)}(\calo).
\end{equation}
$\cV_{{\rm np}| n}^{(g)}(\calo)$ admits a purely perturbative expansion in $1/N$ around the $n^{\rm th}$ exponential correction, and is obtained by resurgence of the perturbative terms in $\cV_{\rm p}^{(g)}(\calo)$.

How does the analysis in this section change if one allows for non-perturbative $q\qb$ contributions to the spectral overlap? It is straightforward to check that a nonzero $\fnp(s)$ leads to additional large $N$ series in (\ref{eq:genusGVariance}) proceeding in half-integer powers of $1/N$ (descending from a cross-term between $\mf_{\rm p}(1-s)$ and $\fnp(s)$) whose coefficients grow more slowly than quadruple-factorially, so the non-perturbative contributions identified in this section survive the presence of nonzero $\fnp(s)$. Whether the cusp forms give their own non-perturbative contributions to the variance cannot be determined analytically in general, but any such terms would leave the genus zero sector unaffected, as discussed in the previous subsection. 

To summarize, we have seen that at large $N$ there are non-perturbative corrections to the variance at \emph{fixed} orders in the genus expansion due to non-Borel summability of the perturbative expansion in $1/N$, with a non-perturbative scale (\ref{eq:LambdaN}) that is independent of the genus. There is also the sum over genera, which may generate yet further non-perturbative corrections. These are beyond the scope of this work but are an interesting target for future work on non-perturbative effects in $\cN = 4$ SYM and the $\sl$ ensemble.

\sec{Remarks on the AdS/CFT Paradigm}\label{sec:remarks}

Sections \ref{sec:averageAndSugra} and \ref{sec:SL2ZStatistics} lend a new perspective on planar $\cN=4$ SYM at strong 't Hooft coupling, its holographic duality with type IIB supergravity on AdS$_5 \x S^5$, and the role of ensemble averaging. In a snippet: ensemble averages capture strongly coupled physics, and holography is fine. Here we make some further comments, and some speculations, on the implications for the AdS/CFT Correspondence. 

\ssec{Wormholes in moduli space and emergent averaged holographic duality}\label{subsec:wormholesAndAverages}

The results of Sections \ref{sec:averageAndSugra} and \ref{sec:SL2ZStatistics} have interesting interpretations relating to the role of connected configurations with multiple disjoint boundaries, i.e. spacetime wormholes, in the conventional holographic paradigm, in which a complicated, UV-complete theory of quantum gravity is dual to a unique (possibly supersymmetric) quantum-mechanical theory without gravity.

In recent years, a new paradigm involving low-dimensional holographic dualities has emerged. There has been a significant accumulation of evidence that certain simple theories of AdS quantum gravity in two and three bulk spacetime dimensions are dual not to unique quantum-mechanical boundary theories but rather to ensembles of such theories. The prototypical example of this paradigm is Jackiw-Teitelboim gravity \cite{Jackiw:1984je,Teitelboim:1983ux,Almheiri:2014cka}, a theory of two-dimensional dilaton gravity which, along with a broad class of generalizations and deformations, has recently been shown to admit a non-perturbative completion in terms of a random matrix integral \cite{Saad:2019lba,Stanford:2019vob,Witten:2020wvy,Maxfield:2020ale,Turiaci:2020fjj}. More recently, a new averaged holographic duality in three spacetime dimensions has been proposed between the ensemble average of Narain's family of free boson CFTs with respect to the Zamolodchikov measure on the conformal manifold and an exotic bulk theory whose perturbative dynamics are equivalent to abelian Chern-Simons theory supplemented by a non-perturbative instruction to sum over certain bulk geometries \cite{Afkhami-Jeddi:2020ezh,Maloney:2020nni,Benjamin:2021wzr,Dong:2021wot,Collier:2021rsn,Datta:2021ftn}. 

Spacetime wormholes play an important role in these dualities, whereupon observables with multiple distinct boundaries do not factorize. On the other hand, in conventional holographic dualities, multi-boundary observables must factorize; even if there are multi-boundary wormhole solutions that are actually stable within string theory\footnote{We are not aware of any existence statements in the literature. Stability is, of course, harder to establish than instability.} at finite $\alpha'$, their contribution to physical observables in the bulk dual of individual members of the ensemble must be cancelled by some stringy mechanism. We want to understand how this picture is informed by our fundamental bulk result, ascertained holographically from S-duality of $\cN = 4$ SYM, that semiclassical type IIB supergravity in AdS$_5 \x S^5$ is both a low-energy limit and an $\sl$ average of type IIB string theory. 

We can phrase the question in the following way. The large $N$ equivalence $\<\O\> = \O_{\rm sugra}$ suggests a gravitational manifestation of the $\sl$ ensemble statistics at large $N$. How much does semiclassical AdS$_5 \x S^5$ string theory ``know'' about these statistics? 

One may probe the $\sl$ ensemble statistics by computing the set of higher moments of $\O(\t)$, as we did in Section \ref{sec:SL2ZStatistics}, which leads us directly to wormholes. The bulk prescription for computing products of boundary observables $\O(\t)$ is to study multi-boundary topologies, with an appropriate boundary condition for bulk fields at each asymptotic infinity. Since the large $N$ average over each individual $\cN=4$ SYM ensemble can be recast as a strong coupling limit, one is then led to ask whether wormholes --- i.e. contributions to the bulk path integral of connected bulk topology --- appear in the multi-boundary string theory calculation. To be clear, our goal here is {\it not} to show that wormholes strictly dominate the semiclassical AdS$_5 \x S^5$ theory: products of boundary observables do factorize, and strongly coupled $\cN=4$ SYM is noisy \c{Cotler:2016fpe,Saad:2018bqo}. Our goal is to understand whether and why wormholes appear as part of the (unaveraged) semiclassical bulk theory, and how they do so in a manner consistent with UV completeness.

Consider the squared observable $\calo(\tau)\calo(\tau)$. In Section \ref{sec:SL2ZStatistics}, we saw that any $\O(\t)$ that depends non-trivially on $\tau$ has a nonzero variance $\cV(\O)$ in the $\sl$ ensemble, with a simple expression in terms of the spectral overlaps given in (\ref{eq:spectralVariance}). The product $\calo(\tau)\calo(\tau)$ is computed by considering two copies of $\cN = 4$ SYM, corresponding to two boundaries of the bulk spacetime. For any fixed $\t$, the squared observable manifestly (tautologically) factorizes as the product. However, an ensemble average over $\cM$ would induce correlations between the two boundaries, leading to the nonzero variance (\ref{eq:spectralVariance}) (and to non-vanishing connected correlations more generally). Our view is that the large $N$ equivalence $\<\O\> = \O_{\rm sugra}$ points to a role for wormholes semiclassically in AdS$_5 \x S^5$, without sacrificing factorization.

To explain further, we find it instructive to refer to other works on this topic in two bulk dimensions \cite{Saad:2021rcu,Blommaert:2021gha,Blommaert:2021fob,Mukhametzhanov:2021nea,Mukhametzhanov:2021hdi,Saad:2021uzi}. A common thread in these works is the identification of additional saddles in the gravitational path integral that are responsible for factorization. Remarkably, in some circumstances these saddles coexist with the wormhole saddles while retaining a semiclassical description. A name that has been given to these effects is ``half-wormholes'' \cite{Saad:2021rcu}; for the purposes of the present discussion, this may be taken as shorthand for ``the part of a single-boundary observable which averages to zero.'' Our work contains the half-wormhole story in disguise. Recall that the spectral decomposition in $\cN = 4$ SYM neatly distinguishes the ensemble average, which is present in the spectral decomposition of $\calo(\tau)$ at all values of the coupling: 
\begin{equation}\label{Ospecdecomp}
  \calo(\tau) = \langle\calo\rangle + \underbrace{{1\over 4\pi i}\int_{\rre s = \half}ds\, \left(\calo,E_s\right)E_s(\tau) + \sum_{n=1}^\infty(\calo,\phi_n)\phi_n(\tau)}_{\calo_{\rm spec}(\tau)}.
\end{equation}
$\calo_{\rm spec}(\tau)$, whose average vanishes, provides the coupling dependence that must be responsible for factorization of products of $\O(\t)$ in the bulk string dual to any particular member of the $\sl$ ensemble. In this sense, one can think of $\calo_{\rm spec}(\t)$ as the half-wormhole. 

To emphasize, the key point of this split vis-\`a-vis holography is that the ensemble average $\<\O\>$ is not, from the bulk point of view, some random constant: on the contrary, it coincides precisely with $\O_{\rm sugra}$ at large $N$. 

This picture extends to higher correlations. Consider squaring $\calo_{\rm spec}(\t)$. The constant term in the spectral decomposition of $\calo_{\rm spec}^2(\t)$ is the variance of $\calo(\t)$; its non-vanishing signals a contribution of connected topologies in the gravitational computation. The remainder depends sensitively on the couplings: 
\begin{equation}\label{eq:OspecSquaredSpectral}
\begin{aligned}
  \calo_{\rm spec}^2(\tau) = \underbrace{{3\over \pi}\left[{1\over 4\pi i}\int_{\rre s= \half}ds\, |(\calo,E_s)|^2 + \sum_{n=1}^\infty (\calo,\phi_n)^2\right]}_{\text{variance }\cV(\calo): \text{ ``wormhole''}} + \underbrace{{{1\over 4\pi i }\int_{\rre s = \half}ds\, (\calo_{\rm spec}^2,E_s)E_s(\tau) + \sum_{n=1}^\infty (\calo_{\rm spec}^2,\phi_n)\phi_n(\tau)}}_{\text{coupling-dependent, noisy}}
  \end{aligned}
\end{equation}
The spectral overlaps may be determined explicitly: taking $(\calo,\phi_n)=0$ for simplicity,
\begin{equation}\label{eq:OspecSquaredOverlaps}
\begin{aligned}
  (\calo_{\rm spec}^2,E_s) &= {1\over (4\pi i )^2}\int_{\rre s_1 = \half}ds_1\int_{\rre s_2 = \half}ds_2 \, (\calo,E_{s_1})(\calo,E_{s_2})K_{1-s,s_1,s_2}\\
  (\calo_{\rm spec}^2,\phi_n) &= {\widetilde a_1^{(n)}\over (4\pi i )^2}\int_{\rre s_1 = \half}ds_1\int_{\rre s_2 = \half}ds_2 \, (\calo,E_{s_1})(\calo,E_{s_2}) K^n_{s_1,s_2}.
\end{aligned}
\end{equation}
The spectral overlaps for $\calo_{\rm spec}^2(\t)$ given above involve those of the individual observables ``linked'' by the Clebsch-Gordan coefficients of \c{zbMATH03796039, watson2008rankin}, quoted in \cite{Benjamin:2021ygh} and Appendix \ref{app:inhomogeneous}. The remainder term in (\ref{eq:OspecSquaredSpectral}) plays a role analogous to the ``linked half-wormholes'' of \cite{Saad:2021rcu}, which combines with the variance (i.e. wormhole) to give the factorized result for $\O_{\rm spec}^2(\t)$, the product of half-wormholes. In this way, there is a direct parallel between each term in \eqr{eq:OspecSquaredOverlaps} and the product of half-wormholes in the collective field description of the toy SYK model of \cite{Saad:2021rcu}. 

Let us again stress the correspondence between bulk quantities and the $\sl$ spectral decompositions. Squaring the half-wormhole term $\O_{\rm spec}(\t)$ yields a spectral decomposition which, analogously to \eqr{Ospecdecomp}, splits into the variance $\cV(\O)$ and fluctuations around $\cV(\O)$. These two pieces can compete, as they do in lower-dimensional models, to give the factorized result for $\O_{\rm spec}^2(\t)$ required by AdS/CFT; but the variance {\it is} there, falling out naturally from the spectral decomposition. This is a concrete way in which a bulk spacetime wormhole can be holographically related to ensemble statistics, even in a fixed instance of $\cN=4$ SYM. Moreover, that the ensemble average $\<\O\>$ and strong coupling limit $\O_{\rm sugra}$ actually coincide for single-boundary observables at large $N$ gives a type of explanation for {\it why} wormholes may exist in the semiclassical gravitational path integral, while still leaving room for AdS$_5 \x S^5$ string theory contributions that restore factorization. We view this as a form of resolution of the factorization puzzle \cite{Witten:1999xp,Maldacena:2004rf}. The extent to which this is addressing the standard factorization puzzle in holography is unclear, though, since our results involve proper averages over exactly marginal couplings: the well-definedness of our ensemble allows us to rigorously derive relations like \eqr{sugraavg} and \eqr{varlargeN}, but the averaging uses an ingredient, the conformal manifold, which is not always present in generic theories dual to gravity.\footnote{Another resolution of the tension between spacetime wormholes and the standard holographic dictionary in higher dimensions, proposing enormous cancellations between topologies due to gauge redundancies in quantum gravity, was recently advocated for on swampland principles in \cite{McNamara:2020uza}, building on earlier work of \cite{Marolf:2020xie}. In the tensionless limit of AdS$_3\times S^3\times T^4$ string theory supported purely by NSNS fluxes, another novel mechanism for factorization was recently discovered in \cite{Eberhardt:2021jvj}, where it was found that the fluctuations around any given background (including Euclidean wormholes) include a sum over geometries. A similar mechanism was found in a toy model for the SYK model in \cite{Mukhametzhanov:2021nea}. We also note the work \c{Heckman:2021vzx} with a different point of view on stringy realizations of ensemble averaging.}

There are various features of the above interpretation that deserve closer examination, centered around the detailed nature of the interplay between the variance and fluctuations, and how much is encoded in supergravity versus the full string theory; a total resolution of the factorization puzzle would make this clear. In connection with that, we have not answered the question originally raised by \cite{Saad:2021rcu}, addressed in low-dimensional toy models but still left open in general, of what bulk configurations in string theory correspond to ``linked half-wormholes.'' A related point is to understand the pattern of $1/N$ suppression of the variance $\cV(\O)$, which is puzzling, and its connection to, and implications for, wormhole amplitudes in AdS$_5 \x S^5$ string theory; indeed, there is some evidence (still developing) that type IIB supergravity on AdS$_5\times S^5$ supports two-boundary wormhole geometries \cite{Maldacena:2004rf,Cotler:2021cqa,Marolf:2021kjc,Mahajan:2021maz}, including some which appear to be stable but suppressed relative to disconnected topologies.

These results may also be interpreted in a more conservative context, that of statistical universality in AdS/CFT.\footnote{The discussion in this paragraph has been enriched by talks by and discussions with Tom Hartman.} Despite the fact that they are subleading topologies, wormhole amplitudes in semiclassical gravity can, in some low-dimensional examples, capture universal aspects of local operator content in chaotic quantum systems. The prototypical example is the linear ramp at late times in the spectral form factor \cite{Cotler:2016fpe,Saad:2018bqo,Cotler:2020ugk}, which signals high-energy level-spacing statistics of RMT. We view this as conceptually similar to the sense in which universal aspects of large $N$ CFT data are captured by certain leading topologies or geometries in gravity (a canonical example is the extended universality \cite{Hartman:2014oaa} of the Cardy formula for the asymptotic density of states in CFT$_2$ \cite{Cardy:1986ie}, which is the Bekenstein-Hawking entropy of the BTZ black hole \cite{Strominger:1997eq}). In this light, perhaps wormholes should be thought of as effective solutions valid within the low-energy gravitational EFT that nevertheless capture universal aspects of operator statistics in chaotic CFTs. (See e.g. proposals in \c{Belin:2020hea,Altland:2021rqn}.) In order to probe their presence, one needs to perform an operation that projects out the fine-grained details of the UV completion. In this paper we have seen that one such way to achieve that is to average over a U-duality group. 

Let us make a few further remarks before moving on. 

One might hope that a resolution of the factorization puzzle for supersymmetric observables could be attained solely at the level of the gravitational path integral of the low-energy effective theory, without the need to invoke UV-sensitive mechanisms for the cancellation of wormhole contributions. Such a mechanism was explicitly demonstrated directly in the bulk in \cite{Iliesiu:2021are} for supersymmetric indices, which are independent of couplings. It is not clear to us if our results have any implications for this mechanism for observables that depend on the coupling, such as the supersymmetric integrated correlator $\cG_N(\t)$. On the one hand, any observable that depends on the coupling will have nontrivial moments in the $\sl$ ensemble, with an average given at large $N$ by its value in supergravity. On the other, it is clear from its explicit expression (\ref{eq:spectralVariance}) that the variance is built from ingredients that, mapped to the bulk, encode more than just supergravity. More generally, it would be good to better understand the bulk degrees of freedom in averaged string theory dual to the $1/N$ corrections of averaged quantities in the $SL(2,\mathbb{Z})$ ensemble.

The structure of the large $N$ expansion of non-trivial moments of CFT observables in the $\sl$ ensemble is somewhat evocative of the duality between Jackiw-Teitelboim gravity and a double-scaled random matrix integral (henceforth RMT) \cite{Saad:2019lba}.  In $\cN =4$ SYM we have seen that there is a hierarchy of both perturbative and non-perturbative corrections to higher moments of CFT observables in the large $N$ limit, and that the variance is parametrically suppressed compared to the mean-squared. In the JT/RMT duality, the connected two-boundary path integral $\langle Z(\beta_1)Z(\beta_2)\rangle_{\rm conn}$ is similarly suppressed
\begin{equation}
  {\langle Z(\beta_1)Z(\beta_2)\rangle_{\rm conn}\over \langle Z(\beta)\rangle^2} \sim L^{-2} \sim e^{-2S_0},
\end{equation}
where $L\sim e^{S_0}$ (where $S_0\sim 1/G_{\rm N}$) is the size of the random matrices. We note, however, that the relationship between $N$ and the gravitational coupling $G_N$ is different in $\cN = 4$ SYM than in the JT/RMT duality. In particular, from the holographic dictionary for AdS$_5\times S^5$, we have 
\begin{equation}\label{eq:NSquaredGNewton}
  N^2 \sim {1\over G_{\rm N}},
\end{equation} 
where $G_{\rm N}$ is the five-dimensional Newton's constant in AdS units.

Observables in RMT are computed in a genus expansion that proceeds in powers of $L^{-2}$. The sum over genera typically diverges like $(2g)!$ and induces non-perturbative contributions of the form $\sim e^{-\# L}$. In $\cN=4$ SYM there is more structure: there are non-perturbative corrections to the variance at large $N$ at fixed orders in the genus expansion from a quadruple-factorially divergent perturbative expansion, potentially in addition to those from the sum over genera, which we have not studied in this work. Also, in the JT/RMT duality the non-perturbative effects in RMT translate into doubly non-perturbative effects in JT gravity of size $e^{-\# e^{S_0}}$; in $\cN=4$ SYM, it is rather unclear to us what the gravitational meaning of the non-perturbative $\sim e^{-\#\sqrt{N}}$ contributions to the variance is. It would be interesting to better understand the physical interpretation of the non-perturbative scale $\L_N$. 

\ssec{Generalizations}

An obvious question is whether a version of the present mechanism in $\cN=4$ SYM extends to other string/M-theory vacua and their dual CFT pairs, in other spacetime dimensions and with less supersymmetry. It seems quite likely to us that this is the case. Let us articulate some possibilities in order of increasing strength. To set up, we consider sequences of CFTs $\mathcal{T}_N$ with conformal manifolds $\cM$ that admit large $N$ limits, with a (possibly trivial) generalized S-duality group, $\mathbb{S}$ \c{Argyres:2007cn}. We do not assume that the exactly marginal couplings are gauge couplings. We take there to be a large $N$ ``double-scaling limit'' that organizes into a sum over genera \`a la 't Hooft, possibly with boundaries, with couplings $\l_i$ on which local observables in the limit theory $\mathcal{T}_\i$ depend. In particular, we are interested in classes of theories where $\mathcal{T}_\i$ includes a ``'t Hooft coupling'' $\l$, constructed from an exactly marginal coupling on $\cM$, that controls the low-energy expansion of the bulk effective field theory; this is characteristic of CFTs with string (rather than M-) theory duals. So as to simplify language, we assume supersymmetry.

The most plausible scenario applies to sequences where $\mathcal{T}_{\i}$ contains a strongly coupled regime, dual to two-derivative AdS supergravity.\foot{The relevant abstract CFT notion of ``strongly coupled'' is that there exists a large $N$ regime with a parametrically large spectral gap to single-trace higher-spin operators \cite{Heemskerk:2009pn, Camanho:2014apa}.} One claim would be that for these sequences of CFTs, the relations \eqr{infavg} and \eqr{sugraavg} continue to hold: 

\begin{quotation}
\ni {\bf Scenario I.} When $\mathcal{T}_{\i}$ contains a strongly coupled regime, dual to an AdS supergravity, the automorphic average over $\mathbb{S}$, i.e. over the U-duality group of the dual string theory, localizes onto a strongly coupled theory at large $N$.
\end{quotation}

\ni Where should we look to test such a scenario? The other maximally supersymmetric duality involving exactly marginal couplings, namely type IIB string theory on AdS$_3 \x S^3 \x T^4/K3$ and its duality to the deformed symmetric orbifold theories, is a prime candidate (discussed further below). It may be technically simpler to consider other 4d SCFTs with AdS$_5 \x M$ type IIB string theory duals and sub-maximal supersymmetry. For example, consider those which are obtainable by deformations of $\cN=4$ SYM, on which the ``mother'' $\sl$ still acts. For example, the conifold theory \c{Klebanov:1998hh}, a $\cN=1$ SCFT dual to AdS$_5 \x T^{1,1}$, is of this type. While the explicit $\sl$ action on $\cM_{\rm conifold}$ is not well understood, one may still be able to identify an averaging mechanism. In particular, an $\sl$-invariant observable is subject to $\sl$ harmonic analysis. 

We emphasize two relevant conceptual points here. First, for $\cN=4$ SYM where $\mathbb{S} = \sl$, the equivalence between the large $N$ ensemble average and the strongly coupled limit follows only from structural features of the spectral decomposition, in particular the general scaling properties of the Eisenstein series and Maass cusp forms. The same should hold for $\mathbb{S}$ for which harmonic analysis is available, and only partial information may be necessary. Second, in $\cN=4$ SYM, the Zamolodchikov measure on $\cM$ and the S-duality-invariant measure on $\cF$ happen to be equal, leading to $\<\O\> = \overline\O$. However, more generally, they are not equal; see e.g. \c{Baggio:2014ioa} for the nice example of 4d $\cN=2$ SQCD. So any generalization of the $\cN=4$ SYM story must specify which ensemble averaging measure is the relevant one. Non-Zamolodchikov choices are certainly allowed. Indeed, the only other natural choice seems to be the $\mathbb{S}$-invariant measure. One may take this as an argument in favor of the automorphic average, i.e. the ensemble average with respect to the $\mathbb{S}$-invariant measure.  

Having said that, we may imagine the slightly stronger possibility: 

\begin{quotation}
\ni {\bf Scenario II.} When $\mathcal{T}_{\i}$ contains a strongly coupled regime, dual to an AdS supergravity, the ensemble average over $\cM$ localizes onto a strongly coupled theory at large $N$, irrespective of any generalized S-duality.
\end{quotation}

\ni In other words, {\it every} AdS supergravity background dual to a limit CFT $\mathcal{T}_{\i}$ may be thought of as an ensemble average.\foot{There are AdS supergravity backgrounds that are not dual to sequences with exact conformal manifolds, e.g. certain 3d Chern-Simons-matter theories. $\mathcal{T}_N$ does not include this class of theories.}  Of course, without S-duality, the only natural choice of measure is the Zamolodchikov measure, and harmonic analysis for $\mathbb{S}$ is not available. As far as we are aware it is an open question whether 4d SCFTs with conformal manifolds must admit a generalized S-duality group action.\foot{This second scenario makes an appealing connection between the non-SUSY AdS conjecture of \cite{Ooguri:2016pdq}, and the folklore that only superconformal theories can possess exactly marginal couplings. The non-SUSY AdS conjecture is that a semiclassical theory of AdS Einstein gravity --- specifically, with all higher-derivative gravitational corrections {\it parametrically} suppressed --- must in fact be a supergravity. The scenario above would give a novel sort of justification for that: non-SUSY AdS Einstein gravity does not exist because non-SUSY conformal manifolds do not exist. This is a caricature as stated, too strong because a sequence of CFTs can develop a continuous coupling that is a large $N$ artifact, as in the ABJM theories where $\l=N/k$ \c{Aharony:2008ug}. However, we expect that there are essential differences between theories obtained from sequences of CFTs with truly continuous parameters, and those with ``emergent'' continuous parameters at large $N$. It would be nice to understand this distinction in general.}

In both of the above scenarios, {\it all} couplings are averaged over. However, perhaps the closest generalization of what happens in $\cN=4$ SYM only requires averaging over the ``gravity direction'' of $\cM$: if $\l:= g N$ is the double-scaled coupling which controls the bulk derivative expansion, where $g$ is an exactly marginal coupling, one may imagine averaging only ``along $g$'', i.e. over a sub-manifold of $\cM$.\foot{$g$ may be packaged in a complexified coupling, as in $\cN=4$ SYM.} This leaves the other couplings, which do not become strong, unaveraged. Such an intermediate scenario may thus be stated as follows:

\begin{quotation}
\ni {\bf Scenario II$'$.} When $\mathcal{T}_{\i}$ contains a strongly coupled regime, dual to an AdS supergravity, the ensemble average over the ``gravity direction'' of $\cM$ localizes onto a strongly coupled theory at large $N$, irrespective of any generalized S-duality.
\end{quotation}

\ni From the bulk point of view, this scenario would draw an equivalence between a large $N$ average and a supergravity with a moduli space, including the functional dependence of the observables on the other moduli. This makes contact with other familiar examples of holographic duality.

As an illustrative and relevant example, consider the large $N$ CFT dual to type IIB string theory on AdS$_3 \x S^3 \x M_4$, where $M_4 = T^4$ or $K3$. Choosing $T^4$ for definiteness, the conformal manifold $\cM$ is 20-dimensional, with a moduli space locally of the form $SO(4,5)/SO(4) \x SO(5)$. One modulus $g$ interpolates to the supergravity regime at large $N$. The $U$-duality group is a subgroup of $SO(5,5;\Z)$, denoted $\mathcal{H}_{\vec q}$ in \c{Larsen:1999uk}, which leaves invariant the five ``fixed scalars'' descending from $T^4$ which are not moduli of the AdS$_3 \x S^3$ theory. So, one may consider doing (at least) three types of averages: average over $\cM$ with respect to the $\mathcal{H}_{\vec q}$\,-invariant measure; average over $\cM$ with respect to the Zamolodchikov measure; or average only over $g$. In the latter two cases, the measures are not known explicitly. At large $N$, each of these cases corresponds to one of the scenarios above, with a putative AdS$_3 \x S^3$ supergravity in the bulk. Only the last scenario leaves some moduli unaveraged. It is an intriguing possibility that the first two scenarios could land on a simple low-energy theory in the bulk.

Finally, an even stronger, and somewhat plausible, possibility --- for which we have no evidence --- is that the above scenarios extend to theories which do not even admit a strongly coupled regime at large $N$:

\begin{quotation}
\ni {\bf Scenario III.} For every sequence $\mathcal{T}_N$, the ensemble average over $\cM$ localizes onto the {\it extremal} theory, $\mathcal{T}_{\i}^*$, at large $N$.
\end{quotation}

\ni There are different things one may mean by ``extremal CFT'', but any definition conveys a notion of optimizing CFT data consistent with consistency conditions. A representative definition is that an extremal CFT maximizes a spectral gap over the space of 't Hooft couplings $\l_i$:
\es{}{\mathcal{T}_{\i}^* \coloneqq \mathcal{T}_\i\,\big|\,\D_{\rm gap}= \D_{\rm gap}(\l_i^*),\quad \text{ where }~~ \D_{\rm gap}(\l_i^*) \coloneqq \max_{\l_i} \D_{\rm gap}(\l_i)\,,}
where $\D_{\rm gap}$ is, say, the single-trace higher-spin gap.\foot{Some other possible definitions include maximizing the gap to the first unprotected primary operator; maximizing sparseness of the light spectrum; maximizing the fraction of OPE data which are extremized over $\l_i$; or optimizing bootstrap constraints in theory space, a concept that may be formed rigorously \cite{El-Showk:2012vjm, El-Showk:2016mxr,Mazac:2018mdx,Mazac:2018ycv,Afkhami-Jeddi:2021iuw}. Note that while is no proof that a single theory must simultaneously extremize {\it all} CFT data, functional and other methods in the conformal bootstrap do justify the expectation of the existence of an extremal CFT at the boundaries of bootstrap functional space.} Then the scenario outlined above is that CFT observables $\O(\l_i)$ obey
\es{}{\lim_{N\rar\i}\<\O\> = \O(\l_i^*)}
In the earlier, less adventurous scenarios, $\D_{\rm gap}(\l_i^*)\rar\i$, while here it remains finite. There are of course many examples of sequences of this type coming from vanilla gauge theories. A canonical example is 4d $\cN=2$ SQCD with gauge group $SU(N)$ and $N_f = 2N\rar\i$, where $a\neq c$ to leading order in large $N$. Others may be found in a classification of 4d $\cN=1$ conformal gauge theories with simple gauge groups and weakly coupled conformal manifolds \c{Razamat:2020pra}.\foot{Those with large $N$ limits were collected in Tables 4.1 -- 4.2 of \c{Perlmutter:2020buo}. The entries with $\a = 1/\sqrt{2}$ have $a\approx c$ at large $c$.} This scenario seems less likely to us, as it abandons the general connection between ensemble averaging and universal gravitational physics. Whether any of these scenarios is correct beyond $\cN=4$ SYM/AdS$_5 \x S^5$ holography deserves further study. 

Let us conclude with further comment on the liminal case of AdS$_3$/CFT$_2$. Certainly we expect the overall picture developed here to hold for AdS$_3$ compactifications, in a sense described above. As for the quest to construct a theory of ``pure'' AdS$_3$ quantum gravity or its 2d CFT dual, we have little concrete to add. It has been speculated that pure AdS$_3$ quantum gravity may be dual to an ``ensemble'' ``average,'' or to a ``random 2d CFT'' \cite{Cotler:2020ugk,Cotler:2020hgz}, in a sense that has not yet been explicitly defined or articulated. There are arguments and suggestive computations, but a more conservative viewpoint casts these results more straightforwardly in terms of the high-energy statistics of local operators in chaotic CFTs. For various reasons, it is well-motivated to consider a slightly generalized quest: find a (S)CFT with large central charge, and a gap to the AdS$_3$ black hole threshold modulo an exactly marginal operator (and its multi-trace composites). This seems both strategically useful and physically interesting. If there is a lesson in our work for the pursuit of semiclassical AdS$_3$ pure gravity, it may be that something morally close to it may be found from an ensemble average in an analogous sense as for AdS$_5 \x S^5$ supergravity, perhaps following one of the scenarios described in this subsection.

\sec{Conclusion}\label{sec:conclusion}

An overarching message of this work is that harmonic analysis for $\sl$ is an insightful and incisive tool for extracting the physical ramifications of S-duality of $\cN=4$ SYM. This approach uncovered a direct role for ensemble averaging in the holographic duality with type IIB string theory on AdS$_5 \x S^5$, revealing supergravity as an emergent large $N$ ensemble average. Our philosophy was simple, and familiar: to manifest the symmetries of the theory. By using an S-duality-invariant formulation from the start, tucking away all functional dependence on the coupling into an $\sl$ eigenbasis, many non-perturbative phenomena presented themselves, often tied up with perturbative physics in a manner characteristic of strong-weak duality in quantum field theory. 

Given their grounding in symmetry considerations --- and how elemental they are mathematically --- we are optimistic that spectral methods for S-duality have more physics in the offing. The results of this paper fell out very efficiently from few ingredients. Our computations are quite different than existing computations of $\cN=4$ SYM observables. It would be productive to reformulate what we know in the spectral language. 

We close by listing some specific future directions of keen interest. 

{\bf More $\cN=4$ SYM observables:} As an aid to future progress we mention some observables one could analyze using $\sl$ spectral methods, in order of increasing ambitiousness. There are some nearby extensions, such as the generalization of $\cG_N(\t)$ to other SYM gauge groups or to the $\sl$-covariant cousin in \c{Dorigoni:2021rdo} and the integrated $\la 22pp\ra$ correlator written in \eqr{22ppdef}, for which we made a tentative prediction of its spectral overlap. A more complicated extension would be to the extremal correlators $\la \O_p \overline \O_p\ra$ in 4d $\cN=2$ SQCD: upon multiplying by an appropriate power of $y$, these become $\sl$ invariant \c{Baggio:2014ioa}. Like $\cG_N(\t)$, they depend only on $\t$ and are derivable from localization. For $p=2$, this is the Zamolodchikov metric on the SQCD conformal manifold. Some preliminary calculations suggest a level of complexity similar to $\cF_N(\t)$. Does the Toda chain relating different extremal correlators play nicely with the spectral decomposition? Perhaps there is a clean recursion relation in $p$ for the respective spectral overlaps.

More substantial (and obvious) targets are {\it unintegrated} correlators, like the $\O_{\mathbf{20'}}$ four-point function. It may be more tractable to compute the average and variance of the unintegrated four-point function than to obtain it across the entire conformal manifold; this would provide access to information about the statistics of unprotected local operators in the $\sl$ ensemble, features that are obscured in the integrated correlators we study in this work. One can also attempt to address information puzzles in holography by studying e.g. thermal two-point functions.

This formalism is not fundamentally restricted to local observables. Consider the thermal free energy on $S^3 \x S^1$. Unlike the observables considered in this paper, this does not appear to have a perturbative expansion in integer powers of $1/y$ \c{Nieto:1999kc}, so modifications of some of our formulas are necessary. It seems plausible that, following the ideas in Section \ref{sec:averageAndSugra} at large $N$, one could derive the famous factor of 3/4 from CFT \c{Gubser:1996de}. 

{\bf Integrated correlators:} We derived the form of the integrated correlator $\cG_N(\t)$, confirming the conjecture of \cite{Dorigoni:2021guq}, as a specific case of our general formulas for $\cN=4$ SYM observables at both finite $N$ and large $N$. This revealed its optimal simplicity in the space of $\cN=4$ SYM observables. What we did {\it not} derive is the Laplace difference equation \eqr{laplacediffeq}, which surely lies at the core of any fundamental understanding of the dynamical content of $\cG_N(\t)$. It would be very interesting to know how common such recursion relations are in $\cN=4$ SYM, relating observables at different values of $N$. We also analyzed the supersymmetric integrated correlator, $\cF_N(\t)$, whose general form would be nice to derive. Such a spectral decomposition would provide an explicit non-trivial example of a CFT observable with non-perturbative (instanton-anti-instanton) physics at finite $N$, for which both $\fnp(s)$ and $(\cF_N,\phi_n)$ are non-vanishing.

{\bf Instantons:} Instantons are redundant, with $k>1$ sectors uniquely determined by the $k=0,1$ sectors. For Borel summable observables at finite $N$, there is a remarkably simple formula \eqr{Rk} for the $k$-instanton radius of convergence (modulo cusp forms, whose contribution we characterized in Subsection \ref{seccusp}). Can we understand these results from traditional instanton calculus? We also derived the existence of nonzero $q\qb$ (instanton-anti-instanton) terms in the conformal dimensions of unprotected operators, including the Konishi dimension $\D_K(\t)$. This result suggests, but does not strictly imply, non-Borel summability of the weak coupling expansion of $\D_K(\t)$. It would be valuable to investigate this question, and to derive the precise \ii terms whose existence we have discovered --- say, for $SU(2)$ --- with direct computations.

{\bf Combining $\sl$ spectral theory with other methods:} We expect synergy from combining these techniques with other preexisting ones in $\cN=4$ SYM, e.g. integrability methods and the superconformal bootstrap. At a more conceptual level, the bootstrap philosophy of imposing abstract constraints could be generalized to spectral overlaps --- that is, to {\it bootstrapping in spectral space}. Recall that for the Eisenstein overlap $(\O,E_s)$, its analyticity properties in the complex spectral parameter $s$ were completely determined in Section \ref{sec:analyticStructureCFTOverlaps}; the cusp form overlap $(\O,\phi_n)$ is less constrained, but for the interesting reason that it signals arithmetic chaos. ``Analyticity in spectral space'' should be investigated as an $\cN=4$ SYM bootstrap constraint. It would also be interesting to pursue the constraints of the functional equation satisfied by the spectral overlap (\ref{spectralCrossing}), viewed as a crossing equation, on the non-perturbative contributions to CFT data. 

{\bf String theory and holography:} We discussed the implications for the AdS/CFT paradigm in Section \ref{sec:remarks}, so let us just briefly mention here a few things. We emphasize again the tantalizing, if uncertain, prospects for CFT derivations of supergravity physics using ensemble averages and the observations of Section \ref{sec:averageAndSugra}. It would also be very interesting to better understand the worldsheet and spacetime perspectives on the non-Borel summability of AdS$_5 \x S^5$ strings, argued holographically in Section \ref{secstringnp}. Finally, our initial exploration of the $\sl$ ensemble statistics and possible connections to RMT would be very interesting to flesh out. Perhaps the spectral formalism can lead to a string/brane description of half-wormholes. 

{\bf Other S-duality groups:} One would like to generalize the whole formalism away from $\sl$ to SCFTs with other S-duality groups $\mathbb{S}$, as discussed in Section \ref{sec:remarks}. Two natural approaches suggest themselves: look for interesting theories, or look for tractable groups for which harmonic analysis is (at least partly) established. In the latter approach, there are some automorphic groups where very tidy results are possible. One inspirational example is $\mathbb{S} = G_q$, the Hecke triangle group, a discrete subgroup of $PSL(2,\mathbb{R})$ generated by $S$ and $T$ elements subject to $S^2=1$ and $(ST)^q=1$. The fundamental domain $\cF_q$ for $G_q$ may be defined as the semi-infinite region of $\HH$ bounded by $|\t|=1$ and $|x| \leq \cos(\pi/q)$. The $L^2(\cF_q)$ eigenspace has continuous and discrete components.\foot{If $q=4,6$, the eigenbasis is directly constructed from that of $G_3 = \sl$ \c{zbMATH00426391}.} It has been conjectured by Phillips and Sarnak \c{Sarnak1985}, and substantially supported by numerics and analytics \c{zbMATH00426391,hjudge}, that there are {\it no} even Maass cusp forms unless $q =3,4,6$, where $G_q$ becomes arithmetic. So observables in SCFTs with $\mathbb{S} = G_{q\neq 3,4,6}$ are especially simple, and non-chaotic. To boot, SCFTs with $\mathbb{S} = G_q$ have been identified as certain deformations of 4d $\cN=2$ SQCD \cite{Ashok:2016oyh}. More generally, arithmeticity is believed to be crucial for the existence of cusp forms; translated into CFT terms, this suggests that arithmeticity of $\mathbb{S}$ is a useful criterion for the classification of SCFTs.

{\bf Arithmetic chaos:} An interesting output of this work is the realization that generic CFT observables in $\cN = 4$ SYM exhibit arithmetic chaos via the presence of the erratic Maass cusp forms in their spectral decomposition.\foot{A recent paper \cite{McLoughlin:2020zew} found evidence for level-spacing statistics of RMT in the GOE ensemble in perturbative $\cN=$4 SYM at finite $N$, though this seems qualitatively different than the chaos and methods being discussed here.} Although we were not able to say much analytically about the contribution of the cusp forms to CFT observables in general (an exception being \eqr{cuspdiag2}) we emphasize that there is a wealth of both established and conjectured knowledge about their distributional and statistical properties in the math literature. An obvious avenue for future work is to leverage these properties to connect the spectral decomposition to more conventional spectral and dynamical notions of quantum chaos in CFTs and holography and to better constrain the chaotic sector of CFT observables. 

\sec*{Acknowledgments}
We thank Ofer Aharony, Chris Beem, Nikolay Bobev, Gabriele di Ubaldo, Daniele Dorigoni, Michael Green, Peter Humphries, Luca Iliesiu, Grisha Korchemsky, Wolfgang Lerche, Hirosi Ooguri, Silviu Pufu, Himanshu Raj, Leonardo Rastelli, Edgar Shaghoulian and Yifan Wang for helpful discussions and questions. This work was initiated at Aspen Center for Physics, which is supported by National Science Foundation grant PHY-1607611. EP is supported by ERC Starting Grant 853507.

\begin{appendix}

\sec{A few details for constraining spectral overlaps}\label{appa}

\ssec{Deriving  \eqr{OEgrowth}}
Here we derive the leading-order bound \eqr{OEgrowth}, ignoring subleading multiplicative corrections. The Maass-Selberg relation (e.g. \cite{Green:2014yxa}, equation C.10) implies that 
\e{}{|E_{\half+i t}|^2 \approx -\half\({\varphi'\o\varphi}\Big(\half+it\Big) + {\varphi'\o\varphi}\Big(\half-it\Big)\)}
where $\varphi(s)$ was defined in \eqr{varphidef}. Then the triangle inequality, together with \cite{Terras_2013} Exercise 3.7.28, gives
\es{}{|E_{\half+i t}|^2
&\leq -\left|{\varphi'\o\varphi}\Big(\half+it\Big)\right|\\
&\lesssim c\log|t|\ \qquad  \qquad (t\rar\i)}
for some constant $c$.\foot{An improved bound, still logarithmic, can be found by using bounds on $|{\z'\o\z}(1+2it)|$ given in \cite{Trudgian_2015}.} Translating this to a bound on $\{\O,E_{\half+it}\}$ using 
\e{}{\Bigg|\G\(\half+it\)\Bigg| \sim e^{-\pi t/2} \qquad (t\rar\i)}
and \cite{Trudgian_2015}
\e{}{|\z(1+it)^{-1}| \sim O(c' \log |t|) \qquad (t\rar\i)}
for another constant\foot{In \cite{Trudgian_2015}, $c'=6.9 \x 10^6$. This is probably not optimal. See also \cite{Gelbart_riemannszeta} for a nice discussion of the Lindelhof conjecture and progress toward its proof.}   $c'$ gives the growth bound \eqr{OEgrowth}. 

\ssec{Proofs for Section \ref{sec:analyticStructureCFTOverlaps}}\label{app:proofs}

Here we provide proofs of the statements made in Section \ref{sec:analyticStructureCFTOverlaps}. 

\begin{itemize}

\item {\it $\fp(s)$ and $\fnp(s)$ are regular for all $s\in\CC$ away from $s=1$ (and its reflection)}

\ni By the property of the RS transform, $\{\O,E_s\}\L(s)^2$ is meromorphic with a possible pole at $s=1$ and, due to the perturbative terms, simple poles at $s>1$ with $\Im s=0$. Since $\L(s)$ is itself meromorphic on $\mathbb{C}$ and regular away from $s=0,\half$, the aforementioned properties apply to $\{\O,E_s\}$ directly. As the poles at $s>1$ are encoded in the explicit factor of $\sin(\pi s)$, holomorphy of $\fp(s)$ and $\fnp(s)$ away from $s=1$ follows.

\item {\it At $s=1$, $\lim_{s\rar 1}\,\{\O,E_{s}\} = 2\, \overline{\O}$.}

\ni A direct computation gives
\es{avfree}{ \lim_{s\rar 1}\{\O,E_{s}\} 
&=  \lim_{s\rar 1}\Big(R_s[\O] \L(1-s)^{-1}\Big) \\&= 2\Res_{s=1} R_s[\O] \\&= 2\,\overline{\O}}

\item {\it On the critical line $s=\half+it$, $\{\O,E_s\}$ is finite for finite $t$. }

\ni $\L(\half+it)^{-1}$ and $(\O,E_{\half+it})$, and hence their product $\{\O,E_{\half+it}\}$, are both finite. These follow from putting together two facts. First, $\z(1+2it) \neq 0$ and $\G(\half+it) \neq 0$ for $t\in\mathbb{R}$ (cf. \cite{Terras_2013}, Exercise 3.5.8). Second, $E_{\half+it}(\t)$ for fixed $\t$ is holomorphic (cf. \cite{Iwaniec2002SpectralMO}, Theorem 6.11).

\item  {\it If $\fnp(s)=0$ (resp. $\fp(s)=0$) then $\fp(s)$ (resp. $\fnp(s)$) is entire.  }

\ni This follows from the preceding results. 

\end{itemize}

\ni Note also a small corollary of these results: {\it if $\O_0(y)$ has a finite number of perturbative terms, it must have non-perturbative terms.} In the language of \eqr{form}, if $\fnp(s)=0$, then $\fp(s)$ is entire. A finite number of perturbative terms means $\fp(s) = \sin \pi s \x (\text{something with poles})$. But $\sin \pi s$ is entire, so there is a contradiction.

\section{Solutions to the inhomogeneous Laplace equation}\label{app:inhomogeneous}

In this appendix we will study the spectral decomposition of the solution to the inhomogeneous Laplace equation sourced by a product of real analytic Eisenstein series
\begin{equation}\label{eq:inhomogeneousLaplaceEquation}
  \left(\Delta_\tau + r(r+1)\right)F_{r;s_1,s_2}(\tau) = E_{s_1}(\tau)E_{s_2}(\tau).
\end{equation}
This function will serve as a prototypical example of a modular invariant that appears in observables in both $\cN=4$ SYM and type IIB string theory that has non-trivial non-perturbative contributions in its spectral decomposition. This function, with $r=3$ and $s_1 = s_2 = {3\over 2}$ was first conjectured to capture the non-perturbative corrections to the $D^6R^4$ interaction, which are mediated by D-instantons, in the low energy expansion of the type IIB effective action in \cite{Green:2005ba}, and was further studied in detail in \cite{Green:2014yxa} (see also \cite{Green:2010kv,Green:2011vz}). They also appear at integer powers of $1/N$ in the large $N$ expansion of the integrated correlator $\cF_N(\tau)$ of Section \ref{subsec:F4} \cite{Chester:2020vyz} for certain positive integer values of $r$ and odd half-integer values of $s_1, s_2$. Such functions were studied at integer values of $s_1, s_2$ in \c{Dorigoni:2021jfr,Dorigoni:2021ngn}, and more generally in \c{klingerlogan2018differential}. 

The homogeneous solution to the Laplace equation is itself a real analytic Eisenstein series $E_{r+1}(\tau)$. However in the physical applications of interest \cite{Green:2014yxa,Chester:2020vyz}, one typically assumes the moderate growth condition
\begin{equation}
  F_{r;s_1,s_2}(\tau) \sim O(y^{s_1+s_2}), \quad y \to \infty.
\end{equation}
In our application (in particular, in the large $N$ expansion of the integrated correlator studied in the very strongly coupled limit in Section \ref{subsec:F4}) we will have $\Re (r+1) > \Re(s_1+s_2)$ so the homogeneous solution, which has the moderate growth $E_{r+1}(\tau) \sim O(y^{r+1})$ at the cusp, is ruled out. 

\subsection{General case}
Here we will follow the approach of \cite{Green:2014yxa} in studying the spectral decomposition of $F_{r;s_1,s_2}(\tau)$. The basic idea is to spectrally decompose the right-hand side of (\ref{eq:inhomogeneousLaplaceEquation}) and then invert the Laplacian. Since the right-hand side of (\ref{eq:inhomogeneousLaplaceEquation}) is a product of eigenfunctions of the Laplacian, its spectral decomposition is straightforwardly facilitated by the ``Clebsch-Gordan'' coefficients of \cite{zbMATH03796039, watson2008rankin, Benjamin:2021ygh}, which describe the triple product of eigenfunctions of the Laplacian.

The starting point is to notice that the right-hand side of (\ref{eq:inhomogeneousLaplaceEquation}) grows as follows at the cusp $y=\infty$, cf. (\ref{eq:eisensteinFourier}):
\begin{equation}\label{eq:eisensteinProduct}
  E_{s_1}(\tau)E_{s_2}(\tau) \sim y^{s_1+s_2} + \varphi(s_1)y^{1-s_1+s_2} + \varphi(s_2)y^{1-s_2+s_1} + \varphi(s_1)\varphi(s_2)y^{2-s_1-s_2}.
\end{equation}
This product is thus of ``slow growth'' at the cusp and hence its spectral decomposition is straightforwardly accommodated by Zagier's adaptation of the Rankin-Selberg method \cite{zbMATH03796039} as described in Section \ref{sec:largeN}. The idea is to subtract from (\ref{eq:eisensteinProduct}) a suitable linear combination of Eisenstein series in order to render it square-integrable. To proceed, we assume $s_2\ne s_1,1-s_1$, which is a special case that needs to be treated separately. We will also assume that $\rre s_1 > \rre s_2 + \half$ although we emphasize that if this is not the case it is trivial to modify the exercise that follows. In this case, the following function is square integrable
\begin{equation}
  H_{s_1,s_2}(\tau) \coloneqq E_{s_1}(\tau)E_{s_2}(\tau) - E_{s_1+s_2}(\tau) - \varphi(s_2)E_{1-s_2+s_1}(\tau) \in L^2(\cF).
\end{equation}
This object admits a straightforward spectral decomposition given by
\begin{equation}\label{eq:HSpecDecomp}
  H_{s_1,s_2}(\tau) = {1\over 4\pi i}\int_{\rre s = \half} ds\, K_{1-s,s_1,s_2}E_s(\tau) + \sum_{n=1}^\infty \widetilde a_1^{(n)}K^n_{s_1,s_2}\phi_n(\tau).
\end{equation}
The spectral coefficients are the Clebsch-Gordan coefficients of \cite{zbMATH03796039, watson2008rankin, Benjamin:2021ygh}, given by
\begin{align}
  K_{s,s_1,s_2} & \coloneqq R_s[E_{s_1}E_{s_2}] = {1\over \Lambda(s)\Lambda(s_1)\Lambda(s_2)}\prod_{\pm_{1,2}}\Lambda\left({s\pm_1 (s_1-\half)\pm_2 (s_2-\half)\over 2}\right)\label{eq:CGEis3}\\
  K^n_{s_1,s_2} & \coloneqq R_{s_1}[\nu_nE_{s_2}] = {1\over 4\Lambda(s_1)\Lambda(s_2)}\widetilde L^{(n)}\left(s_1+s_2-\half\right)\widetilde L^{(n)}(s_1-s_2+\half)\label{eq:CGEis2Cusp}
\end{align}
and
\begin{equation}
  \widetilde a_1^{(n)} = {1\over\sqrt{(\nu_n,\nu_n)}}
\end{equation}
is the first Fourier coefficient of $\phi_n(\tau)$, alternatively given in terms of the norm of the Maass form in the Hecke normalization (cf. (\ref{eq:L2vsHeckeNorms})). Here, $\widetilde L^{(n)}$ is a symmetrized $L$-function associated with the cusp form, given by a Dirichlet series in terms of the cusp form Fourier coefficients (cf. (\ref{eq:cuspFourier})) 
\begin{align}
  \widetilde L^{(n)}(s) &\coloneqq \pi^{-s} \Gamma\left(\half(s+it_n)\right)\Gamma\left(\half(s-it_n)\right)\sum_{k=1}^\infty {a_k^{(n)}\over k^s}\label{eq:cuspLFunction} \\
  &= L^{(n)}(1-s)\label{eq:cuspLFunctionalEquation}.
\end{align}
Although the sum in (\ref{eq:cuspLFunction}) only converges for $\rre s > 1$, the $L$-function inherits a meromorphic continuation to the entire complex $s$ plane from the Eisenstein series and thus satisfies the functional equation (\ref{eq:cuspLFunctionalEquation}). Note in particular the absence of a constant term in (\ref{eq:HSpecDecomp}); this is due to the fact that for $s_2 \ne s_1, 1-s_1$, $K_{s,s_1,s_2}$ is non-singular at $s=1$. In the main text we will encounter a rescaled solution of the inhomogeneous Laplace equation $\widetilde F_{r;s_1,s_2} = \chi(s_1)\chi(s_2)F_{r;s_1,s_2}$ with $\chi(s)$ given by (\ref{eq:chiDefinition}), so it will be convenient to work with rescaled versions of these spectral coefficients
\begin{align}
  \widetilde K_{s,s_1,s_2} &\coloneqq \chi(s_1)\chi(s_2)K_{s,s_1,s_2}\label{eq:KtildeEis3} \\
  \widetilde K^n_{s_1,s_2} & \coloneqq \widetilde a_1^{(n)}\chi(s_1)\chi(s_2)K^n_{s_1,s_2}\label{eq:KtildeEis2Cusp}.
\end{align}

The spectral decomposition of $F_{r;s_1,s_2}$ is then given by inverting the action of the Laplacian on $E_{s_1}E_{s_2}$:
\begin{equation}\label{eq:Frs1s2SpecDecomp}
\begin{aligned}
  F_{r;s_1,s_2}(\tau) =& \, (\Delta_\tau-\mu(r+1))^{-1}\left[H_{s_1,s_2}(\tau) + E_{s_1+s_2}(\tau) + \varphi(s_2)E_{1-s_2+s_1}(\tau)\right]\\
  =& \, {E_{s_1+s_2}(\tau)\over \mu(s_1+s_2)-\mu(r+1)} + {\varphi(s_2)E_{1-s_2+s_1}(\tau)\over\mu(1-s_2+s_1)-\mu(r+1)} + {1\over 4\pi i}\int_{\rre s=\half} ds {K_{1-s,s_1,s_2}\over \mu(s)-\mu(r+1)}E_s(\tau)\\
  & \, +\sum_{n=1}^\infty {\widetilde a_1^{(n)}K^n_{s_1,s_2}\over \mu_n-\mu(r+1)}\phi_n(\tau).
\end{aligned}
\end{equation}

We pause to make some comments on the structure of the spectral decomposition (\ref{eq:Frs1s2SpecDecomp}), which will also broadly apply to the result (\ref{eq:FrssSpecDecomp}) in the special case of $s_2 = s_1,1-s_1$ that we consider next.

In the main text, we see that these solutions to the inhomogeneous Laplace equation appear at integer powers of $1/N$ in the large $N$ expansion of the integrated four-point function $\cF_N(\t)$ of Section \ref{subsec:F4}. Thus the spectral coefficients in (\ref{eq:Frs1s2SpecDecomp}) furnish the large $N$ expansion of the spectral coefficients of the integrated correlator. For example, the Eisenstein series on the first line are associated with the perturbative part of the overlap, $\fp(s)$. In particular, they arise as certain residues of $N^{2-2g-s}{\pi s(1-s)\over \sin(\pi s)}\fp^{(g)}(1-s)E_s^*$ in the spectral decomposition in the large $N$ limit. In the application to $\cF_N(\t)$, $s_1$ and $s_2$ are odd half-integers, and the Eisenstein series in (\ref{eq:Frs1s2SpecDecomp}) arise as the residues of the perturbative part of the spectral integrand (at different genera) at certain integer values of $s$. On the other hand, the third term on the first line of (\ref{eq:Frs1s2SpecDecomp}) corresponds to a genuinely non-perturbative contribution to the spectral overlap $(\calo,E_s)$. In particular, we would have $\Lambda(s)\fnp^{(g,m)}(s) \propto {K_{1-s,s_1,s_2}\over \mu(s)-\mu(r+1)}$ for appropriate values of $g,m$ in the notation of Section \ref{subsec:fnp}. Indeed, this term cleanly exhibits $|s|\to\infty$ asymptotics that exemplify the factorial growth requirement (\ref{fnpgrowth}) of non-perturbative contributions to the spectral overlap
\begin{equation}
  K_{1-s,s_1,s_2} \sim \Gamma(|s|),\quad |s|\to\infty.
\end{equation} 
The combination $K_{1-s,s_1,s_2}\over\mu(s)-\mu(r+1)$ is also characterized by a finite number of spurious poles away from the critical contour in the complex $s$-plane. This is due to the fact that the solutions to the inhomogeneous Laplace equations are not themselves strictly square-integrable as they exhibit power-law growth at the cusp (as discussed in Section \ref{sec:VSC}, this is not inconsistent because they arise as coefficients in the large $N$ expansion of square-integrable observables rather than at finite $N$). Finally, the last term in \eqr{eq:Frs1s2SpecDecomp} represents the contribution of the Maass cusp forms to the spectral decomposition: $(\calo,\phi_n) \propto {K^n_{s_1,s_2}\over\mu_n-\mu(r+1)}$.

\subsection{$s_1 = s_2$}

We now consider the special case of $s_2 = s_1 = s$. The reflected case of $s_2 = 1-s_1$ can be treated essentially identically. In this case, the right-hand side of (\ref{eq:inhomogeneousLaplaceEquation}) has the following growth at the cusp
\begin{equation}\label{Escusp}
  E_s(\tau)^2 \sim y^{2s} + 2\varphi(s) y + \varphi(s)^2 y^{2-2s}.
\end{equation}
We proceed assuming that $\rre s > \half$, so that the following combination is square integrable
\begin{equation}\label{Hs}
  H_s(\tau) \coloneqq E_s(\tau)^2 - E_{2s}(\tau) - 2\varphi(s) \widehat E_1(\tau).
\end{equation}
where $\widehat E_1(\tau)$ is the regular part of the Eisenstein series at $s=1$,
\begin{equation}
  \widehat E_1(\tau) \coloneqq \lim_{s\to 1}\left(E_s(\tau) - {3\over \pi(s-1)}\right)
\end{equation}
It is not an eigenfunction of the Laplacian
\begin{equation}
  \Delta_\tau \widehat E_1(\tau) = -{3\over \pi}
\end{equation}
and has the following Fourier decomposition
\begin{equation}\label{eq:Ehat1Fourier}
  \widehat E_1(\tau) = y - {3\over \pi}\log y + {6\over \pi}\left(12\log A - \log 4\pi\right) + {12\over \pi}\sum_{k=1}^\infty {\cos(2\pi k x)\sigma_1(k)e^{-2\pi k y}\over k},
\end{equation}
where $A$ is Glaisher's constant, $\log A = {1\over 12} - \zeta'(-1)$. Its role in \eqr{Hs} is to subtract the linear term in \eqr{Escusp}. $H_s(\tau)$ then has the following spectral decomposition
\begin{equation}
  H_s(\tau) = K_s + {1\over 4\pi i} \int_{\rre s = \half} ds' \, K_{1-s',s,s}E_{s'}(\tau) + \sum_{n=1}^\infty \widetilde a_1^{(n)}K^n_{s,s}\nu_n(\tau).
\end{equation}
We note that there is now a constant term in the spectral decomposition, due to the fact that $K_{s',s,s}$ is now singular at $s'=1$ and that there are constants in the Fourier decomposition of $\widehat E_1(\tau)$. The constant term is given by
\begin{equation}
  K_s = \Res_{s'=1}K_{s',s,s} - 2\varphi(s)\omega = -{3\over \pi}\varphi'(s),
\end{equation}
where $\omega \coloneqq {6\over\pi}\left(12\log A-\log(4\pi)\right)$ is the constant in the Fourier decomposition of $\widehat E_1(\tau)$ (cf. (\ref{eq:Ehat1Fourier})).

Assembling the pieces, this leads to the following spectral decomposition for $F_{r;s,s}(\tau)$
\begin{equation}
\begin{aligned}\label{eq:FrssSpecDecomp}
  F_{r;s,s}(\tau) =& \, \left(\Delta_\tau-\mu(r+1)\right)^{-1}E_s(\tau)^2\\
  =& \, {E_{2s}(\tau)\over \mu(2s)-\mu(r+1)} + 2\varphi(s)\left({3\over \pi \mu(r+1)^2} - {1\over \mu(r+1)}\widehat E_1(\tau)\right) - {K_s\over \mu(r+1)}\\
  & \, + {1\over 4\pi i} \int_{\rre s'=\half} ds'{K_{1-s',s,s} \over \mu(s')-\mu(r+1)}E_{s'}(\tau) + \sum_{n=1}^\infty {\widetilde a_1^{(n)}K^n_{s,s}\over \mu_n -\mu(r+1)}\phi_n(\tau).
\end{aligned}
\end{equation}

\subsection{A worked example: $r = 3$, $s_1 = s_2 = {3\over 2}$}\label{appb3}
To elucidate the sense in which the perturbative and non-perturbative contributions (in $y^{-1}$) to $F_{r;s_1,s_2}$ are contained in its spectral decomposition, we will study the example of $r =3$, $s_1 = s_2 = {3\over 2}$ in detail. In particular, we will study the modular invariant
\begin{equation}
  \cE_{3;{3\over 2},{3\over 2}}(\tau) = 4\zeta(3)^2 F_{3;{3\over 2},{3\over 2}}(\tau)
\end{equation}
that appears as the coefficient of the $D^6R^4$ interaction in the low-energy expansion of the effective action of type IIB string theory and at order $1/N$ in the large $N$ expansion of the integrated $\langle 2222 \rangle$ correlator $\cF_N$ of Section \ref{subsec:F4}.

It will turn out that reproducing the perturbative expansion of the zero-instanton sector of this function from the spectral decomposition is an instructive exercise. In \cite{Green:2014yxa}, it is shown that the zero mode of $\mathcal{E}_{3;{3\over 2},{3\over 2}}$ is given by
\begin{equation}
\begin{aligned}\label{eq:calEZeromode}
  \left(\mathcal{E}_{3;{3\over 2},{3\over 2}}\right)_0(y) =& \, {2\over 3}\zeta(3)^2 y^3 + {4\over 3}\zeta(3)\zeta(2)y + 4\zeta(4)y^{-1} + {4\over 27}\zeta(6)y^{-3}  + 2\sum_{k=1}^\infty f^P_{k,-k}(y)
\end{aligned}
\end{equation}
where
\begin{equation}\label{eq:fpkk}
  f^P_{k,-k}(y) = {32\pi\over 315 k^3}\sigma_2(k)^2 \sum_{i,j=0}^1 q_3^{i,j}(\pi k y) K_i(2\pi k y) K_j(2\pi k y).
\end{equation}
The first four terms in (\ref{eq:calEZeromode}) are the perturbative contributions to the zero mode, while the sum over terms involving (\ref{eq:fpkk}) capture non-perturbative $(q\bar q)^k$ contributions.
Here the coefficients $q_3^{i,j}$ are given by \cite{Green:2014yxa}
\begin{equation}
\begin{aligned}
  q_3^{0,0}(z) &= z(-512 z^4 + 48 z^2 - 15) \\
  q_3^{0,1}(z) &= -128 z^4 - 12 z^2 - 15 = q_3^{1,0}(z) \\
  q_3^{1,1}(z) &= 512 z^5 + 16 z^3 + 33 z - 15 z^{-1}.
\end{aligned}
\end{equation}
We would like to understand how (\ref{eq:calEZeromode}) is reproduced by the spectral decomposition
\begin{equation}
\begin{aligned}
  \mathcal{E}_{3;{3\over 2},{3\over 2}}(\tau) =& \, 4\zeta(3)^2\bigg[ {1\over 6} E_3(\tau) + 2\varphi(3/2)\left({3\over 144\pi}+{1\over 12}\widehat E_1(\tau)\right) + {1\over 12}K_{3/2}\\
  & \, + {1\over 4\pi i}\int_{\rre s = \half}ds \, {K_{1-s,{3\over 2},{3\over 2}}\over \mu(s) + 12}E_s(\tau) + \sum_{n=1}^\infty {\widetilde a_1^{(n)}K^n_{{3\over 2},{3\over 2}}\over \mu_n+12}\phi_n(\tau)\bigg].
\end{aligned}
\end{equation}
In particular, the zero mode is given by
\begin{equation}
\begin{aligned}\label{eq:calEZeromodeSpectral}
  \left(\mathcal{E}_{3;{3\over 2},{3\over 2}}\right)_0(y) =& \, 4\zeta(3)^2\bigg[ {1\over 6}\left(y^3 + \varphi(3)y^{-2}\right) + 2\varphi(3/2)\left({3\over 144\pi}+ {1\over 12}\left(y - {3\over \pi}\log y + \omega \right)\right) + {1\over 12}K_{3/2}\\
  & \, + {1\over 4\pi i} \int_{\rre s = \half}ds\, {K_{1-s,{3\over 2},{3\over 2}}\over \mu(s)+12}\left(y^s + \varphi(s)y^{1-s}\right)\bigg].
\end{aligned}
\end{equation}

It is not immediately clear how (\ref{eq:calEZeromodeSpectral}) is consistent with (\ref{eq:calEZeromode}). For instance, the latter has terms of order $y^{-1}$ and $y^{-3}$ that are not obviously present in the former, while the former has terms of order $\log y$, $y^{-2}$ and $y^0$ that must be cancelled in order to reproduce the latter. The resolution is of course contained in the spectral integral in the second line of (\ref{eq:calEZeromodeSpectral}) --- the spectral integrand has poles in the complex $s$ plane with residues that precisely resolve these perturbative discrepancies upon deformation of the spectral contour so that all such poles lie to one side of the contour. In particular, the integrand ${K_{1-s,{3\over 2},{3\over 2}}\over \mu(s)+12}\varphi(s)y^{1-s}$ has poles at $s = 1,2,3,4$ in the $s$ half-plane to the right of the critical contour (along with reflected poles in the left half-plane). We now imagine deforming the contour to $\rre s >4$. In doing so, we pick up residues at the intervening poles. The residue at $s = 1$ precisely cancels the spurious constant terms and $\log y$ term. The residue at $s = 3$ exactly cancels the spurious term of order $y^{-2}$. And finally, the residues at $s = 2$ and $s =4$ exactly reproduce the missing terms of order $y^{-1}$ and $y^{-3}$, respectively. Thus the deformed spectral integral precisely accounts for the non-perturbative (instanton-anti-instanton) contributions to the zero-instanton sector of $\cE$. In particular, we have
\begin{equation}
  {4\zeta(3)^2\over 2\pi i} \int_{\rre s> 4} ds\, {K_{1-s,{3\over 2},{3\over 2}}\over \mu(s) + 12}\varphi(s)y^{1-s} \stackrel{!}{=} 2\sum_{k=1}^\infty f^P_{k,-k}(y),
\end{equation}
which we have checked numerically.

\end{appendix}

\bibliographystyle{JHEP}
\bibliography{spec_4d_draft_bib}

\providecommand{\href}[2]{#2}\begingroup\raggedright\begin{thebibliography}{100}

\bibitem{Dorigoni:2021guq}
D.~Dorigoni, M.~B. Green, and C.~Wen, {\it {Exact properties of an integrated
  correlator in $ \mathcal{N} $ = 4 SU(N) SYM}},  {\em JHEP} {\bf 05} (2021)
  089, [\href{http://arxiv.org/abs/2102.09537}{{\tt arXiv:2102.09537}}].

\bibitem{Montonen:1977sn}
C.~Montonen and D.~I. Olive, {\it {Magnetic Monopoles as Gauge Particles?}},
  {\em Phys. Lett. B} {\bf 72} (1977) 117--120.

\bibitem{Witten:1978mh}
E.~Witten and D.~I. Olive, {\it {Supersymmetry Algebras That Include
  Topological Charges}},  {\em Phys. Lett. B} {\bf 78} (1978) 97--101.

\bibitem{Osborn:1979tq}
H.~Osborn, {\it {Topological Charges for N=4 Supersymmetric Gauge Theories and
  Monopoles of Spin 1}},  {\em Phys. Lett. B} {\bf 83} (1979) 321--326.

\bibitem{Argyres:2006qr}
P.~C. Argyres, A.~Kapustin, and N.~Seiberg, {\it {On S-duality for
  non-simply-laced gauge groups}},  {\em JHEP} {\bf 06} (2006) 043,
  [\href{http://arxiv.org/abs/hep-th/0603048}{{\tt hep-th/0603048}}].

\bibitem{Aharony:2013hda}
O.~Aharony, N.~Seiberg, and Y.~Tachikawa, {\it {Reading between the lines of
  four-dimensional gauge theories}},  {\em JHEP} {\bf 08} (2013) 115,
  [\href{http://arxiv.org/abs/1305.0318}{{\tt arXiv:1305.0318}}].

\bibitem{Iwaniec2002SpectralMO}
H.~Iwaniec, {\it Spectral methods of automorphic forms},  2002.

\bibitem{Bianchi:1999ge}
M.~Bianchi, S.~Kovacs, G.~Rossi, and Y.~S. Stanev, {\it {On the logarithmic
  behavior in N=4 SYM theory}},  {\em JHEP} {\bf 08} (1999) 020,
  [\href{http://arxiv.org/abs/hep-th/9906188}{{\tt hep-th/9906188}}].

\bibitem{Alday:2016tll}
L.~F. Alday and G.~P. Korchemsky, {\it {Revisiting instanton corrections to the
  Konishi multiplet}},  {\em JHEP} {\bf 12} (2016) 005,
  [\href{http://arxiv.org/abs/1605.06346}{{\tt arXiv:1605.06346}}].

\bibitem{Alday:2016jeo}
L.~F. Alday and G.~P. Korchemsky, {\it {Instanton corrections to twist-two
  operators}},  {\em JHEP} {\bf 06} (2017) 008,
  [\href{http://arxiv.org/abs/1609.08164}{{\tt arXiv:1609.08164}}].

\bibitem{Alday:2016bkq}
L.~F. Alday and G.~P. Korchemsky, {\it {On instanton effects in the operator
  product expansion}},  {\em JHEP} {\bf 05} (2017) 049,
  [\href{http://arxiv.org/abs/1610.01425}{{\tt arXiv:1610.01425}}].

\bibitem{Sen:2013oza}
A.~Sen, {\it {S-duality Improved Superstring Perturbation Theory}},  {\em JHEP}
  {\bf 11} (2013) 029, [\href{http://arxiv.org/abs/1304.0458}{{\tt
  arXiv:1304.0458}}].

\bibitem{Beem:2013hha}
C.~Beem, L.~Rastelli, A.~Sen, and B.~C. van Rees, {\it {Resummation and
  S-duality in N=4 SYM}},  {\em JHEP} {\bf 04} (2014) 122,
  [\href{http://arxiv.org/abs/1306.3228}{{\tt arXiv:1306.3228}}].

\bibitem{Alday:2013bha}
L.~F. Alday and A.~Bissi, {\it {Modular interpolating functions for N=4 SYM}},
  {\em JHEP} {\bf 07} (2014) 007, [\href{http://arxiv.org/abs/1311.3215}{{\tt
  arXiv:1311.3215}}].

\bibitem{Chowdhury:2016hny}
A.~Chowdhury, M.~Honda, and S.~Thakur, {\it {S-duality invariant perturbation
  theory improved by holography}},  {\em JHEP} {\bf 04} (2017) 137,
  [\href{http://arxiv.org/abs/1607.01716}{{\tt arXiv:1607.01716}}].

\bibitem{Binder:2019jwn}
D.~J. Binder, S.~M. Chester, S.~S. Pufu, and Y.~Wang, {\it {$ \mathcal{N} $ = 4
  Super-Yang-Mills correlators at strong coupling from string theory and
  localization}},  {\em JHEP} {\bf 12} (2019) 119,
  [\href{http://arxiv.org/abs/1902.06263}{{\tt arXiv:1902.06263}}].

\bibitem{Dorigoni:2021bvj}
D.~Dorigoni, M.~B. Green, and C.~Wen, {\it {Novel Representation of an
  Integrated Correlator in $\mathcal N$ = 4 Supersymmetric Yang-Mills Theory}},
   {\em Phys. Rev. Lett.} {\bf 126} (2021), no.~16 161601,
  [\href{http://arxiv.org/abs/2102.08305}{{\tt arXiv:2102.08305}}].

\bibitem{sarnak}
P.~Sarnak, ``{Arithmetic Quantum Chaos}.''
  {http://web.math.princeton.edu/sarnak/Arithmetic\%20Quantum\%20Chaos.pdf},
  May, 1993.

\bibitem{kimsarn}
H.~H. Kim, D.~Ramakrishnan, and P.~Sarnak, {\it Functoriality for the exterior
  square of gl4 and the symmetric fourth of gl2},  {\em Journal of the American
  Mathematical Society} {\bf 16} (2003), no.~1 139--183.

\bibitem{Azeyanagi:2013fla}
T.~Azeyanagi, M.~Hanada, M.~Honda, Y.~Matsuo, and S.~Shiba, {\it {A new look at
  instantons and large-N limit}},  {\em JHEP} {\bf 05} (2014) 008,
  [\href{http://arxiv.org/abs/1307.0809}{{\tt arXiv:1307.0809}}].

\bibitem{Chester:2019jas}
S.~M. Chester, M.~B. Green, S.~S. Pufu, Y.~Wang, and C.~Wen, {\it {Modular
  invariance in superstring theory from $ \mathcal{N} $ = 4 super-Yang-Mills}},
   {\em JHEP} {\bf 11} (2020) 016, [\href{http://arxiv.org/abs/1912.13365}{{\tt
  arXiv:1912.13365}}].

\bibitem{Chester:2020vyz}
S.~M. Chester, M.~B. Green, S.~S. Pufu, Y.~Wang, and C.~Wen, {\it {New modular
  invariants in $ \mathcal{N} $ = 4 Super-Yang-Mills theory}},  {\em JHEP} {\bf
  04} (2021) 212, [\href{http://arxiv.org/abs/2008.02713}{{\tt
  arXiv:2008.02713}}].

\bibitem{Alday:2021vfb}
L.~F. Alday, S.~M. Chester, and T.~Hansen, {\it {Modular invariant holographic
  correlators for $ \mathcal{N} $ = 4 SYM with general gauge group}},  {\em
  JHEP} {\bf 12} (2021) 159, [\href{http://arxiv.org/abs/2110.13106}{{\tt
  arXiv:2110.13106}}].

\bibitem{Chester:2020dja}
S.~M. Chester and S.~S. Pufu, {\it {Far beyond the planar limit in
  strongly-coupled $ \mathcal{N} $ = 4 SYM}},  {\em JHEP} {\bf 01} (2021) 103,
  [\href{http://arxiv.org/abs/2003.08412}{{\tt arXiv:2003.08412}}].

\bibitem{Saad:2019lba}
P.~Saad, S.~H. Shenker, and D.~Stanford, {\it {JT gravity as a matrix
  integral}},  \href{http://arxiv.org/abs/1903.11115}{{\tt arXiv:1903.11115}}.

\bibitem{Stanford:2019vob}
D.~Stanford and E.~Witten, {\it {JT gravity and the ensembles of random matrix
  theory}},  {\em Adv. Theor. Math. Phys.} {\bf 24} (2020), no.~6 1475--1680,
  [\href{http://arxiv.org/abs/1907.03363}{{\tt arXiv:1907.03363}}].

\bibitem{Afkhami-Jeddi:2020ezh}
N.~Afkhami-Jeddi, H.~Cohn, T.~Hartman, and A.~Tajdini, {\it {Free partition
  functions and an averaged holographic duality}},  {\em JHEP} {\bf 01} (2021)
  130, [\href{http://arxiv.org/abs/2006.04839}{{\tt arXiv:2006.04839}}].

\bibitem{Maloney:2020nni}
A.~Maloney and E.~Witten, {\it {Averaging over Narain moduli space}},  {\em
  JHEP} {\bf 10} (2020) 187, [\href{http://arxiv.org/abs/2006.04855}{{\tt
  arXiv:2006.04855}}].

\bibitem{Saad:2021rcu}
P.~Saad, S.~H. Shenker, D.~Stanford, and S.~Yao, {\it {Wormholes without
  averaging}},  \href{http://arxiv.org/abs/2103.16754}{{\tt arXiv:2103.16754}}.

\bibitem{Terras_2013}
A.~Terras, {\em Harmonic Analysis on Symmetric Spaces{\textemdash}Euclidean
  Space, the Sphere, and the Poincar{\'{e}} Upper Half-Plane}.
\newblock Springer New York, 2013.

\bibitem{Sarnak_1987}
P.~Sarnak, {\it {Statistical Properties of Eigenvalues of the Hecke
  Operators}},  in {\em {Analytic Number Theory and Diophantine Problems}},
  pp.~321--331.
\newblock Birkh{\"a}user Boston, 1987.

\bibitem{hejrack}
D.~A. Hejhal and B.~N. Rackner, {\it On the topography of maass waveforms for
  psl(2, z)},  {\em Experimental Mathematics} {\bf 1} (1992), no.~4 275--305,
  [\href{http://arxiv.org/abs/https://doi.org/10.1080/10586458.1992.10504562}{{\tt
  https://doi.org/10.1080/10586458.1992.10504562}}].

\bibitem{sarnakk}
P.~Sarnak, {\it Spectra of hyperbolic surfaces},  {\em Contents} {\bf 40} (10,
  2003).

\bibitem{lmfdb}
{The LMFDB Collaboration}, ``The {L}-functions and modular forms database.''
  {http://www.lmfdb.org}, 2022.
\newblock [Online; accessed 9 January 2022].

\bibitem{1983}
D.~A. Hejhal, {\em The Selberg Trace Formula for {PSL}(2,R)}.
\newblock Springer Berlin Heidelberg, 1983.

\bibitem{MR3287209}
A.~Selberg, {\em Collected papers. {I}}.
\newblock Springer Collected Works in Mathematics. Springer, Heidelberg, 2014.
\newblock With a foreword by K. Chandrasekharan, Reprint of the 1989 edition
  [MR1117906].

\bibitem{rankin_1939}
R.~A. Rankin, {\it {Contributions to the theory of Ramanujan's function
  $\tau(n)$ and similar arithmetical functions: II. The order of the Fourier
  coefficients of integral modular forms}},  {\em Mathematical Proceedings of
  the Cambridge Philosophical Society} {\bf 35} (1939), no.~3 357--372.

\bibitem{selberg1940bemerkungen}
A.~Selberg, {\em Bemerkungen {\"u}ber eine Dirichletsche Reihe, die mit der
  Theorie der Modulformen nahe verbunden ist}.
\newblock Archiv for Mathematik og Naturvidenskab. Cammermeyer, 1940.

\bibitem{Aharony:2002hx}
O.~Aharony, B.~Kol, and S.~Yankielowicz, {\it {On exactly marginal deformations
  of N=4 SYM and type IIB supergravity on AdS(5) x S**5}},  {\em JHEP} {\bf 06}
  (2002) 039, [\href{http://arxiv.org/abs/hep-th/0205090}{{\tt
  hep-th/0205090}}].

\bibitem{Whittaker}
R.~Szmytkowski and S.~Bielski, {\it {An orthogonality relation for the
  Whittaker functions of the second kind of imaginary order}},  {\em Integral
  Transforms and Special Functions} {\bf 21} (2010), no.~10 739--744,
  [\href{http://arxiv.org/abs/0910.1492}{{\tt arXiv:0910.1492}}].

\bibitem{cusp82}
J.-M. Deshouillers, H.~Iwaniec, R.~S. Phillips, and P.~Sarnak, {\it Maass cusp
  forms},  {\em Proceedings of the National Academy of Sciences} {\bf 82}
  (1985), no.~11 3533--3534.

\bibitem{hejh}
D.~A. Hejhal and B.~N. Rackner, {\it {On the topography of Maass waveforms for
  {${\rm PSL}(2,{\bf Z})$}}},  {\em Experimental Mathematics} {\bf 1} (1992),
  no.~4 275 -- 305.

\bibitem{sque}
P.~Sarnak, {\it Recent progress on the quantum unique ergodicity conjecture},
  {\em Bulletin (New Series) of the American Mathematical Society} {\bf 48}
  (05, 2011).

\bibitem{Benjamin:2021ygh}
N.~Benjamin, S.~Collier, A.~L. Fitzpatrick, A.~Maloney, and E.~Perlmutter, {\it
  {Harmonic analysis of 2d CFT partition functions}},
  \href{http://arxiv.org/abs/2107.10744}{{\tt arXiv:2107.10744}}.

\bibitem{Pestun:2007rz}
V.~Pestun, {\it {Localization of gauge theory on a four-sphere and
  supersymmetric Wilson loops}},  {\em Commun. Math. Phys.} {\bf 313} (2012)
  71--129, [\href{http://arxiv.org/abs/0712.2824}{{\tt arXiv:0712.2824}}].

\bibitem{Russo:2013kea}
J.~G. Russo and K.~Zarembo, {\it {Massive N=2 Gauge Theories at Large N}},
  {\em JHEP} {\bf 11} (2013) 130, [\href{http://arxiv.org/abs/1309.1004}{{\tt
  arXiv:1309.1004}}].

\bibitem{FLAJOLET19953}
P.~Flajolet, X.~Gourdon, and P.~Dumas, {\it Mellin transforms and asymptotics:
  Harmonic sums},  {\em Theoretical Computer Science} {\bf 144} (1995), no.~1
  3--58.

\bibitem{Kozcaz:2016wvy}
C.~Kozcaz, T.~Sulejmanpasic, Y.~Tanizaki, and M.~\"Unsal, {\it {Cheshire Cat
  resurgence, Self-resurgence and Quasi-Exact Solvable Systems}},  {\em Commun.
  Math. Phys.} {\bf 364} (2018), no.~3 835--878,
  [\href{http://arxiv.org/abs/1609.06198}{{\tt arXiv:1609.06198}}].

\bibitem{Dorigoni:2017smz}
D.~Dorigoni and P.~Glass, {\it {The grin of Cheshire cat resurgence from
  supersymmetric localization}},  {\em SciPost Phys.} {\bf 4} (2018), no.~2
  012, [\href{http://arxiv.org/abs/1711.04802}{{\tt arXiv:1711.04802}}].

\bibitem{Dorigoni:2019kux}
D.~Dorigoni and P.~Glass, {\it {Picard-Lefschetz decomposition and Cheshire Cat
  resurgence in 3D $ \mathcal{N} $ = 2 field theories}},  {\em JHEP} {\bf 12}
  (2019) 085, [\href{http://arxiv.org/abs/1909.05262}{{\tt arXiv:1909.05262}}].

\bibitem{Grassi:2014cla}
A.~Grassi, M.~Marino, and S.~Zakany, {\it {Resumming the string perturbation
  series}},  {\em JHEP} {\bf 05} (2015) 038,
  [\href{http://arxiv.org/abs/1405.4214}{{\tt arXiv:1405.4214}}].

\bibitem{Green:2014yxa}
M.~B. Green, S.~D. Miller, and P.~Vanhove, {\it {$SL(2, \mathbb{Z})$-invariance
  and D-instanton contributions to the $D^6 R^4$ interaction}},  {\em Commun.
  Num. Theor. Phys.} {\bf 09} (2015) 307--344,
  [\href{http://arxiv.org/abs/1404.2192}{{\tt arXiv:1404.2192}}].

\bibitem{Arutyunov:2016etw}
G.~Arutyunov, D.~Dorigoni, and S.~Savin, {\it {Resurgence of the dressing phase
  for AdS$_{5} ×$ S$^{5}$}},  {\em JHEP} {\bf 01} (2017) 055,
  [\href{http://arxiv.org/abs/1608.03797}{{\tt arXiv:1608.03797}}].

\bibitem{zbMATH03796039}
D.~{Zagier}, {\it {The Rankin-Selberg method for automorphic functions which
  are not of rapid decay}},  {\em {J. Fac. Sci., Univ. Tokyo, Sect. I A}} {\bf
  28} (1981) 415--437.

\bibitem{Honda:2017qdb}
M.~Honda, {\it {Supersymmetric solutions and Borel singularities for N=2
  supersymmetric Chern-Simons theories}},  {\em Phys. Rev. Lett.} {\bf 121}
  (2018), no.~2 021601, [\href{http://arxiv.org/abs/1710.05010}{{\tt
  arXiv:1710.05010}}].

\bibitem{Aniceto:2014hoa}
I.~Aniceto, J.~G. Russo, and R.~Schiappa, {\it {Resurgent Analysis of
  Localizable Observables in Supersymmetric Gauge Theories}},  {\em JHEP} {\bf
  03} (2015) 172, [\href{http://arxiv.org/abs/1410.5834}{{\tt
  arXiv:1410.5834}}].

\bibitem{Dunne:2012ae}
G.~V. Dunne and M.~Unsal, {\it {Resurgence and Trans-series in Quantum Field
  Theory: The CP(N-1) Model}},  {\em JHEP} {\bf 11} (2012) 170,
  [\href{http://arxiv.org/abs/1210.2423}{{\tt arXiv:1210.2423}}].

\bibitem{Maloney:2007ud}
A.~Maloney and E.~Witten, {\it {Quantum Gravity Partition Functions in Three
  Dimensions}},  {\em JHEP} {\bf 02} (2010) 029,
  [\href{http://arxiv.org/abs/0712.0155}{{\tt arXiv:0712.0155}}].

\bibitem{Keller:2014xba}
C.~A. Keller and A.~Maloney, {\it {Poincare Series, 3D Gravity and CFT
  Spectroscopy}},  {\em JHEP} {\bf 02} (2015) 080,
  [\href{http://arxiv.org/abs/1407.6008}{{\tt arXiv:1407.6008}}].

\bibitem{Dorigoni:2014hea}
D.~Dorigoni, {\it {An Introduction to Resurgence, Trans-Series and Alien
  Calculus}},  {\em Annals Phys.} {\bf 409} (2019) 167914,
  [\href{http://arxiv.org/abs/1411.3585}{{\tt arXiv:1411.3585}}].

\bibitem{Gerchkovitz:2016gxx}
E.~Gerchkovitz, J.~Gomis, N.~Ishtiaque, A.~Karasik, Z.~Komargodski, and S.~S.
  Pufu, {\it {Correlation Functions of Coulomb Branch Operators}},  {\em JHEP}
  {\bf 01} (2017) 103, [\href{http://arxiv.org/abs/1602.05971}{{\tt
  arXiv:1602.05971}}].

\bibitem{1993MaCom..61..245H}
D.~A. {Hejhal} and S.~{Arno}, {\it {On Fourier coefficients of Maass waveforms
  for PSL(2, Z)}},  {\em Mathematics of Computation} {\bf 61} (1993), no.~203
  245--267.

\bibitem{Steil:1994ue}
G.~Steil, {\it {Eigenvalues of the Laplacian and of the Hecke operators for
  PSL(2,Z)}},  1994.

\bibitem{Wang:2015jna}
Y.~Wang and X.~Yin, {\it {Constraining Higher Derivative Supergravity with
  Scattering Amplitudes}},  {\em Phys. Rev. D} {\bf 92} (2015), no.~4 041701,
  [\href{http://arxiv.org/abs/1502.03810}{{\tt arXiv:1502.03810}}].

\bibitem{Wang:2015aua}
Y.~Wang and X.~Yin, {\it {Supervertices and Non-renormalization Conditions in
  Maximal Supergravity Theories}},  \href{http://arxiv.org/abs/1505.05861}{{\tt
  arXiv:1505.05861}}.

\bibitem{Aharony:2016dwx}
O.~Aharony, L.~F. Alday, A.~Bissi, and E.~Perlmutter, {\it {Loops in AdS from
  Conformal Field Theory}},  {\em JHEP} {\bf 07} (2017) 036,
  [\href{http://arxiv.org/abs/1612.03891}{{\tt arXiv:1612.03891}}].

\bibitem{cmp/1103901558}
E.~Br\'ezin, C.~Itzykson, G.~Parisi, and J.~B. Zuber, {\it {Planar diagrams}},
  {\em Communications in Mathematical Physics} {\bf 59} (1978), no.~1 35 -- 51.

\bibitem{Beisert:2006ez}
N.~Beisert, B.~Eden, and M.~Staudacher, {\it {Transcendentality and Crossing}},
   {\em J. Stat. Mech.} {\bf 0701} (2007) P01021,
  [\href{http://arxiv.org/abs/hep-th/0610251}{{\tt hep-th/0610251}}].

\bibitem{Alday:2007mf}
L.~F. Alday and J.~M. Maldacena, {\it {Comments on operators with large spin}},
   {\em JHEP} {\bf 11} (2007) 019, [\href{http://arxiv.org/abs/0708.0672}{{\tt
  arXiv:0708.0672}}].

\bibitem{Marino:2020ggm}
M.~Marino and T.~Reis, {\it {Three roads to the energy gap}},
  \href{http://arxiv.org/abs/2010.16174}{{\tt arXiv:2010.16174}}.

\bibitem{Gross:1988ib}
D.~J. Gross and V.~Periwal, {\it {String Perturbation Theory Diverges}},  {\em
  Phys. Rev. Lett.} {\bf 60} (1988) 2105.

\bibitem{Shenker:1990uf}
S.~H. Shenker, {\it {The Strength of nonperturbative effects in string
  theory}},  in {\em {Cargese Study Institute: Random Surfaces, Quantum Gravity
  and Strings}}, pp.~809--819, 8, 1990.

\bibitem{Drukker:2011zy}
N.~Drukker, M.~Marino, and P.~Putrov, {\it {Nonperturbative aspects of ABJM
  theory}},  {\em JHEP} {\bf 11} (2011) 141,
  [\href{http://arxiv.org/abs/1103.4844}{{\tt arXiv:1103.4844}}].

\bibitem{Marino:2012zq}
M.~Mari\~no, {\it {Lectures on non-perturbative effects in large $N$ gauge
  theories, matrix models and strings}},  {\em Fortsch. Phys.} {\bf 62} (2014)
  455--540, [\href{http://arxiv.org/abs/1206.6272}{{\tt arXiv:1206.6272}}].

\bibitem{Green:2008bf}
M.~B. Green, J.~G. Russo, and P.~Vanhove, {\it {Modular properties of two-loop
  maximal supergravity and connections with string theory}},  {\em JHEP} {\bf
  07} (2008) 126, [\href{http://arxiv.org/abs/0807.0389}{{\tt
  arXiv:0807.0389}}].

\bibitem{Alday:2018pdi}
L.~F. Alday, A.~Bissi, and E.~Perlmutter, {\it {Genus-One String Amplitudes
  from Conformal Field Theory}},  {\em JHEP} {\bf 06} (2019) 010,
  [\href{http://arxiv.org/abs/1809.10670}{{\tt arXiv:1809.10670}}].

\bibitem{Coronado:2018ypq}
F.~Coronado, {\it {Perturbative four-point functions in planar $ \mathcal{N}=4
  $ SYM from hexagonalization}},  {\em JHEP} {\bf 01} (2019) 056,
  [\href{http://arxiv.org/abs/1811.00467}{{\tt arXiv:1811.00467}}].

\bibitem{Coronado:2018ypq2}
F.~Coronado, {\it {Bootstrapping the Simplest Correlator in Planar $\mathcal N
  = 4$ Supersymmetric Yang-Mills Theory to All Loops}},  {\em Phys. Rev. Lett.}
  {\bf 124} (2020), no.~17 171601, [\href{http://arxiv.org/abs/1811.03282}{{\tt
  arXiv:1811.03282}}].

\bibitem{Kostov:2019stn}
I.~Kostov, V.~B. Petkova, and D.~Serban, {\it {Determinant Formula for the
  Octagon Form Factor in $N$=4 Supersymmetric Yang-Mills Theory}},  {\em Phys.
  Rev. Lett.} {\bf 122} (2019), no.~23 231601,
  [\href{http://arxiv.org/abs/1903.05038}{{\tt arXiv:1903.05038}}].

\bibitem{Kostov:2019auq}
I.~Kostov, V.~B. Petkova, and D.~Serban, {\it {The Octagon as a Determinant}},
  {\em JHEP} {\bf 11} (2019) 178, [\href{http://arxiv.org/abs/1905.11467}{{\tt
  arXiv:1905.11467}}].

\bibitem{Bargheer:2019kxb}
T.~Bargheer, F.~Coronado, and P.~Vieira, {\it {Octagons I: Combinatorics and
  Non-Planar Resummations}},  {\em JHEP} {\bf 08} (2019) 162,
  [\href{http://arxiv.org/abs/1904.00965}{{\tt arXiv:1904.00965}}].

\bibitem{Belitsky:2019fan}
A.~V. Belitsky and G.~P. Korchemsky, {\it {Exact null octagon}},  {\em JHEP}
  {\bf 05} (2020) 070, [\href{http://arxiv.org/abs/1907.13131}{{\tt
  arXiv:1907.13131}}].

\bibitem{Bargheer:2019exp}
T.~Bargheer, F.~Coronado, and P.~Vieira, {\it {Octagons II: Strong Coupling}},
  \href{http://arxiv.org/abs/1909.04077}{{\tt arXiv:1909.04077}}.

\bibitem{Belitsky:2020qrm}
A.~V. Belitsky and G.~P. Korchemsky, {\it {Octagon at finite coupling}},  {\em
  JHEP} {\bf 07} (2020) 219, [\href{http://arxiv.org/abs/2003.01121}{{\tt
  arXiv:2003.01121}}].

\bibitem{Beem:2016wfs}
C.~Beem, L.~Rastelli, and B.~C. van Rees, {\it {More ${\mathcal N}=4$
  superconformal bootstrap}},  {\em Phys. Rev. D} {\bf 96} (2017), no.~4
  046014, [\href{http://arxiv.org/abs/1612.02363}{{\tt arXiv:1612.02363}}].

\bibitem{Cordova:2016emh}
C.~Cordova, T.~T. Dumitrescu, and K.~Intriligator, {\it {Multiplets of
  Superconformal Symmetry in Diverse Dimensions}},  {\em JHEP} {\bf 03} (2019)
  163, [\href{http://arxiv.org/abs/1612.00809}{{\tt arXiv:1612.00809}}].

\bibitem{Beem:2013qxa}
C.~Beem, L.~Rastelli, and B.~C. van Rees, {\it {The $\mathcal N=4$
  Superconformal Bootstrap}},  {\em Phys. Rev. Lett.} {\bf 111} (2013) 071601,
  [\href{http://arxiv.org/abs/1304.1803}{{\tt arXiv:1304.1803}}].

\bibitem{Beem:2013sza}
C.~Beem, M.~Lemos, P.~Liendo, W.~Peelaers, L.~Rastelli, and B.~C. van Rees,
  {\it {Infinite Chiral Symmetry in Four Dimensions}},  {\em Commun. Math.
  Phys.} {\bf 336} (2015), no.~3 1359--1433,
  [\href{http://arxiv.org/abs/1312.5344}{{\tt arXiv:1312.5344}}].

\bibitem{Alday:2013opa}
L.~F. Alday and A.~Bissi, {\it {The superconformal bootstrap for structure
  constants}},  {\em JHEP} {\bf 09} (2014) 144,
  [\href{http://arxiv.org/abs/1310.3757}{{\tt arXiv:1310.3757}}].

\bibitem{Bissi:2020jve}
A.~Bissi, A.~Manenti, and A.~Vichi, {\it {Bootstrapping mixed correlators in $
  \mathcal{N} $ = 4 super Yang-Mills}},  {\em JHEP} {\bf 05} (2021) 111,
  [\href{http://arxiv.org/abs/2010.15126}{{\tt arXiv:2010.15126}}].

\bibitem{Chester:2021aun}
S.~M. Chester, R.~Dempsey, and S.~S. Pufu, {\it {Bootstrapping $\mathcal{N}=4$
  super-Yang-Mills on the conformal manifold}},
  \href{http://arxiv.org/abs/2111.07989}{{\tt arXiv:2111.07989}}.

\bibitem{Korchemsky:2015cyx}
G.~P. Korchemsky, {\it {On level crossing in conformal field theories}},  {\em
  JHEP} {\bf 03} (2016) 212, [\href{http://arxiv.org/abs/1512.05362}{{\tt
  arXiv:1512.05362}}].

\bibitem{DHoker:1999kzh}
E.~D'Hoker, D.~Z. Freedman, S.~D. Mathur, A.~Matusis, and L.~Rastelli, {\it
  {Graviton exchange and complete four point functions in the AdS / CFT
  correspondence}},  {\em Nucl. Phys. B} {\bf 562} (1999) 353--394,
  [\href{http://arxiv.org/abs/hep-th/9903196}{{\tt hep-th/9903196}}].

\bibitem{Heemskerk:2009pn}
I.~Heemskerk, J.~Penedones, J.~Polchinski, and J.~Sully, {\it {Holography from
  Conformal Field Theory}},  {\em JHEP} {\bf 10} (2009) 079,
  [\href{http://arxiv.org/abs/0907.0151}{{\tt arXiv:0907.0151}}].

\bibitem{Rastelli:2016nze}
L.~Rastelli and X.~Zhou, {\it {Mellin amplitudes for $AdS_5\times S^5$}},  {\em
  Phys. Rev. Lett.} {\bf 118} (2017), no.~9 091602,
  [\href{http://arxiv.org/abs/1608.06624}{{\tt arXiv:1608.06624}}].

\bibitem{Rastelli:2017udc}
L.~Rastelli and X.~Zhou, {\it {How to Succeed at Holographic Correlators
  Without Really Trying}},  {\em JHEP} {\bf 04} (2018) 014,
  [\href{http://arxiv.org/abs/1710.05923}{{\tt arXiv:1710.05923}}].

\bibitem{Goncalves:2014ffa}
V.~Goncalves, {\it {Four point function of $\mathcal{N}=4$ stress-tensor
  multiplet at strong coupling}},  {\em JHEP} {\bf 04} (2015) 150,
  [\href{http://arxiv.org/abs/1411.1675}{{\tt arXiv:1411.1675}}].

\bibitem{Aprile:2017xsp}
F.~Aprile, J.~M. Drummond, P.~Heslop, and H.~Paul, {\it {Unmixing
  Supergravity}},  {\em JHEP} {\bf 02} (2018) 133,
  [\href{http://arxiv.org/abs/1706.08456}{{\tt arXiv:1706.08456}}].

\bibitem{Aprile:2018efk}
F.~Aprile, J.~Drummond, P.~Heslop, and H.~Paul, {\it {Double-trace spectrum of
  $N=4$ supersymmetric Yang-Mills theory at strong coupling}},  {\em Phys. Rev.
  D} {\bf 98} (2018), no.~12 126008,
  [\href{http://arxiv.org/abs/1802.06889}{{\tt arXiv:1802.06889}}].

\bibitem{Caron-Huot:2018kta}
S.~Caron-Huot and A.-K. Trinh, {\it {All tree-level correlators in
  AdS$_{5}\times$ S$_{5}$ supergravity: hidden ten-dimensional conformal
  symmetry}},  {\em JHEP} {\bf 01} (2019) 196,
  [\href{http://arxiv.org/abs/1809.09173}{{\tt arXiv:1809.09173}}].

\bibitem{Drummond:2019odu}
J.~M. Drummond, D.~Nandan, H.~Paul, and K.~S. Rigatos, {\it {String corrections
  to AdS amplitudes and the double-trace spectrum of $ \mathcal{N} $ = 4 SYM}},
   {\em JHEP} {\bf 12} (2019) 173, [\href{http://arxiv.org/abs/1907.00992}{{\tt
  arXiv:1907.00992}}].

\bibitem{Chester:2019pvm}
S.~M. Chester, {\it {Genus-2 holographic correlator on AdS$_{5}\times$ S$^{5}$
  from localization}},  {\em JHEP} {\bf 04} (2020) 193,
  [\href{http://arxiv.org/abs/1908.05247}{{\tt arXiv:1908.05247}}].

\bibitem{Aprile:2020mus}
F.~Aprile, J.~M. Drummond, H.~Paul, and M.~Santagata, {\it {The
  Virasoro-Shapiro amplitude in AdS$_{5}\times$ S$^{5}$ and level splitting of
  10d conformal symmetry}},  {\em JHEP} {\bf 11} (2021) 109,
  [\href{http://arxiv.org/abs/2012.12092}{{\tt arXiv:2012.12092}}].

\bibitem{Marboe:2018ugv}
C.~Marboe and D.~Volin, {\it {The full spectrum of AdS$_5$/CFT$_4$ II: Weak
  coupling expansion via the quantum spectral curve}},  {\em J. Phys. A} {\bf
  54} (2021), no.~5 055201, [\href{http://arxiv.org/abs/1812.09238}{{\tt
  arXiv:1812.09238}}].

\bibitem{Fiamberti:2008sh}
F.~Fiamberti, A.~Santambrogio, C.~Sieg, and D.~Zanon, {\it {Anomalous dimension
  with wrapping at four loops in N=4 SYM}},  {\em Nucl. Phys. B} {\bf 805}
  (2008) 231--266, [\href{http://arxiv.org/abs/0806.2095}{{\tt
  arXiv:0806.2095}}].

\bibitem{Bajnok:2008qj}
Z.~Bajnok, R.~A. Janik, and T.~Lukowski, {\it {Four loop twist two, BFKL,
  wrapping and strings}},  {\em Nucl. Phys. B} {\bf 816} (2009) 376--398,
  [\href{http://arxiv.org/abs/0811.4448}{{\tt arXiv:0811.4448}}].

\bibitem{Velizhanin:2009gv}
V.~N. Velizhanin, {\it {The Non-planar contribution to the four-loop universal
  anomalous dimension in N=4 Supersymmetric Yang-Mills theory}},  {\em JETP
  Lett.} {\bf 89} (2009) 593--596, [\href{http://arxiv.org/abs/0902.4646}{{\tt
  arXiv:0902.4646}}].

\bibitem{Kniehl:2021ysp}
B.~A. Kniehl and V.~N. Velizhanin, {\it {Non-planar universal anomalous
  dimension of twist-two operators with general Lorentz spin at four loops in
  $N=4$ SYM theory}},  {\em Nucl. Phys. B} {\bf 968} (2021) 115429,
  [\href{http://arxiv.org/abs/2103.16420}{{\tt arXiv:2103.16420}}].

\bibitem{Tseytlin:2003ac}
A.~A. Tseytlin, {\it {On semiclassical approximation and spinning string vertex
  operators in AdS(5) x S**5}},  {\em Nucl. Phys. B} {\bf 664} (2003) 247--275,
  [\href{http://arxiv.org/abs/hep-th/0304139}{{\tt hep-th/0304139}}].

\bibitem{watson2008rankin}
T.~C. Watson, {\it Rankin triple products and quantum chaos},  2008.

\bibitem{Schlenker:2022dyo}
J.-M. Schlenker and E.~Witten, {\it {No Ensemble Averaging Below the Black Hole
  Threshold}},  \href{http://arxiv.org/abs/2202.01372}{{\tt arXiv:2202.01372}}.

\bibitem{Jackiw:1984je}
R.~Jackiw, {\it {Lower Dimensional Gravity}},  {\em Nucl. Phys. B} {\bf 252}
  (1985) 343--356.

\bibitem{Teitelboim:1983ux}
C.~Teitelboim, {\it {Gravitation and Hamiltonian Structure in Two Space-Time
  Dimensions}},  {\em Phys. Lett. B} {\bf 126} (1983) 41--45.

\bibitem{Almheiri:2014cka}
A.~Almheiri and J.~Polchinski, {\it {Models of AdS$_{2}$ backreaction and
  holography}},  {\em JHEP} {\bf 11} (2015) 014,
  [\href{http://arxiv.org/abs/1402.6334}{{\tt arXiv:1402.6334}}].

\bibitem{Witten:2020wvy}
E.~Witten, {\it {Matrix Models and Deformations of JT Gravity}},  {\em Proc.
  Roy. Soc. Lond. A} {\bf 476} (2020), no.~2244 20200582,
  [\href{http://arxiv.org/abs/2006.13414}{{\tt arXiv:2006.13414}}].

\bibitem{Maxfield:2020ale}
H.~Maxfield and G.~J. Turiaci, {\it {The path integral of 3D gravity near
  extremality; or, JT gravity with defects as a matrix integral}},  {\em JHEP}
  {\bf 01} (2021) 118, [\href{http://arxiv.org/abs/2006.11317}{{\tt
  arXiv:2006.11317}}].

\bibitem{Turiaci:2020fjj}
G.~J. Turiaci, M.~Usatyuk, and W.~W. Weng, {\it {Dilaton-gravity, deformations
  of the minimal string, and matrix models}},
  \href{http://arxiv.org/abs/2011.06038}{{\tt arXiv:2011.06038}}.

\bibitem{Benjamin:2021wzr}
N.~Benjamin, C.~A. Keller, H.~Ooguri, and I.~G. Zadeh, {\it {Narain to
  Narnia}},  \href{http://arxiv.org/abs/2103.15826}{{\tt arXiv:2103.15826}}.

\bibitem{Dong:2021wot}
J.~Dong, T.~Hartman, and Y.~Jiang, {\it {Averaging over moduli in deformed WZW
  models}},  \href{http://arxiv.org/abs/2105.12594}{{\tt arXiv:2105.12594}}.

\bibitem{Collier:2021rsn}
S.~Collier and A.~Maloney, {\it {Wormholes and Spectral Statistics in the
  Narain Ensemble}},  \href{http://arxiv.org/abs/2106.12760}{{\tt
  arXiv:2106.12760}}.

\bibitem{Datta:2021ftn}
S.~Datta, S.~Duary, P.~Kraus, P.~Maity, and A.~Maloney, {\it {Adding Flavor to
  the Narain Ensemble}},  \href{http://arxiv.org/abs/2102.12509}{{\tt
  arXiv:2102.12509}}.

\bibitem{Cotler:2016fpe}
J.~S. Cotler, G.~Gur-Ari, M.~Hanada, J.~Polchinski, P.~Saad, S.~H. Shenker,
  D.~Stanford, A.~Streicher, and M.~Tezuka, {\it {Black Holes and Random
  Matrices}},  {\em JHEP} {\bf 05} (2017) 118,
  [\href{http://arxiv.org/abs/1611.04650}{{\tt arXiv:1611.04650}}]. [Erratum:
  JHEP 09, 002 (2018)].

\bibitem{Saad:2018bqo}
P.~Saad, S.~H. Shenker, and D.~Stanford, {\it {A semiclassical ramp in SYK and
  in gravity}},  \href{http://arxiv.org/abs/1806.06840}{{\tt
  arXiv:1806.06840}}.

\bibitem{Blommaert:2021gha}
A.~Blommaert and J.~Kruthoff, {\it {Gravity without averaging}},
  \href{http://arxiv.org/abs/2107.02178}{{\tt arXiv:2107.02178}}.

\bibitem{Blommaert:2021fob}
A.~Blommaert, L.~V. Iliesiu, and J.~Kruthoff, {\it {Gravity factorized}},
  \href{http://arxiv.org/abs/2111.07863}{{\tt arXiv:2111.07863}}.

\bibitem{Mukhametzhanov:2021nea}
B.~Mukhametzhanov, {\it {Half-wormholes in SYK with one time point}},
  \href{http://arxiv.org/abs/2105.08207}{{\tt arXiv:2105.08207}}.

\bibitem{Mukhametzhanov:2021hdi}
B.~Mukhametzhanov, {\it {Factorization and complex couplings in SYK and in
  Matrix Models}},  \href{http://arxiv.org/abs/2110.06221}{{\tt
  arXiv:2110.06221}}.

\bibitem{Saad:2021uzi}
P.~Saad, S.~Shenker, and S.~Yao, {\it {Comments on wormholes and
  factorization}},  \href{http://arxiv.org/abs/2107.13130}{{\tt
  arXiv:2107.13130}}.

\bibitem{Witten:1999xp}
E.~Witten and S.-T. Yau, {\it {Connectedness of the boundary in the AdS / CFT
  correspondence}},  {\em Adv. Theor. Math. Phys.} {\bf 3} (1999) 1635--1655,
  [\href{http://arxiv.org/abs/hep-th/9910245}{{\tt hep-th/9910245}}].

\bibitem{Maldacena:2004rf}
J.~M. Maldacena and L.~Maoz, {\it {Wormholes in AdS}},  {\em JHEP} {\bf 02}
  (2004) 053, [\href{http://arxiv.org/abs/hep-th/0401024}{{\tt
  hep-th/0401024}}].

\bibitem{McNamara:2020uza}
J.~McNamara and C.~Vafa, {\it {Baby Universes, Holography, and the Swampland}},
   \href{http://arxiv.org/abs/2004.06738}{{\tt arXiv:2004.06738}}.

\bibitem{Marolf:2020xie}
D.~Marolf and H.~Maxfield, {\it {Transcending the ensemble: baby universes,
  spacetime wormholes, and the order and disorder of black hole information}},
  {\em JHEP} {\bf 08} (2020) 044, [\href{http://arxiv.org/abs/2002.08950}{{\tt
  arXiv:2002.08950}}].

\bibitem{Eberhardt:2021jvj}
L.~Eberhardt, {\it {Summing over Geometries in String Theory}},  {\em JHEP}
  {\bf 05} (2021) 233, [\href{http://arxiv.org/abs/2102.12355}{{\tt
  arXiv:2102.12355}}].

\bibitem{Heckman:2021vzx}
J.~J. Heckman, A.~P. Turner, and X.~Yu, {\it {Disorder Averaging and its UV
  (Dis)Contents}},  \href{http://arxiv.org/abs/2111.06404}{{\tt
  arXiv:2111.06404}}.

\bibitem{Cotler:2021cqa}
J.~Cotler and K.~Jensen, {\it {Wormholes and black hole microstates in
  AdS/CFT}},  {\em JHEP} {\bf 09} (2021) 001,
  [\href{http://arxiv.org/abs/2104.00601}{{\tt arXiv:2104.00601}}].

\bibitem{Marolf:2021kjc}
D.~Marolf and J.~E. Santos, {\it {AdS Euclidean wormholes}},  {\em Class.
  Quant. Grav.} {\bf 38} (2021), no.~22 224002,
  [\href{http://arxiv.org/abs/2101.08875}{{\tt arXiv:2101.08875}}].

\bibitem{Mahajan:2021maz}
R.~Mahajan, D.~Marolf, and J.~E. Santos, {\it {The double cone geometry is
  stable to brane nucleation}},  {\em JHEP} {\bf 09} (2021) 156,
  [\href{http://arxiv.org/abs/2104.00022}{{\tt arXiv:2104.00022}}].

\bibitem{Cotler:2020ugk}
J.~Cotler and K.~Jensen, {\it {AdS$_{3}$ gravity and random CFT}},  {\em JHEP}
  {\bf 04} (2021) 033, [\href{http://arxiv.org/abs/2006.08648}{{\tt
  arXiv:2006.08648}}].

\bibitem{Hartman:2014oaa}
T.~Hartman, C.~A. Keller, and B.~Stoica, {\it {Universal Spectrum of 2d
  Conformal Field Theory in the Large c Limit}},  {\em JHEP} {\bf 09} (2014)
  118, [\href{http://arxiv.org/abs/1405.5137}{{\tt arXiv:1405.5137}}].

\bibitem{Cardy:1986ie}
J.~L. Cardy, {\it {Operator Content of Two-Dimensional Conformally Invariant
  Theories}},  {\em Nucl. Phys. B} {\bf 270} (1986) 186--204.

\bibitem{Strominger:1997eq}
A.~Strominger, {\it {Black hole entropy from near horizon microstates}},  {\em
  JHEP} {\bf 02} (1998) 009, [\href{http://arxiv.org/abs/hep-th/9712251}{{\tt
  hep-th/9712251}}].

\bibitem{Belin:2020hea}
A.~Belin and J.~de~Boer, {\it {Random statistics of OPE coefficients and
  Euclidean wormholes}},  {\em Class. Quant. Grav.} {\bf 38} (2021), no.~16
  164001, [\href{http://arxiv.org/abs/2006.05499}{{\tt arXiv:2006.05499}}].

\bibitem{Altland:2021rqn}
A.~Altland, D.~Bagrets, P.~Nayak, J.~Sonner, and M.~Vielma, {\it {From operator
  statistics to wormholes}},  {\em Phys. Rev. Res.} {\bf 3} (2021), no.~3
  033259, [\href{http://arxiv.org/abs/2105.12129}{{\tt arXiv:2105.12129}}].

\bibitem{Iliesiu:2021are}
L.~V. Iliesiu, M.~Kologlu, and G.~J. Turiaci, {\it {Supersymmetric indices
  factorize}},  \href{http://arxiv.org/abs/2107.09062}{{\tt arXiv:2107.09062}}.

\bibitem{Argyres:2007cn}
P.~C. Argyres and N.~Seiberg, {\it {S-duality in N=2 supersymmetric gauge
  theories}},  {\em JHEP} {\bf 12} (2007) 088,
  [\href{http://arxiv.org/abs/0711.0054}{{\tt arXiv:0711.0054}}].

\bibitem{Camanho:2014apa}
X.~O. Camanho, J.~D. Edelstein, J.~Maldacena, and A.~Zhiboedov, {\it {Causality
  Constraints on Corrections to the Graviton Three-Point Coupling}},  {\em
  JHEP} {\bf 02} (2016) 020, [\href{http://arxiv.org/abs/1407.5597}{{\tt
  arXiv:1407.5597}}].

\bibitem{Klebanov:1998hh}
I.~R. Klebanov and E.~Witten, {\it {Superconformal field theory on three-branes
  at a Calabi-Yau singularity}},  {\em Nucl. Phys. B} {\bf 536} (1998)
  199--218, [\href{http://arxiv.org/abs/hep-th/9807080}{{\tt hep-th/9807080}}].

\bibitem{Baggio:2014ioa}
M.~Baggio, V.~Niarchos, and K.~Papadodimas, {\it {tt$^{*}$ equations,
  localization and exact chiral rings in 4d $ \mathcal{N} $ =2 SCFTs}},  {\em
  JHEP} {\bf 02} (2015) 122, [\href{http://arxiv.org/abs/1409.4212}{{\tt
  arXiv:1409.4212}}].

\bibitem{Ooguri:2016pdq}
H.~Ooguri and C.~Vafa, {\it {Non-supersymmetric AdS and the Swampland}},  {\em
  Adv. Theor. Math. Phys.} {\bf 21} (2017) 1787--1801,
  [\href{http://arxiv.org/abs/1610.01533}{{\tt arXiv:1610.01533}}].

\bibitem{Aharony:2008ug}
O.~Aharony, O.~Bergman, D.~L. Jafferis, and J.~Maldacena, {\it {N=6
  superconformal Chern-Simons-matter theories, M2-branes and their gravity
  duals}},  {\em JHEP} {\bf 10} (2008) 091,
  [\href{http://arxiv.org/abs/0806.1218}{{\tt arXiv:0806.1218}}].

\bibitem{Larsen:1999uk}
F.~Larsen and E.~J. Martinec, {\it {U(1) charges and moduli in the D1 - D5
  system}},  {\em JHEP} {\bf 06} (1999) 019,
  [\href{http://arxiv.org/abs/hep-th/9905064}{{\tt hep-th/9905064}}].

\bibitem{El-Showk:2012vjm}
S.~El-Showk and M.~F. Paulos, {\it {Bootstrapping Conformal Field Theories with
  the Extremal Functional Method}},  {\em Phys. Rev. Lett.} {\bf 111} (2013),
  no.~24 241601, [\href{http://arxiv.org/abs/1211.2810}{{\tt
  arXiv:1211.2810}}].

\bibitem{El-Showk:2016mxr}
S.~El-Showk and M.~F. Paulos, {\it {Extremal bootstrapping: go with the flow}},
   {\em JHEP} {\bf 03} (2018) 148, [\href{http://arxiv.org/abs/1605.08087}{{\tt
  arXiv:1605.08087}}].

\bibitem{Mazac:2018mdx}
D.~Mazac and M.~F. Paulos, {\it {The analytic functional bootstrap. Part I: 1D
  CFTs and 2D S-matrices}},  {\em JHEP} {\bf 02} (2019) 162,
  [\href{http://arxiv.org/abs/1803.10233}{{\tt arXiv:1803.10233}}].

\bibitem{Mazac:2018ycv}
D.~Mazac and M.~F. Paulos, {\it {The analytic functional bootstrap. Part II.
  Natural bases for the crossing equation}},  {\em JHEP} {\bf 02} (2019) 163,
  [\href{http://arxiv.org/abs/1811.10646}{{\tt arXiv:1811.10646}}].

\bibitem{Afkhami-Jeddi:2021iuw}
N.~Afkhami-Jeddi, {\it {Conformal Bootstrap Deformations}},
  \href{http://arxiv.org/abs/2111.01799}{{\tt arXiv:2111.01799}}.

\bibitem{Razamat:2020pra}
S.~S. Razamat, E.~Sabag, and G.~Zafrir, {\it {Weakly coupled conformal
  manifolds in 4d}},  {\em JHEP} {\bf 06} (2020) 179,
  [\href{http://arxiv.org/abs/2004.07097}{{\tt arXiv:2004.07097}}].

\bibitem{Perlmutter:2020buo}
E.~Perlmutter, L.~Rastelli, C.~Vafa, and I.~Valenzuela, {\it {A CFT Distance
  Conjecture}},  \href{http://arxiv.org/abs/2011.10040}{{\tt
  arXiv:2011.10040}}.

\bibitem{Cotler:2020hgz}
J.~Cotler and K.~Jensen, {\it {AdS$_3$ wormholes from a modular bootstrap}},
  {\em JHEP} {\bf 11} (2020) 058, [\href{http://arxiv.org/abs/2007.15653}{{\tt
  arXiv:2007.15653}}].

\bibitem{Dorigoni:2021rdo}
D.~Dorigoni, M.~B. Green, and C.~Wen, {\it {Exact expressions for n-point
  maximal U(1)$_{Y}$-violating integrated correlators in SU(N) $ \mathcal{N} $
  = 4 SYM}},  {\em JHEP} {\bf 11} (2021) 132,
  [\href{http://arxiv.org/abs/2109.08086}{{\tt arXiv:2109.08086}}].

\bibitem{Nieto:1999kc}
A.~Nieto and M.~H.~G. Tytgat, {\it {Effective field theory approach to N=4
  supersymmetric Yang-Mills at finite temperature}},
  \href{http://arxiv.org/abs/hep-th/9906147}{{\tt hep-th/9906147}}.

\bibitem{Gubser:1996de}
S.~S. Gubser, I.~R. Klebanov, and A.~W. Peet, {\it {Entropy and temperature of
  black 3-branes}},  {\em Phys. Rev. D} {\bf 54} (1996) 3915--3919,
  [\href{http://arxiv.org/abs/hep-th/9602135}{{\tt hep-th/9602135}}].

\bibitem{zbMATH00426391}
D.~A. {Hejhal}, {\it {On eigenvalues of the Laplacian for Hecke triangle
  groups}},  in {\em Zeta functions in geometry}, pp.~359--408.
\newblock Tokyo: Kinokuniya Company Ltd., 1992.

\bibitem{Sarnak1985}
P.~Sarnak and R.~Phillips, {\it On cusp forms for co-finite subgroups of psl(2,
  $\mathbb{R}$)},  {\em Inventiones mathematicae} {\bf 80} (1985) 339--364.

\bibitem{hjudge}
L.~Hillairet and C.~Judge, {\it Hyperbolic triangles without embedded
  eigenvalues},  {\em Annals of Mathematics} {\bf 187} (2018), no.~2 301--377.

\bibitem{Ashok:2016oyh}
S.~K. Ashok, E.~Dell'Aquila, A.~Lerda, and M.~Raman, {\it {S-duality, triangle
  groups and modular anomalies in $ \mathcal{N}=2 $ SQCD}},  {\em JHEP} {\bf
  04} (2016) 118, [\href{http://arxiv.org/abs/1601.01827}{{\tt
  arXiv:1601.01827}}].

\bibitem{McLoughlin:2020zew}
T.~McLoughlin, R.~Pereira, and A.~Spiering, {\it {Quantum Chaos in Perturbative
  super-Yang-Mills Theory}},  \href{http://arxiv.org/abs/2011.04633}{{\tt
  arXiv:2011.04633}}.

\bibitem{Trudgian_2015}
T.~Trudgian, {\it Explicit bounds on the logarithmic derivative and the
  reciprocal of the riemann zeta-function},  {\em Functiones et Approximatio
  Commentarii Mathematici} {\bf 52} (mar, 2015).

\bibitem{Gelbart_riemannszeta}
S.~S. Gelbart, Stephen, and D.~Miller, {\it Riemann's zeta function and
  beyond},  {\em Bull. Amer. Math. Soc. (N.S)} {\bf 41} 015--450.

\bibitem{Green:2005ba}
M.~B. Green and P.~Vanhove, {\it {Duality and higher derivative terms in M
  theory}},  {\em JHEP} {\bf 01} (2006) 093,
  [\href{http://arxiv.org/abs/hep-th/0510027}{{\tt hep-th/0510027}}].

\bibitem{Green:2010kv}
M.~B. Green, S.~D. Miller, J.~G. Russo, and P.~Vanhove, {\it {Eisenstein series
  for higher-rank groups and string theory amplitudes}},  {\em Commun. Num.
  Theor. Phys.} {\bf 4} (2010) 551--596,
  [\href{http://arxiv.org/abs/1004.0163}{{\tt arXiv:1004.0163}}].

\bibitem{Green:2011vz}
M.~B. Green, S.~D. Miller, and P.~Vanhove, {\it {Small representations, string
  instantons, and Fourier modes of Eisenstein series}},  {\em J. Number Theor.}
  {\bf 146} (2015) 187--309, [\href{http://arxiv.org/abs/1111.2983}{{\tt
  arXiv:1111.2983}}].

\bibitem{Dorigoni:2021jfr}
D.~Dorigoni, A.~Kleinschmidt, and O.~Schlotterer, {\it {Poincar\'e series for
  modular graph forms at depth two. I. Seeds and Laplace systems}},
  \href{http://arxiv.org/abs/2109.05017}{{\tt arXiv:2109.05017}}.

\bibitem{Dorigoni:2021ngn}
D.~Dorigoni, A.~Kleinschmidt, and O.~Schlotterer, {\it {Poincar\'e series for
  modular graph forms at depth two. II. Iterated integrals of cusp forms}},
  \href{http://arxiv.org/abs/2109.05018}{{\tt arXiv:2109.05018}}.

\bibitem{klingerlogan2018differential}
K.~Klinger-Logan, {\it {Differential equations in automorphic forms}},
  \href{http://arxiv.org/abs/1801.00838}{{\tt arXiv:1801.00838}}.

\end{thebibliography}\endgroup

\end{document}